**Physics-Based Machine Learning Approach for Modeling**

**the Temperature-Dependent Yield Strength of Superalloys**

B. Steingrimsson [*1, 2], X. Fan[3], B. Adam[4], and P. K. Liaw [5]

1.     Imagars LLC, 2062 Thorncroft Drive Suite 1214, Hillsboro, OR 97124, USA. Email: baldur@imagars.com

2.     Department of Materials Science and Engineering, The University of Tennessee, Knoxville, TN, 37996, USA. Email: bsteingr@vols.utk.edu

3.     Department of Materials Science and Engineering, The University of Tennessee, Knoxville, TN, 37996, USA.

4.     School of Mechanical, Industrial and Manufacturing Engineering, Oregon State University, Corvallis, OR 97331, USA.

5.     Department of Materials Science and Engineering, The University of Tennessee, Knoxville, TN, 37996, USA.

[*] Corresponding authors: baldur@imagars.com, bsteingr@vols.utk.edu.

**Abstract**

In the pursuit of developing high-temperature alloys with improved properties for meeting the performance requirements of next-generation energy and aerospace demands, integrated computational materials engineering (ICME) has played a crucial role. In this paper a machine learning (ML) approach is presented, capable of predicting the temperature-dependent yield strengths of superalloys, utilizing a bilinear log model. Importantly, the model introduces the parameter break temperature, $T_{break}$, which serves as an upper boundary for operating conditions, ensuring acceptable mechanical performance. In contrast to conventional black-box approaches, our model is based on the underlying fundamental physics, directly built into the model. We present a technique of global optimization, one allowing the concurrent optimization of model parameters over the low-temperature and high-temperature regimes. The results presented extend previous work on high-entropy alloys (HEAs) and offer further support for the bilinear log model and its applicability for modeling the temperature-dependent strength behavior of superalloys as well as HEAs.

## Introduction

The conventional approach to alloy design assumes selecting a main component, based on the primary property needed in the alloy, with one or more additions to provide secondary properties. For over six decades, superalloys have provided reliable and cost-effective means for achieving high operating temperatures and stress conditions, first in aircraft and later in industrial gas turbines [1]. The zenith of superalloy development was in the 1960s, as columnar grain alloys and single crystals were then made feasible, and many polycrystalline alloys matured to commercial reality [1]. Since then, advances in the powder metallurgy (P/M) processing and oxide dispersion strengthening have ensued, along with advances in additive manufacturing (AM) and integrated computational materials engineering (ICME).

In the case of Nickel-based superalloys, Nickel is selected for the main component, because of its high melting point. Nevertheless, chromium is added to prevent corrosion, and titanium and aluminum are included to increase strength. The Ni-based superalloys can accommodate γ, γ', γ″, η, carbide, or topologically-closed-packaged (TCP) phases, as explained in the Supplementary Manuscript. Co-based superalloys can similarly contain γ, γ', carbide, or TCP phases. The Co-based superalloys are usually strengthened by a combination of carbides and solid-solution hardeners [1]. The use of steels for applications necessitating superalloys may also be of interest, as further explained in the Supplementary Manuscript, because certain steels have exhibited creep and oxidation resistance analogous to that of the Ni-based superalloys, while being much less expensive to manufacture. Similar to the Ni- and Co-based superalloys, the Fe-based superalloys can accommodate a γ matrix phase along with a γ' phase. The γ' phase is introduced in the form of precipitates to strengthen the Fe-based superalloys.

Machine learning (ML) and data analytics, coupled with physics-based modeling, can assist with the effective search of the compositional space of superalloys and with the optimization of thermomechanical post-processes. Figure 1 outlines the multiple sources that affect the mechanical properties of superalloys. As opposed to specifically applying ML, narrowly-defined in terms of multi-variate regression, single-layer or multi-layer neural networks, Bayesian graphical models, support vector machines, or decision trees, to the identification of superalloy compositions of interest, we reformulate the task in the broader context of engineering optimization [2], [3], [4]. We recommend selecting an optimization technique suitable for the application at hand and the data available, but we certainly include ML in the consideration. Our rational is based in part on observations from Agrawal et al. [5]. According to Table 2 and Figure 5 of [5], there is at most a difference of a few percentages between the techniques applied to model the fatigue strengths of stainless steels. For further background information on the application of ML to modeling of mechanical properties of alloys, and the motivation for coupling with physics-based modeling approaches, refer to [2], [3], [4].

This paper introduces a bilinear log model for predicting the yield strength of superalloys across temperatures. It consists of separate exponentials, for a low-temperature and a high-temperature regime, with a break temperature, $T_{\text{break}}$, in between. The underlying physics are accounted for, e.g., through diffusion processes required to initiate phase transformations in the high-temperature regime [6]. Moreover, we show in [3] how piecewise linear regression can be employed to extend the model beyond two exponentials and yield accurate fit, in case of a non-convex objective function caused by hump(s) in the data. Previous models for the temperature dependence of yield strengths (YS) only accounted for a single exponential [7], [8]. There was, therefore, no break temperature, $T_{\text{break}}$. We regard the break temperature very important for the optimization of the high-temperature properties of alloys. For the reliable operation, the temperature of turbine blades or disks (rotors) made from refractory superalloys may need to stay below $T_{break}$. Once the temperature increases beyond $T_{break}$, the superalloys can lose strengths rapidly due to rapid diffusion, leading to the easy dislocation motion and dissolution of strengthening phases [6]. We consider $T_{break}$ a very important parameter for the design of superalloys with attractive high-temperature properties, one warranting the inclusion in the superalloy specifications. Hence, it is of paramount importance to be able to accurately estimate $T_{break}$, e.g., using the global-optimization approach presented in [3].

In terms of impacts, this paper addresses a physics-based, i.e., not a black box, approach to ML and data analytics. Preference to such approaches, especially for applications involving materials science, may primarily stem from the fact that they involve fewer parameters than the black-box models, and hence require less input data. With experiments often being time consuming and expensive, in the case of such applications, high-quality input data tends to be a scarce resource. Moreover, the physics-based modeling approaches help establish causal links between output observations and the behavior of the underlying material system.

**Results**

Figure 1 summarizes the multiple sources that impact the mechanical properties of superalloys. It bears noting that improvements in the yield strength may come at the expense of other properties. Hence, we have suggested a framework for *joint optimization* [2], [3], [4]. For instance, there typically is a trade-off between the ductility and the strength of the superalloys. Figure 1 resembles an inverse design process, where components of the design are explicitly calculated from the target performance metrics provided.

For forty superalloy compositions, Table 1 compares the performance of the bilinear log model to that of a model based on a single exponential. When averaged over these forty compositions, the bilinear log model results in a mean squared error (MSE) of 0.00175 in the logarithmic domain, compared to the MSE of 0.0525 for the model based on a single exponential. This difference is much exacerbated in the linear domain, where the bilinear log model yields the MSE of only 4,151 MPa$^2$, compared to the MSE of 25,836 MPa$^2$ for the model based on a single exponential. Analogous to Wu et al. [7], we employ a model with a single exponential for the overall YS, $\sigma_y$ ($T$), of the form

$$\sigma_y(T) = \sigma_a \exp\left(\frac{-T}{C}\right) + \sigma_b, \tag{1}$$

where $\sigma_a$, $C$, and $\sigma_b$ represent fitting coefficients.

In an effort to identify superalloy compositions exhibiting the ability to retain strengths at high temperatures, we present Figure 2 - Figure 3. The strengths of most metals tend to decrease, as the temperature is increased, since as the temperature is increased, it becomes easier for dislocations to surmount obstacles, through the support from thermal activation [9]. In the case of high-temperature applications, one may look to derive a model of the form

$$YS = h\,(\text{composition}, T) \tag{2}$$

for the prediction of the YS of the superalloys across temperature.

Looking at Figure 2 - Figure 4, one first observes that the strength vs. temperature data definitely does not look linear. Therefore, a multi-variate linear regression may not constitute a preferred approach. Second, the dependence of the strength on temperature does come across as approximately exponential, but not exactly. Figure 2 -Figure 4 seem to feature a high-temperature and a low-temperature regime. Third, one may refrain from employing an automated ML suite, such as the Tree-Based Pipeline Optimization Tool (T-POT) [10], because of the limited ability of such black-box models to provide much needed insights into the underlying physics. Instead, one is motivated to make the most of the limited data available, by incorporating important a priori information about the underlying physics into the model structure, for the purpose of deriving such insights. For specifics on the algorithmic aspects of the incorporation of a priori information into the bilinear log model, refer to the Method section. The results in Figure 2 - Figure 4 were derived, using the global-optimization approach, applied separately to individual alloys, for the purpose of obtaining a tighter fit and more accurate estimation of $T_{break}$, than for separate optimization over the low and high temperature regimes.

In regards to the underlying physics, Ni-based superalloys containing γ' phases, tend to be particularly resistant to temperature with reference to their in-service mechanical properties [9]. Strengthening generally occurs through both alloying elements acting as solutes in the face-centered-cubic (FCC) austenitic matrix phase, and through the solute precipitation strengthening, such as the γ' phase. This phase is responsible for the yield strength of many nickel-based superalloys being relatively insensitive to temperature. The effect arises from dislocation locking on secondary {100} slip planes caused by cross-slipping of primary dislocations on {111} slip planes in the γ' phases during plastic deformation. Ordinary dislocation slip occurs on the {111}<110> plane, both for the γ and the γ' phases. There is, however, a tendency for dislocations in the γ' phase to cross-slip on to secondary{100} planes, where they can have a lower anti-phase domain boundary energy (APB). The activation of these secondary slip planes is thermally activated as is the APB, crucial for maximizing the γ' strength, where both decrease with an increase in temperature. This trend explains then the signature peak in the flow stress, seen for γ'-strengthened superalloys [11], [12].

When great strength is required at lower temperatures, e.g., for turbine discs, precipitation strengthening in Ni-based superalloys can be still be utilized, either by employing other phases, such as the γ″-phase in Nb-bearing alloys or multimodal size distribution of the γ′-phase. The γ″ phase occurs in Ni-based superalloys, such as Inconel 718, which have significant addition of Niobium or Vanadium. The composition of the γ″ phase is $Ni_3Nb$ or $Ni_3V$. The γ″ phase consists of a body-centered-tetragonal lattice with an ordered arrangement of Nickel and Niobium atoms. Strengthening occurs primarily through coherency-hardening and order-hardening mechanisms [9]. Further, in regards to dislocation dynamics and diffusion mechanisms, ordinary slips occur on the {111}<110> plane, both for the γ and the γ′ phases. There is, however, a tendency for dislocations in the γ′ phase to cross-slip on to {100} planes, where they can have a lower anti-phase domain boundary energy. Situations can arise where an extended dislocation is partly on a close-packed plane and partly on a cube plane. Such a dislocation can become locked and result in an increase in strength (can help retain strength in the low-temperature regime). In the elevated-temperature regime, thermal activations tend to be sufficiently violent to allow dislocations to overcome obstacles [9].

Figure 5 offers a comparison of temperature-dependent yield strengths across superalloy compositions. Similar to [1], we consider the three primary categories of superalloys: (1) Fe-Ni-, (2) Ni-, and (3) Co-based superalloys. Fe-Ni-based superalloys, such as the popular Inconel 718, came into existence as an extension of the stainless-steel technology and are generally heat treated (wrought). Co- and Ni-based superalloys may be heat treated or not (wrought or cast), depending on the application and composition involved. Figure 5 illustrates the following:

1. The precipitation-strengthened superalloys seem to result in higher YS than the solid-solution-strengthened ones, in case of both the Ni-based and the Fe-Ni-based superalloys.

2. The Ni-based superalloys tend to produce higher YS than the Fe-Ni-based superalloys, in case of both precipitation and solid-solution strengthening.

As far as the superior YS of the precipitation-strengthened superalloys is concerned, Figure 5 exhibits resemblance and consistency with Figure 1.1 of [1]. Superalloys can be strengthened through solid-solution strengthening (and its chemistry), but also through the presence of strengthening phases (usually precipitates) [1]. Co-based superalloys can exhibit a tendency towards the transformation of the FCC matrix phase to stable lower-temperature phases. In the case of Ni-based and Fe-Ni-based superalloys, the FCC matrices can exhibit favorable characteristics for precipitation of uniquely effective strengthening phases [1].

**Discussion and Conclusions**

When looking at the superalloy compositions from the Supplementary Manuscript, the following observations stand out:

1. Many of the superalloy specimens, such as Astroloy, Hastelloy S, Hastelloy X, Waspaloy. Nimonic PE16, Rene 41, Nimonic 80a, Nimonic 90, Rene 95, Nickel 200, Nimonic 263, Udimet 520, Udimet 500, Haynes 556, Inconel 600, Inconel 601, Incoloy 825 and Inconel 807, present fairly straight forward cases, where the bilinear log model offers much better performance than the model based on a single exponent.

2. For some of the superalloy specimen, such as Nimonic PE 13, Nickel 201, Udimet 720, Inconel 801, Inconel X750, and Hastelloy C22, accurate modeling seems to be precluded by a lack of data points in the high-temperature regime.

3. Some of the superalloy specimen, such as Udimet D979, Inconel 802, Inconel 800, Inconel 625, Inconel 617 (sheet), Haynes 188 and Nimonic 75, seem to exhibit behavior featuring a flat intermediate regime.

4. At least one of the superalloy specimens (Inconel 718) may exhibit a trilinear log behavior.

5. Both configurations of the Haynes 230 superalloys, as well as the L 605 superalloy, seem to exhibit an anomalous yield-strength phenomenon (a hump between the low-temperature and high-temperature regimes) [3].

The good match of the bilinear log model with the temperature-dependent strength data for Astroloy, Hastelloy S, Hastelloy X, Waspaloy, Nimonic PE16, Rene 41, Nimonic 80a, Nimonic 90, Rene 95, Nickel 200, Nimonic 263, Udimet 520, Udimet 500, Haynes 556, Inconel 600, Inconel 601, Incoloy 825, and Inconel 807 is consistent with the observations from Figure 2 - Figure 3, from [3] and from [13]. The models utilized here are presented in the Method section.

The experimental data for the superalloy specimens of Nimonic PE 13, Nickel 201, Udimet 720, Inconel 801, Inconel X750, and Hastelloy C22 did not allow for the estimation of an exponent for the high-temperature regime. In order to estimate the high-temperature exponent, one needs at least two data points above $T_{break}$.

A flat intermediate regime, such as observed for Udimet D979, Inconel 802, Inconel 800, Inconel 625, Inconel 617 (sheet), Haynes 188, and Nimonic 75, has also been reported for medium- and high-entropy alloys (MEAs and HEAs) [13]. These cases may necessitate the application of the tri-linear log model from Eqs. (13) – (16). Reference [14] offers an explanation for a tri-linear behavior for intrinsic temperature-dependent yield strengths in BCC metals and substitutional solid solutions:

1. Plastic deformation of the BCC metals and alloys is believed to be mediated by the thermally-activated formation and movement of kink pairs, at low temperatures [14].

2. Above a certain "critical temperature," a flat strength plateau is achieved, where the strength becomes virtually temperature or strain-rate independent [14].

3. When the test temperature exceeds approx. 40% of the melting temperature, diffusional processes can lead to a rapid decrease of yield stress, and the YS becomes strain-rate dependent again [14].

We are looking at two break temperatures, $T_{break1}$ and $T_{break2}$, in the case of the tri-linear log model [3]. But Ref. [14] only references one such temperature, and refers to it as the "critical temperature" or "knee temperature", which apparently corresponds to $T_{break1}$ in our model.

The flat intermediate plateau seems less prevalent in alloys with FCC crystal structures. The FCC-crystal structure possesses 12 slip planes with 4 closest packed planes of {111} and 3 closest packed directions per plane of <110>. The HCP crystal structure, on the other hand, only has 3 slip systems. The BCC crystal structure does not have truly closest packed planes, so the slips must be thermally activated in the BCC metals. For the further analysis of slip systems, Peierls barriers, dislocation mechanisms, flow stress, and temperature-dependent stress behavior, in BCC vs. FCC metals, refer to the work of Seeger [15], [16]. In the specific case of Inconel 718, the need for a tri-linear log model may be hard to ascertain, due to the possible uncertainty in the experimentally measured data points.

The hump between the low-temperature and the high-temperature regimes (the anomalous yield-strength phenomenon) is characteristic for the γ'- or γ''-strengthened superalloys [3], [12]. The phenomenon manifests itself as a positive temperature dependence of the total yield strength, and is found in superalloys strengthened by $L1_2$ ordered intermetallics [17]. The increased strength of the γ' phase with temperature is explained by a thermally-activated cross slip of dislocations from {111} planes to {100} planes, where the partial dislocation locking occurs [18], [19]. The influence of the γ' strengthening increases with increasing the γ' volume fraction and increasing temperature below the peak yield strength, as seen in Figure S5 [11], [20]. Additional factors involve the γ' size, γ' solute content, γ-γ' mismatch, and both the stacking fault energy (SF) as well as the anti-phase boundary (APB) energy. The latter two factors are the major contributions to hardening for both the matrix phase as well as the precipitate phase, and are both strongly affected by alloying additions [21].

Supplementary datasets for the above-mentioned factors were collected for all alloys, seen in Table S1, where possible, for the solution-annealed and peak-aged condition, maximizing the precipitation-strengthening effect. In reviewing the data on (if present) the strengthening-precipitate type, volume fraction, and average size as well as the lattice misfit for all alloys, the significant scatter becomes apparent, resulting from the lack of data, lack in consistency in the methodology and processing. While this feature provides challenges to data extraction and comparability, it also shows the complex precipitation and growth as function of prior processing and heat treatment, as well as the need for incorporating physics-based prediction models that can account for multimodal-precipitate distributions [22], [23].

For example, this trend is illustrated by the alloy CMSX-4, a $\gamma'$-strengthened, single-crystal, Nickel-based superalloy, where three exponentials are needed for accurate modeling in case of Heat Treatment A, but four exponentials in case of Heat Treatment B, according to the Supplementary Figure 4 [3], [17].

In conclusion, we proposed a bilinear log model for predicting the YS of superalloys across temperatures and studied its effectiveness for 38 distinct compositions (40 configurations). We considered the break temperature, $T_{break}$, an important parameter for the design of superalloys materials with favorable elevated-temperature properties, one warranting the inclusion in the superalloy alloy specifications. For reliable operations, and not accounting for coatings, the operating temperatures for the corresponding superalloys may need to stay below $T_{break}$. Earlier models for the temperature dependence of the yield strength only accounted for a single exponential, and thus did not feature a break temperature. Across these 40 configurations, we demonstrated much superior performance of the bilinear log model, compared to the one comprising a single exponential. In the logarithmic domain, the bilinear log model resulted in MSE of 0.00175, compared to the MSE of 0.0525 for the single-exponent model. But in the linear domain, the bilinear log model produced MSE of 4,151 $MPa^2$, averaged across these 40 configurations, compared to the MSE of 25,836 $MPa^2$ for the single-exponent model. Our observations regarding the superior YS behavior of precipitation-strengthened superalloys, compared to solid-solution strengthened superalloys, for both the Ni-based and Fe-Ni-based superalloys, were consistent with those of Donachie et al. [1]. Similarly, our observations concerning superior YS properties of Ni-based superalloys, compared to Fe-Ni-based superalloys, in the case of both precipitation and solid-solution hardening, were also consistent with those of Donachie et al. [1].

**Methods**

Although the primary emphasis here is on the yield strength, the optimization of the mechanical properties is assumed to take place within a framework for joint optimization [3], [13].

*Methodology for Maximization of the YS*

Our approach accurately captures the input sources that contribute to variations in the YS observed, relative to variations in the output. The input parameters for the model used in the present work are [3]:

$$\text{Input} = (\text{composition}, T, \text{process}, \text{defects}, \text{grain size}, \text{microstructure}). \quad (3)$$

In this context, "defects" are defined broadly, such as to include inhomogeneities, impurities, dislocations, or further undesirable features, while "$T$" represents temperature. Analogously, the term "microstructure" broadly represents microstructures here, at the nano or micro scale, and phase properties, and the term "process" broadly refers to manufacturing processes and post-processing. Similarly, the term "grain size" generally refers to the distribution of grain sizes. Section 4.4 of [4] accounts for the dependence between input sources, and Section 4.5 of [4] addresses the expected dependence of the US on the individual input sources listed. The prediction model can be summarized as [3], [13]

$$US = h[ \text{ composition}, T, \text{process}, \text{defects}(\text{process}, T), \text{grains}(\text{process}, T), \text{microstructure}(\text{process}, T) ]. \quad (4)$$

If the YS corresponding to a certain input combination is known, one can simply look up the known value. However, if the YS corresponding to a given input combination is not known, then a prediction step can be applied (e.g., interpolation or extrapolation) [13].

*Methodology for Modeling YS at Elevated Temperature*

Motivated by Figures 3(c) and (d) of [3], along with physics-based insights from [6], we model the temperature dependence of the YS($T$), in terms of a bilinear log model, parametrized by the melting temperature, $T_m$, as follows [13]:

$$YS(T) = \min( \log(YS_1(T)), \log(YS_2(T)) ), \quad (5)$$

$$YS_1(T) = \exp( - C_1 * T / T_m + C_2), \qquad 0 < T < T_{break}, \quad (6)$$

$$YS_2(T) = \exp( - C_3 * T / T_m + C_4), \qquad T_{break} < T < T_m. \quad (7)$$

$T_{break}$ is subjected to a separate physics (diffusion)-induced constraint [6]:

$$0.35 \lesssim T_{break} / T_m \lesssim 0.55, \quad (8)$$

as well as a continuity constraint between the low-temperature and high-temperature regimes:

$$YS_1(T_{break}) = YS_2(T_{break}); \quad (9)$$

$$T_{break} = \frac{( C_4 - C_2 )}{(C_3 - C_1)} T_m. \quad (10)$$

As elucidated in [3], [13], a conceptually simple approach for fitting the model in Eqs. (5) – (10) to the YS data available consists of first deriving the constant coefficients, $C_1$ and $C_2$, by applying linear regression to the data points available to the lowest temperature region ($0 < T < 0.35\ T_m$) as well as to the intermediate region ($0.35\ T_m \leq T \leq 0.55\ T_m$). One can then obtain the constants, $C_3$ and $C_4$, by applying linear regression to the data points available to the intermediate ($0.35\ T_m \leq T \leq 0.55\ T_m$) and high-temperature ($T > 0.55\ T_m$) regions. Note that $T_{break}$ does not have to be known in advance. As supported by Eq. (10), $T_{break}$ is an inherent property of a given alloy that emerges from the model as the break point between the two linear regions. The model in Eqs. (5) – (7) comprises of only four (4) independent parameters, $C_1$, $C_2$, $C_3$, and $C_4$, which simply can be estimated by applying linear regression separately to low-temperature and high-temperature regimes, even to a fairly small data set. Note, furthermore, that for a new alloy system, $T_m$, does not have to be known experimentally in advance either; a rough estimate for $T_m$ may be obtained, by using "the rule of mixing" and a more refined estimate obtained, by employing the Calculation of Phase Diagram (CALPHAD) simulations [4].

A superior method for deriving the coefficients, $C_1$, $C_2$, $C_3$, and $C_4$, entails the concurrent optimization over the low-temperature and high-temperature regimes, using the global optimization. Here, one looks to minimize [13]

$$\min_{C_1, C_2, C_3, C_4} \operatorname*{norm}_2 \left( (YS(T_i) - y_i)^2 \right), \tag{11}$$

where $y_i$ represents the YS values measured,

$$YS(T_i) = \min(\ \log(YS_1(T_i)),\ \log(YS_2(T_i))\ ), \tag{12}$$

and $YS_1(T_i)$ and $YS_2(T_i)$ are modeled, employing Eqs. (6) and (7), respectively. Matlab provides a function, fminunc(), for solving this sort of unconstrained minimization over a generic function. Depending on the grain sizes and the compositions comprising the alloys, a tri-linear log model may offer a better fit for specific alloys [3], [13], [24]:

$$YS(T) = \min(\ YS_1(T),\ YS_2(T),\ YS_3(T)\ ), \tag{13}$$

$$\text{YS}_1(T) = \exp(-C_1 * T / T_m + C_2), \qquad 0 < T < T_{\text{break1}}, \qquad (14)$$

$$\text{YS}_2(T) = \exp(-C_3 * T / T_m + C_4), \qquad T_{\text{break1}} < T < T_{\text{break2}}, \qquad (15)$$

$$\text{YS}_3(T) = \exp(-C_5 * T / T_m + C_6), \qquad T_{\text{break2}} < T < T_m. \qquad (16)$$

## Data Availability

The data in this paper, including those in the Supplementary Figures, can be requested by contacting the corresponding author (baldur@imagars.com, bsteingr@vols.utk.edu).

## Code Availability

Matlab comprises the software package primarily used for this study. The Supplementary Figures 7 and 8 of [3] contain an Matlab source code for the objective functions optimized in case of the bilinear or trilinear log model, respectively.

## Acknowledgements

XF and PKL very much appreciate the support of the U.S. Army Research Office Project (W911NF-13-1-0438 and W911NF-19-2-0049) with the program managers, Drs. M. P. Bakas, S. N. Mathaudhu, and D. M. Stepp. PKL thanks the support from the National Science Foundation (DMR-1611180, 1809640, and 2226508) with the program directors, Drs. J. Madison, J. Yang, G. Shiflet, and D. Farkas. XF and PKL also appreciate the support from the Bunch Fellowship. XF and PKL would like to acknowledge funding from the State of Tennessee and Tennessee Higher Education Commission (THEC) through their support of the Center for Materials Processing (CMP). BS very much appreciates the support from the National Science Foundation (IIP-1447395 and IIP-1632408), with the program directors, Drs. G. Larsen and R. Mehta, from the U.S. Air Force (FA864921P0754), with J. Evans as the program manager, and from the U.S. Navy (N6833521C0420), with Drs. D. Shifler and J. Wolk as the program managers.

The authors also want to thank Dr. Graham Tewksbury for offering insightful comments on strengthening mechanisms in superalloys during an October 21, 2021, material science graduate seminar presentation, conducted by BS at Oregon State University. The authors similarly want to thank Dr. Chanho Lee for bringing precursors to Supplementary Figures S1 – S3 to their attention.

## Author Contributions

B.S. and P.K.L conceived the project. B.S. performed the modeling of the temperature-dependent yield strength and put together the Supplementary Manuscript. X.F. helped prepare the database. R.F. contributed towards analytical modeling. All authors edited and proofread the final manuscript and participated in discussions.

## Competing Interests

The authors declare no competing interests (financial or non-financial).

**Table 1:** Quantification of ability of superalloy compositions to retain yield strengths at high temperatures. $T_{break}$ refers to the breaking point between bilinear log models, defined in Eq. (10).

| No. | Alloy | Category | Solvus Temperature [°C] | $T_{break}$ [°C] | MSE: Two exponentials | | MSE: Single exponential | |
|---|---|---|---|---|---|---|---|---|
| | | | | | Log | Linear | Log | Linear |
| 1 | Waspaloy | Ni-base (precipit. hard.) | 1,330 | 755.2 | 6.367e-04 | 5,118 | 0.0070 | 44,260 |
| 2 | Udimet D979 | Fe-Ni-base (precipit. hard.) | 1,390 | 717.7 | 0.005391 | 7,678 | 0.0138 | 36,204 |
| 3 | Udimet 720 | Ni-base (precipit. hard.) | 1,371 | Not enough high-temp data pts. | | | 0.00056 | 243,720 |
| 4 | Udimet 520 | Ni-base (precipit. hard.) | 1,371 | 728.6 | 5.663e-05 | 2,221 | 0.0162 | 50,701 |
| 5 | Udimet 500 | Ni-base (precipit. hard.) | 1,360 | 748.9 | 7.901e-05 | 1,331 | 0.0181 | 48,108 |
| 6 | Astroloy | Ni-base (precipit. hard.) | 1,399 | 760.6 | 5.759e-04 | 6,643 | 0.0091 | 80,097 |
| 7 | Incoloy 825 | Fe-Ni-base (solid sol.) | 1,370 | 727.6 | 3.196e-07 | 14 | 0.0367 | 2,198 |
| 8 | Inconel 807 | Fe-Ni-base (solid sol.) | 1,275 | 774.1 | 1.069e-04 | 1,358 | 0.00136 | 3,442 |
| 9 | Inconel 802 | Fe-Ni-base (solid sol.) | 1,372 | 760.0 | 0.00252 | 1,049 | 0.00495 | 2,248 |
| 10 | Inconel 801 | Fe-Ni-base (solid sol.) | 1,372 | Not enough high-temp data pts. | | | 0.00009 | 10,504 |
| 11 | Inconel 800 | Fe-Ni-base (solid sol.) | 1,357 | 675.8 | 4.131e-04 | 637 | 0.00163 | 2,364 |
| 12 | Incoloy 800 | Fe-Ni-base (solid sol.) | 1,350 | Not enough high-temp data pts. | | | 0.00623 | 699.8 |
| 13 | Inconel X750 | Ni-base (precipit. hard.) | 1,393 | Not enough high-temp data pts. | | | 0.00000 | 90,438 |
| 14 | Inconel 718 | Ni-base (precipit. hard.) | 1,260 | 717.3 | 0.01707 | 7,622 | 0.00819 | 58,230 |
| 15 | Inconel 625 | Ni-base (solid sol.) | 1,290 | 766.8 | 2.911e-04 | 598 | 0.01615 | 14,501 |
| 16 | Inconel 617 (sheet) | Ni-base (solid sol.) | 1,332 | 893.4 | 0.00201 | 2,790 | 0.00235 | 4,010 |
| 17 | Inconel 601 | Ni-base (solid sol.) | 1,360 | 735.0 | 7.862e-04 | 99 | 0.18572 | 7,360 |
| 18 | Inconel 600 | Ni-base (solid sol.) | 1,354 | 760.1 | 1.639e-04 | 10 | 0.22850 | 4,314 |
| 19 | Ti6Al4V | α and β phases | 1,604 | 501.6 | 1.793e-04 | 134 | 0.01299 | 12,367 |
| 20 | Haynes 556 | Fe-Ni-base (solid sol.) | 1,371 | 775.6 | 1.039e-04 | 1,390 | 0.00107 | 3,771 |
| 21 | Haynes 230 (a) | Ni-base (solid sol.) | 1,288 | 798.4 | 0.002124 | 1,591 | 0.00376 | 6,176 |
| 22 | Haynes 230 (b) | Ni-base (solid sol.) | 1,288 | 829.3 | 0.002035 | 260 | 0.12314 | 7,681 |
| 23 | Haynes 188 | Co-base (solid sol.) | 1,315 | 932.8 | 0.001307 | 4,807 | 0.00138 | 6,002 |
| 24 | Hastelloy C22 | Ni-base (solid sol.) | 1,357 | Not enough high-temp data pts. | | | 0.00020 | 4,647 |
| 25 | Hastelloy S | Ni-base (solid sol.) | 1,335 | 765.2 | 8.509e-05 | 454 | 0.00859 | 6,211 |
| 26 | Hastelloy X | Ni-base (solid sol.) | 1,260 | 837.6 | 0.002745 | 212 | 0.13881 | 5,298 |
| 27 | Hastelloy X (10 μm) | Ni-base (solid sol.) | 1,260 | 744.4 | 0.003165 | 137 | 0.24225 | 11,620 |
| 28 | Rene 95 | Ni-base (precipit. hard.) | 1,343 | 627.6 | 0.000001 | 80,757 | 0.00135 | 217,771 |
| 29 | Rene 41 | Ni-base (precipit. hard.) | 1,316 | 748.1 | 0.000009 | 663 | 0.03624 | 79,716 |
| 30 | Pure Nickel | Ni-base | 1,455 | Not enough high-temp data pts. | | | 0.00892 | 64 |
| 31 | Nickel 201 | Ni-base (solid sol.) | 1,435 | Not enough high-temp data pts. | | | 0.00246 | 132 |
| 32 | Nickel 200 | Ni-base (solid sol.) | 1,435 | 349.4 | 0.001337 | 22 | 0.01677 | 258 |
| 33 | Nimonic 263 | Ni-base (precipit. hard.) | 1,300 | 758.8 | 0.000262 | 82 | 0.08506 | 15,601 |
| 34 | Nimonic 90 | Ni-base (precipit. hard.) | 1,310 | 724.8 | 7.875e-05 | 74 | 0.08111 | 29,514 |
| 35 | Nimonic 80a | Ni-base (precipit. hard.) | 1,320 | 754.3 | 0.000351 | 156 | 0.05091 | 19,933 |
| 36 | Nimonic 75 | Fe-Ni-base (solid sol.) | 1,340 | 731.2 | 0.001561 | 92 | 0.06840 | 1,391 |
| 37 | Nimonic PE16 | Fe-Ni-base (precipit. hard.) | 1,310 | 731.6 | 2.512e-05 | 7 | 0.1399 | 16,557 |
| 38 | Nimonic PE13 | Fe-Ni-base (solid sol.) | 1,310 | Not enough high-temp data pts. | | | 0.00131 | 6,424 |
| 39 | N155 | Fe-Ni-base (solid sol.) | 1,288 | 847.5 | 0.008876 | 687 | 0.0670 | 2,998 |
| 40 | L605 | Co-base (solid sol.) | 1,330 | Not enough high-temp data pts. | | | 0.00782 | 3,666 |

**Average** (No. 1-2, 4-9, 11, 14-23, 25-30, 32-37 and 39): 0.00175 | 4,151 | 0.0525 | 25,836

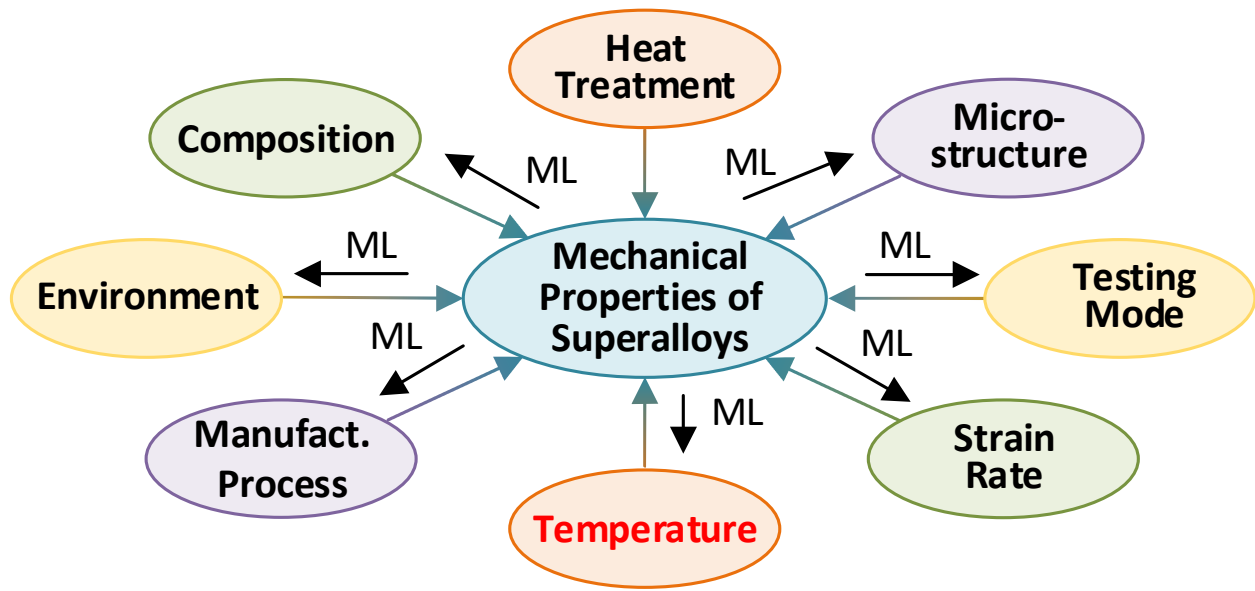

**Figure 1:** High-level depiction of the role of ML and optimization inferring the features giving rise to the superalloy properties observed, including the yield strength [2], [4].

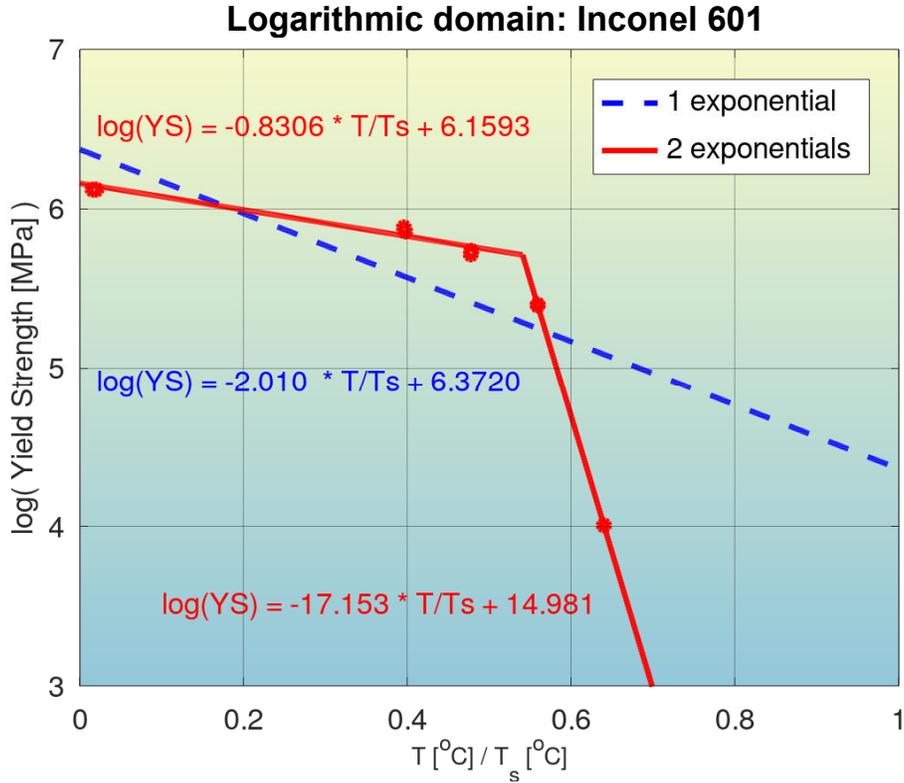

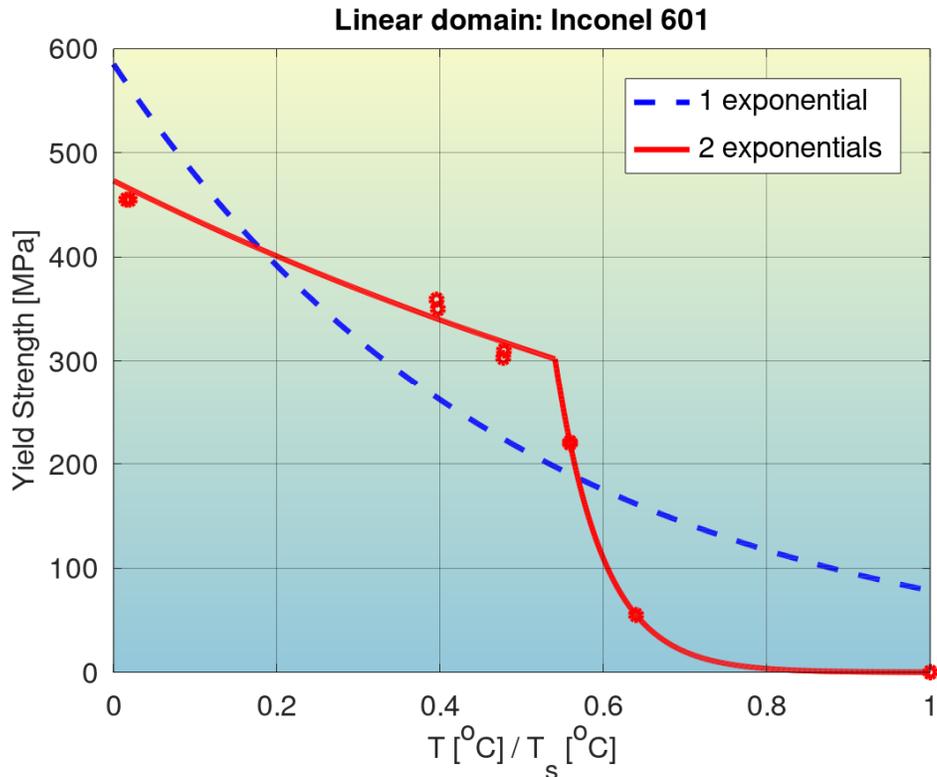

**Figure 2**: Quantification of modeling accuracy of the bilinear log model, for the composition No. 17 from Table 1 (Inconel 601), and comparison to that of a model with a single exponential. One outlier has been excluded from the modeling.

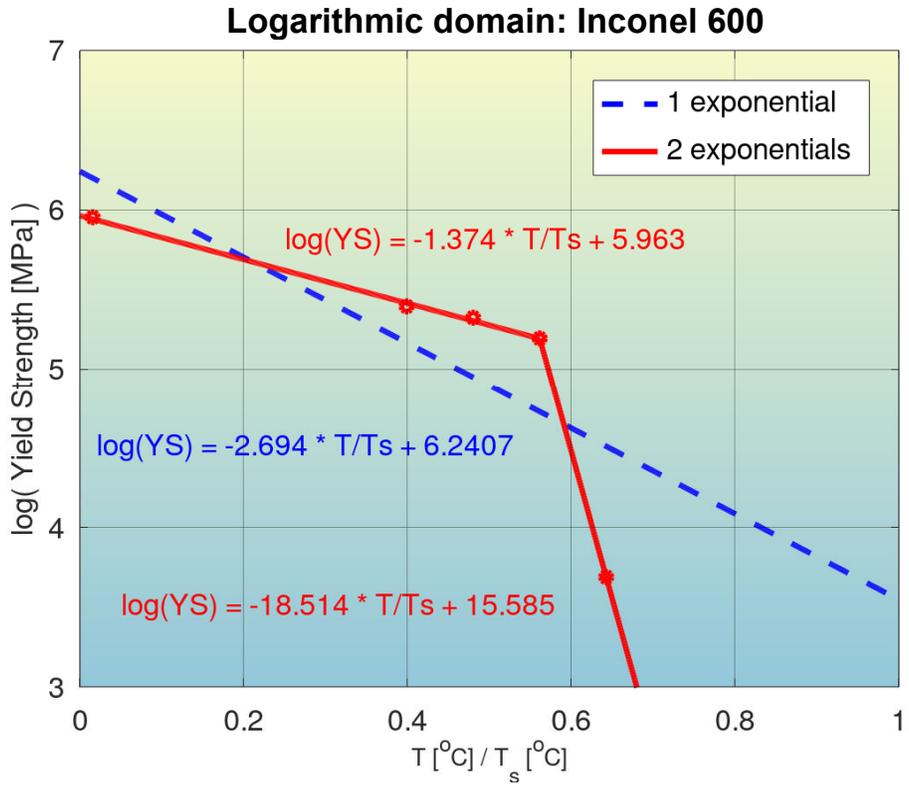

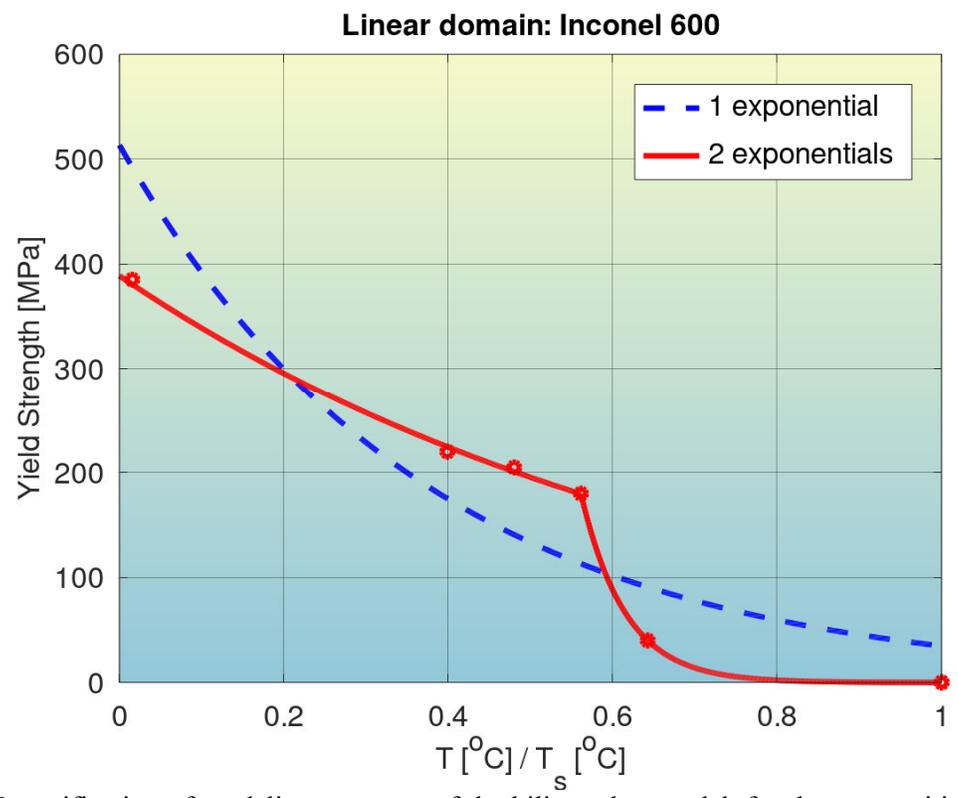

**Figure 3**: Quantification of modeling accuracy of the bilinear log model, for the composition No. 18 from Table 1 (Inconel 600), and comparison to that of a model with a single exponential. One outlier has been excluded from the modeling.

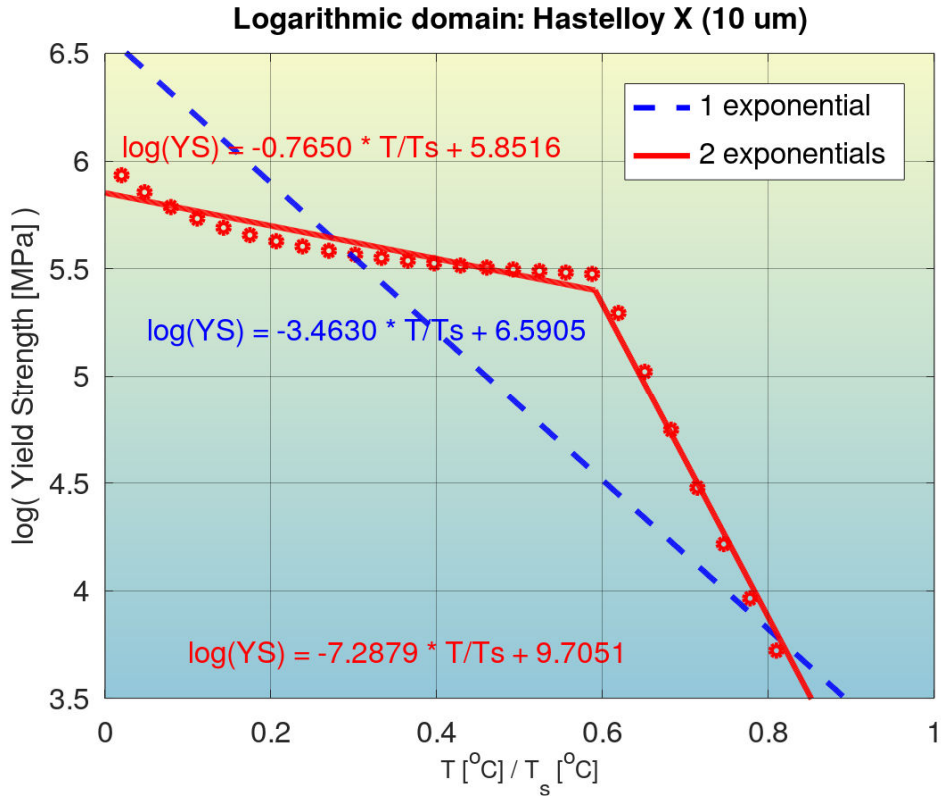

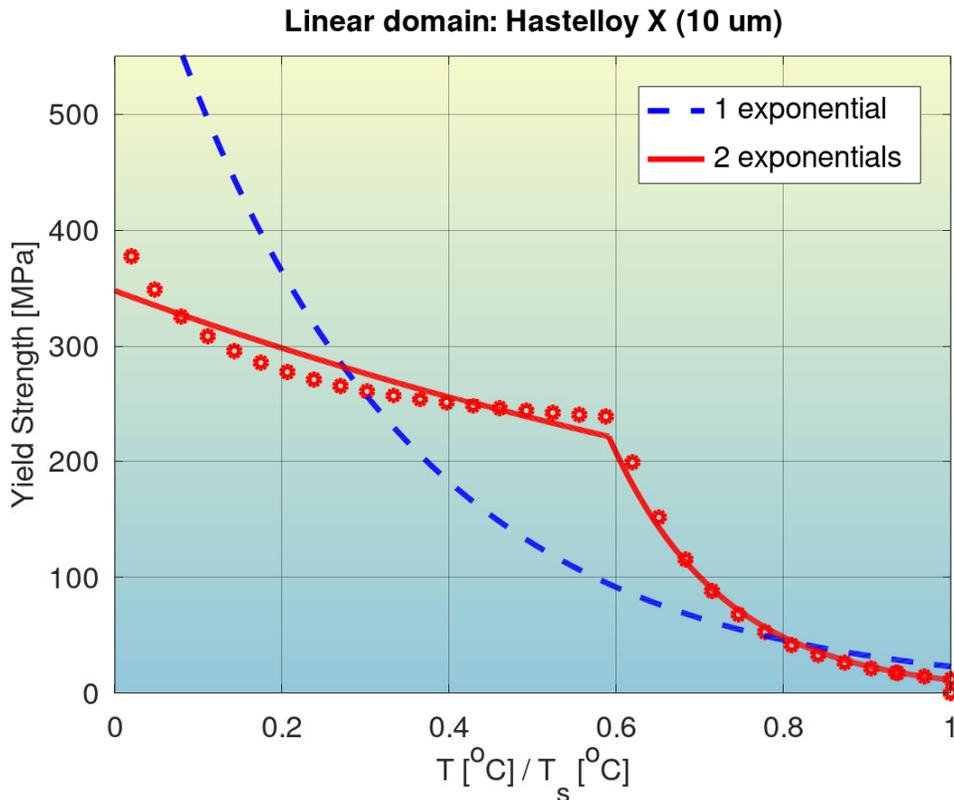

**Figure 4**: Quantification of modeling accuracy of the bilinear log model, for the composition No. 27 from Table 1 (Hastelloy X 10 μm), and comparison to that of a model with a single exponential. One outlier has been excluded from the modeling.

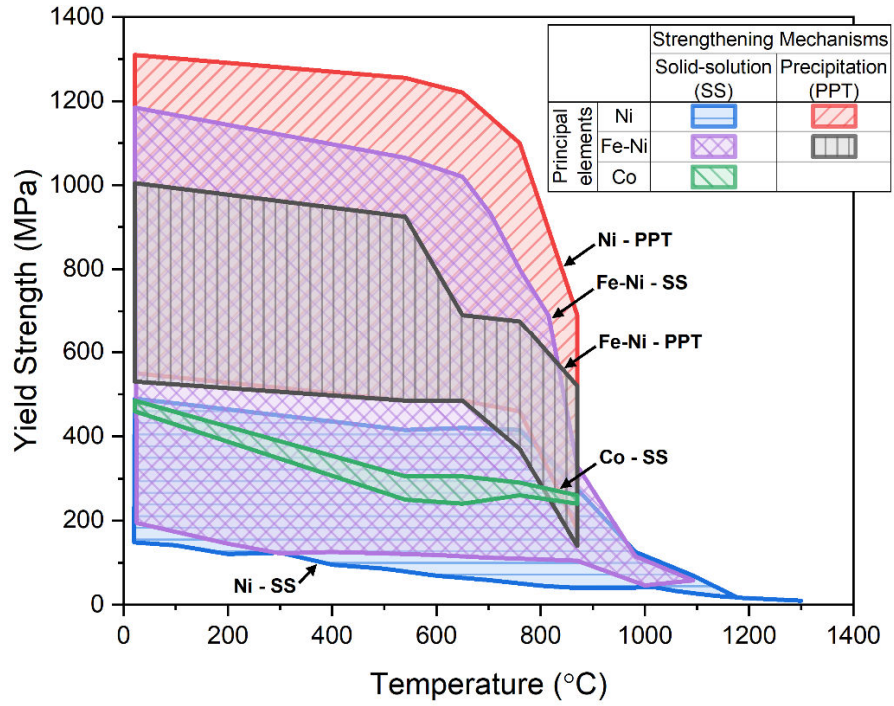

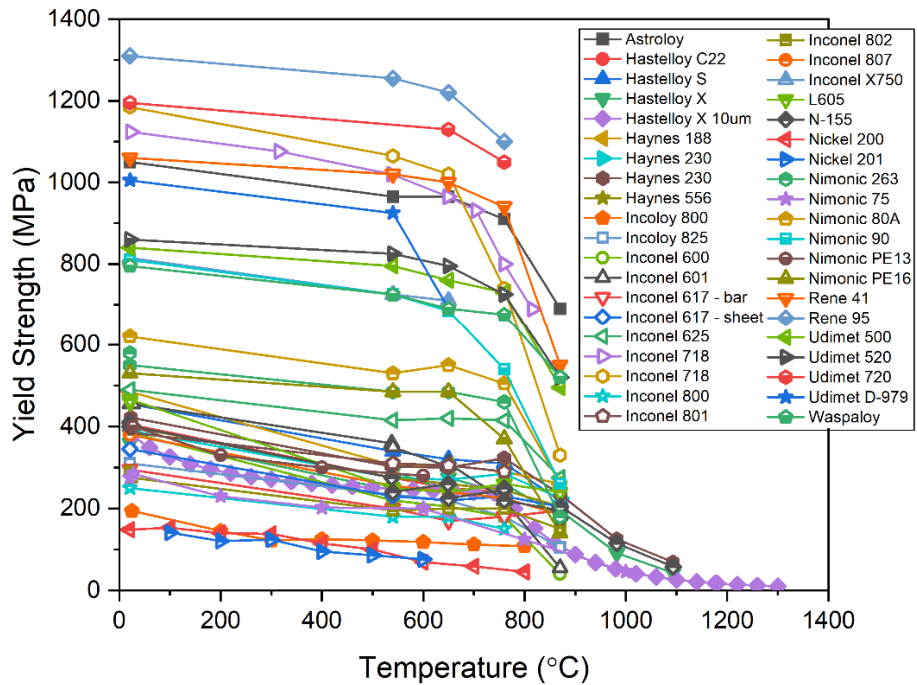

**Figure 5**: Comparison of temperature-dependent yield-strength properties of Fe-Ni-based, Ni-based, and Co-based superalloys.

# Supplementary Manuscript

**Table S1:** Key properties impacting high-temperature strengths for the superalloys under study. [1] Unless otherwise noted, all values are for a wrought, standard annealed and aged state, where applicable. [2] Equivalent Circle Diameter (ECD). [3] Total volume fraction in case of a multimodal distribution. [4] Primary precipitate [5] Secondary/aged precipitate form; [6] Lattice-mismatch austenite matrix and precipitate phase at a constant temperature.

| No. | Alloy | Category | Precipitate / dispersoid [1] | | | | | Matrix grain size [μm] [1,2] | Ref. |
| | | | Type | Volume Fraction [3] | Avg. size [μm] [4] | Avg. size [μm] [5] | Lattice mismatch [%] [6] | | |
|---|---|---|---|---|---|---|---|---|---|
| 1 | Waspaloy | Ni-base (precipit. hard.) | $\gamma'$ | 0.2 - 0.5; 0.3-0.4 | 0.025 | - | 0.5 | 25-35 | [1], [2] |
| 2 | Udimet D979 | Fe-Ni-base (precipit. hard.) | $\gamma'$ | - | 0.02-0.04 | - | - | 20-30 | [3], [4] |
| 3 | Udimet 720 | Ni-base (precipit. hard.) | $\gamma'$ | 0.3 - 0.4 | 0.05-0.1 | 1 | 0.02 | $325 - 350$ | [5], [6], [7] |
| 4 | Udimet 520 | Ni-base (precipit. hard.) | $\gamma'$ | 0.3 | 0.02-0.05 | - | - | 60 | [8] |
| 5 | Udimet 500 | Ni-base (precipit. hard.) | $\gamma'$ | 0.25 | 0.08-0.1 | - | 0.3 | 60-90 | [2], [9], [10] |
| 6 | Astroloy | Ni-base (precipit. hard.) | $\gamma'$ | - | - | - | - | - | - |
| 7 | Incoloy 825 | Fe-Ni-base (solid sol.) | MC | 0.002 | 0.04 | - | - | 122 | [11] |
| 8 | Inconel 807 | Fe-Ni-base (solid sol.) | - | - | - | - | - | - | - |
| 9 | Inconel 802 | Fe-Ni-base (solid sol.) | - | - | - | - | - | - | - |
| 10 | Inconel 801 | Fe-Ni-base (solid sol.) | - | - | - | - | - | - | - |
| 11 | Inconel 800 | Fe-Ni-base (solid sol.) | MC | 0.002-0.003 | 0.1 | - | - | 120 | [12] |
| 12 | Incoloy 800 | Fe-Ni-base (solid sol.) | - | - | - | - | - | - | - |
| 13 | Inconel X750 | Ni-base (precipit. hard.) | $\gamma'$ | 0.25 | 0.02-0.03 | - | 0.8-1.0 | 170 | [13] |
| 14 | Inconel 718 | Ni-base (precipit. hard.) | $\gamma'$, $\gamma''$ | 0.15-0.2 | 0.01-0.03 | 0.02 | - | 300 | [14], [15] |
| 15 | Inconel 625 | Ni-base (solid sol.) | M(C,N), M23C6 | - | - | - | - | 70-90 | [16], [17] |
| 16 | Inconel 617 | Ni-base (solid sol.) | M(C,N), M23C6 M6C | - | - | - | - | 45-100 | [18], [19] |
| 17 | Inconel 601 | Ni-base (solid sol.) | M(C,N) | - | - | - | - | 70-90 | Special Metals Infosheet |
| 18 | Inconel 600 | Ni-base (solid sol.) | M(C,N) | - | - | - | - | 30-40 | [20] |
| 19 | Ti6Al4V | $\alpha$ and $\beta$ phases | - | - | - | - | - | - | - |

Table S1 (continued): Key properties impacting high-temperature strengths for the superalloys under the study. [1] Unless otherwise noted, all values are for a wrought, standard annealed and aged state, where applicable. [2] Equivalent Circle Diameter (ECD). [3] Total volume fraction in case of multimodal distribution. [4] Primary precipitate [5] Secondary/aged precipitate form; [6] Lattice mismatch of the austenite matrix and precipitate phase at a constant temperature.

| No. | Alloy | Category | Precipitate / dispersoid [1] | | | | | Matrix grain size [μm] [1,2] | Ref. |
|---|---|---|---|---|---|---|---|---|---|
| | | | Type | Volume Fraction [3] | Avg. size [μm] [4] | Avg. size [μm] [5] | Lattice mismatch [%] [6] | | |
| 20 | Haynes 556 | Fe-Ni-base (solid sol.) | - | - | - | - | - | 45 | [21] |
| 21 | Haynes 230 (a) | Ni-base (solid sol.) | - | - | - | - | - | 70-90 | [18], [22] |
| 22 | Haynes 230 (b) | | - | - | - | - | - | - | - |
| 23 | Haynes 188 | Co-base (solid sol.) | - | - | - | - | - | 40-50 | [23], [24] |
| 24 | Hastelloy C22 | Ni-base (solid sol.) | Ni2Cr, M6C, M23C6 | - | - | - | - | 50-100 | [25], Haynes Data Sheet |
| 25 | Hastelloy S | Ni-base (solid sol.) | - | - | - | - | - | - | - |
| 26 | Hastelloy X | Ni-base (solid sol.) | - | - | - | - | - | 50 | [26] |
| 27 | Hastelloy X (10 μm) | | - | - | - | - | - | - | - |
| 28 | Rene 95 | Ni-base (precipit. hard.) | γ′, MC | 0.4-0.5 | 0.2 | - | - | 10-20 | [27], [28] |
| 29 | Rene 41 | Ni-base (precipit. hard.) | γ′ | 0.3-0.4 | 0.1 | - | - | 40-60 | [29], [30] |
| 30 | Pure Nickel | Ni-base | - | - | - | - | - | - | - |
| 31 | Nickel 201 | Ni-base (solid sol.) | - | - | - | - | - | - | - |
| 32 | Nickel 200 | Ni-base (solid sol.) | - | - | - | - | - | - | - |
| 33 | Nimonic 263 | Ni-base (precipit. hard.) | γ′, η | 0.1-0.15 | 0.01-0.03 | 0.02-0.1 | - | 100-200 | [31], [32], [33] |
| 34 | Nimonic 90 | Ni-base (precipit. hard.) | γ′ | - | - | - | 0.34 | - | [7], [34], [35] |
| 35 | Nimonic 80a | Ni-base (precipit. hard.) | γ′ | - | 0.02-0.05 | - | 0.32 | 90 | [7], [34], [35] |
| 36 | Nimonic 75 | Fe-Ni-base (solid sol.) | M23C6 | 0.7-1.2 | - | - | - | 10-20 | [34], [35], [36] |
| 37 | Nimonic PE16 | Fe-Ni-base (precipit. hard.) | γ′ | 0.08-0.12 | 0.01-0.05 | - | 0.03 | 15-20 | [37], Special Metals Datasheet |
| 38 | Nimonic PE13 | Fe-Ni-base (solid sol.) | - | - | - | - | - | - | - |
| 39 | N155 | Fe-Ni-base (solid sol.) | - | - | - | - | - | 20-30 | [38] |
| 40 | L605 | Co-base (solid sol.) | - | - | - | - | - | 60-90 | [39] |

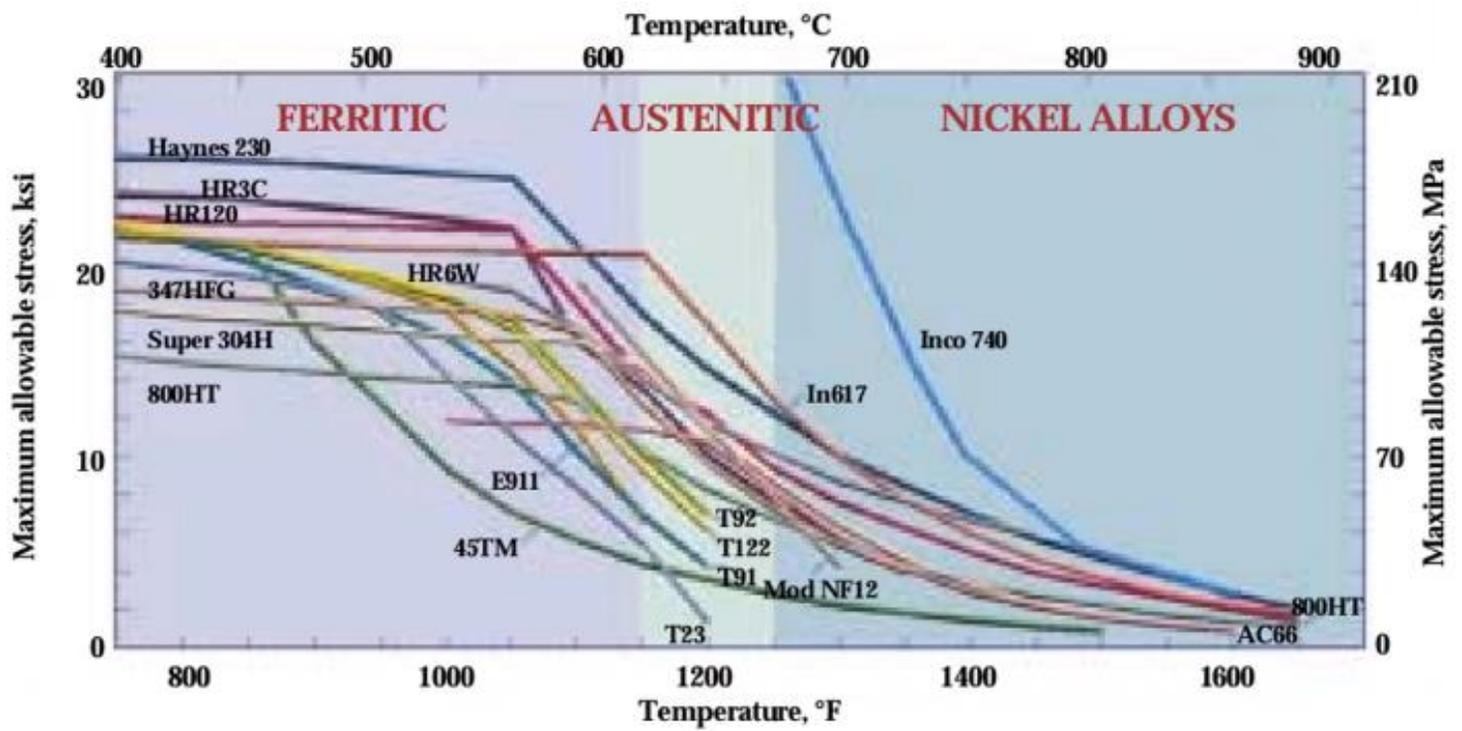

**Figure S1**: Allowable stress for various classes of alloys (adapted from [40]).

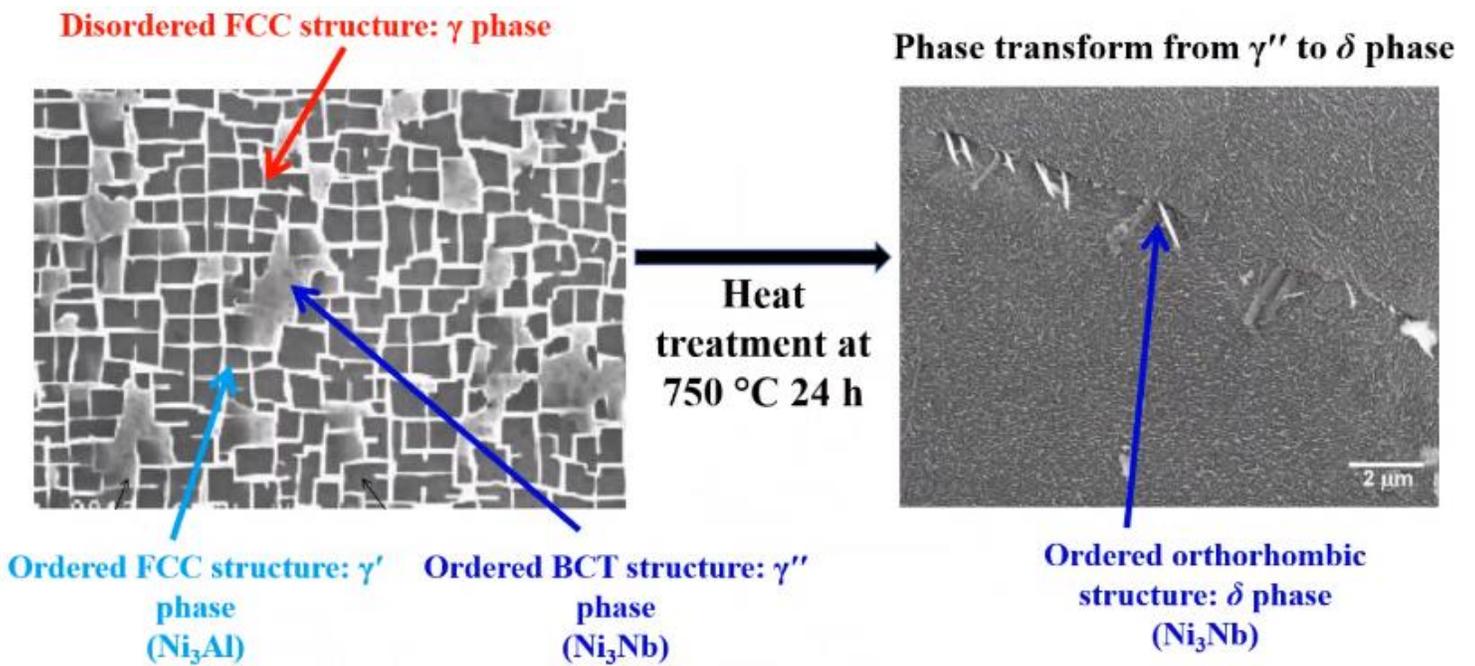

**Figure S2**: Phase transformations in Ni-based superalloys, caused by dissolution of γ″ phase, at the elevated temperature of 750 ᵒC (adapted from [41]).

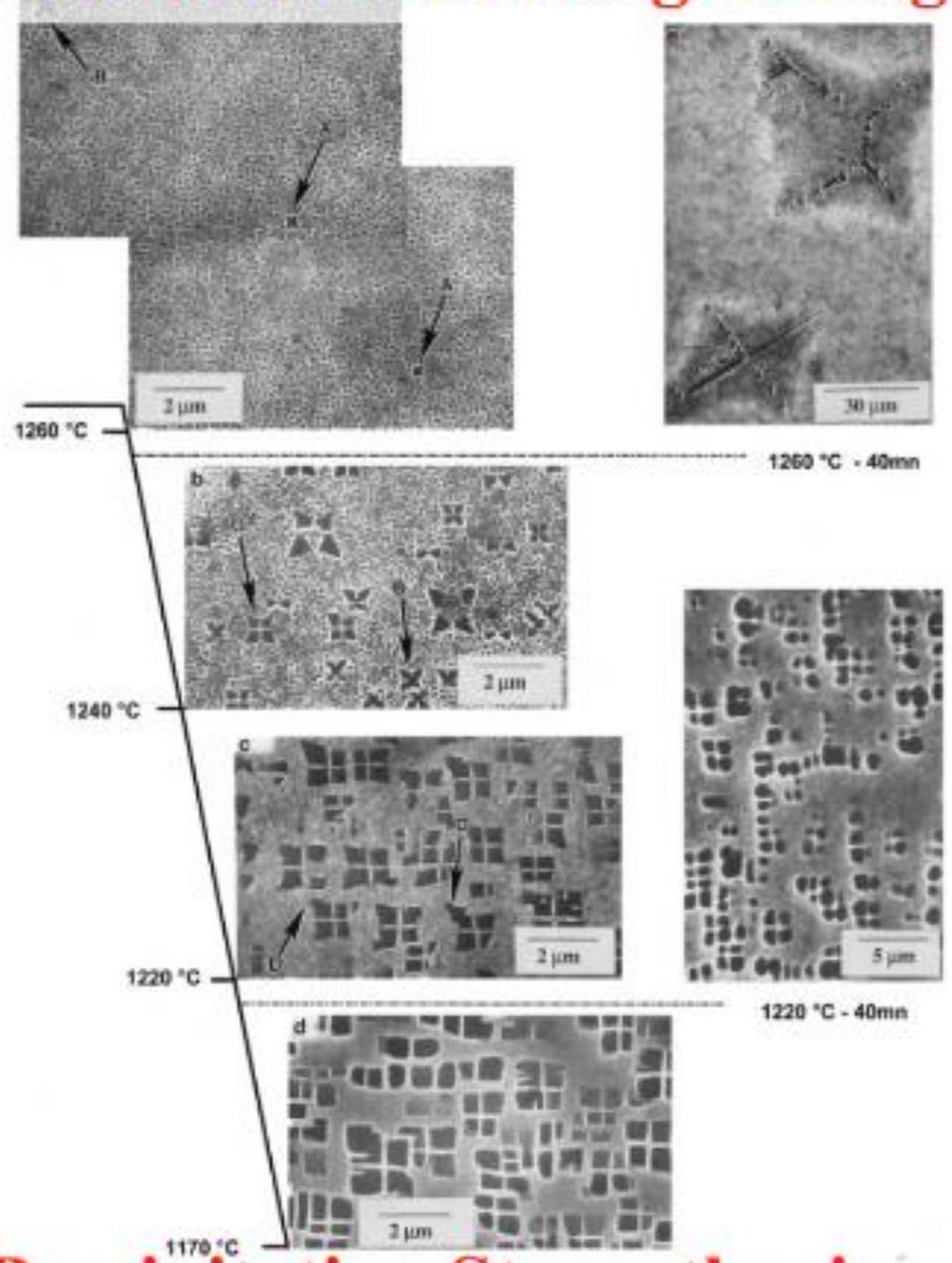

**Figure S3**: For Ni-based superalloys, precipitation strengthening tends to be the main strengthening mechanism at low-to-medium temperatures, but solid-solution strengthening at high temperatures, i.e., at temperatures above 1,260 ºC (adapted from [42]).

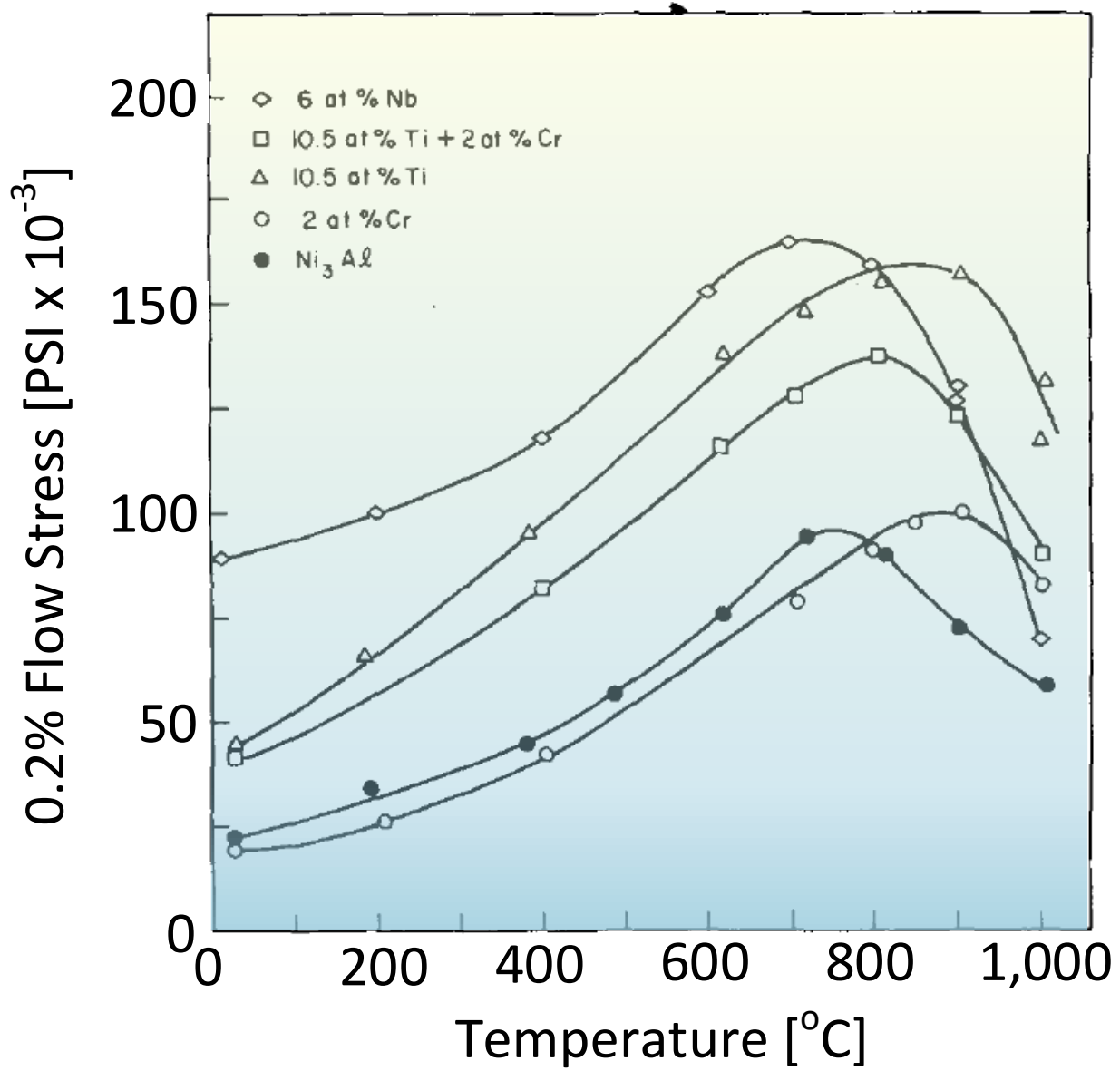

**Figure S4**: Yield stress vs. temperature for γ′ showing the yield stress peak and the influence of solutes on Ni₃Al (adapted from [43]).

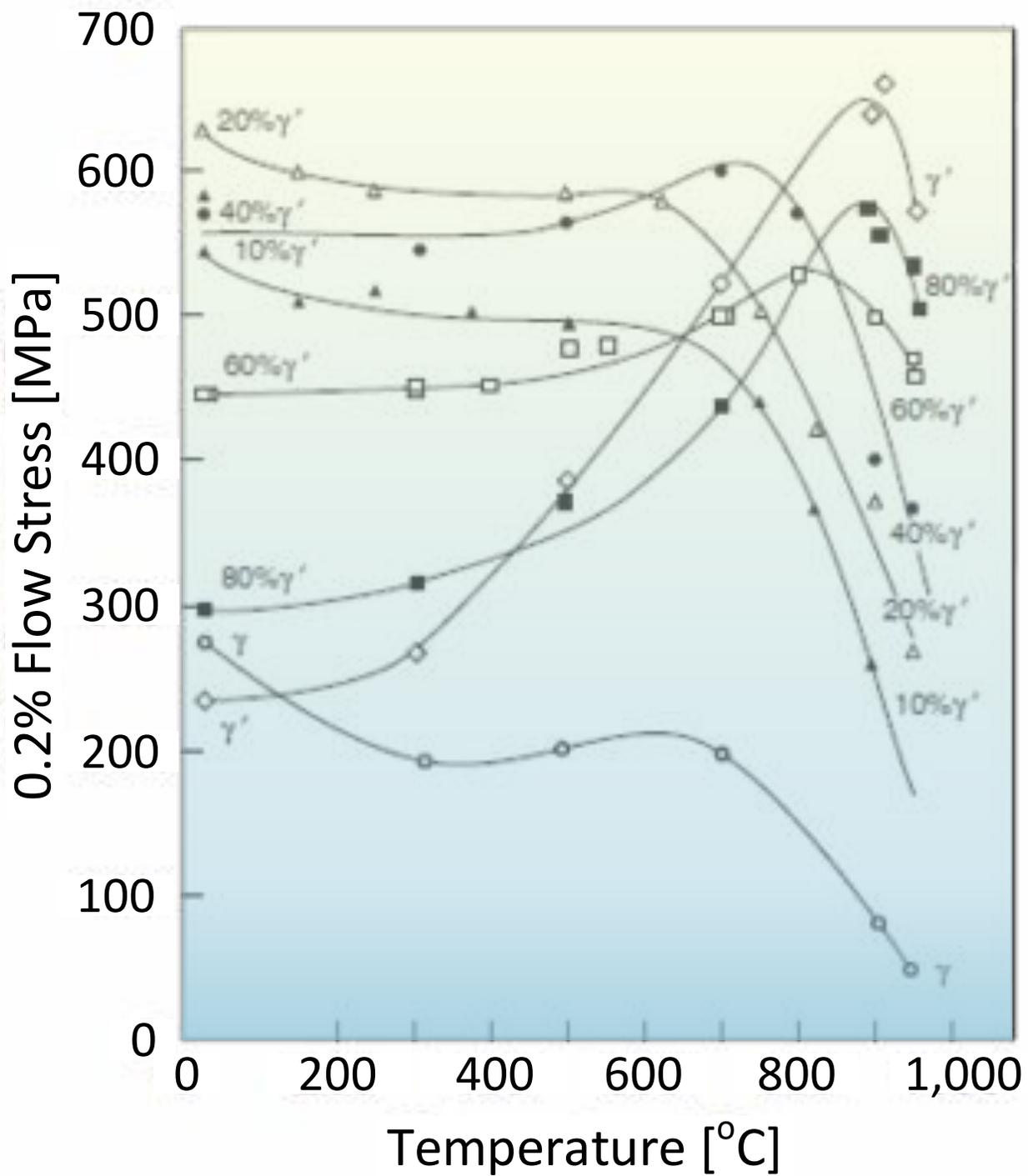

**Figure S5**: Effects of γ′ volume fraction and temperature on yield strength (adapted from [44]).

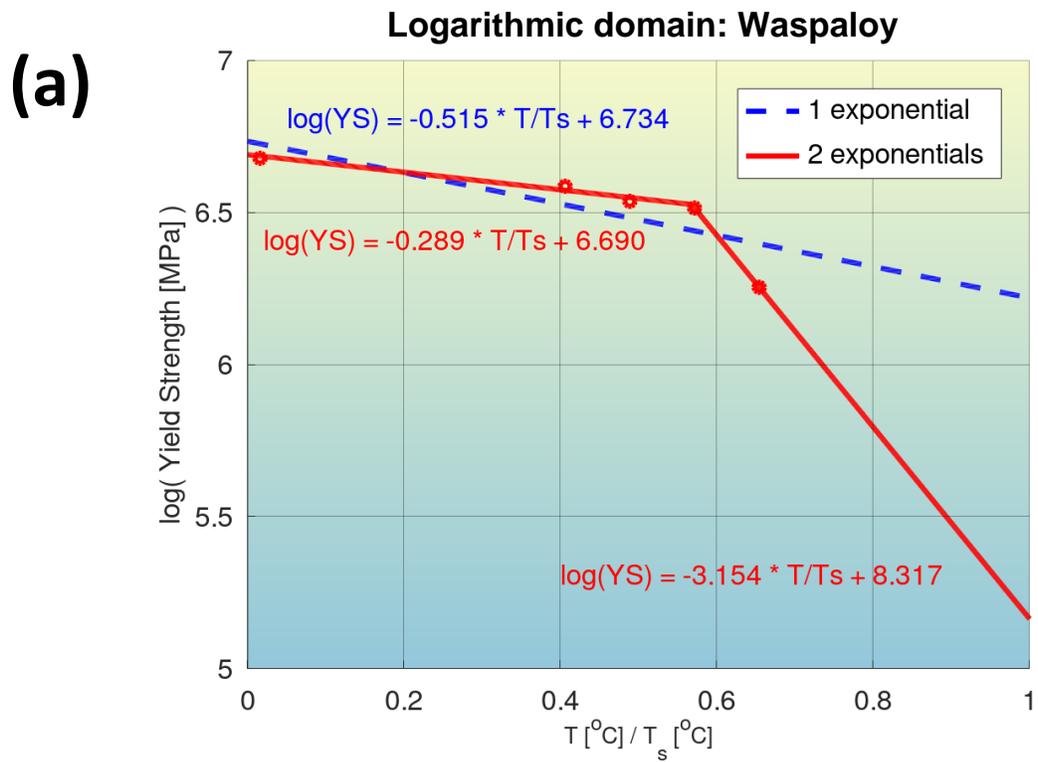

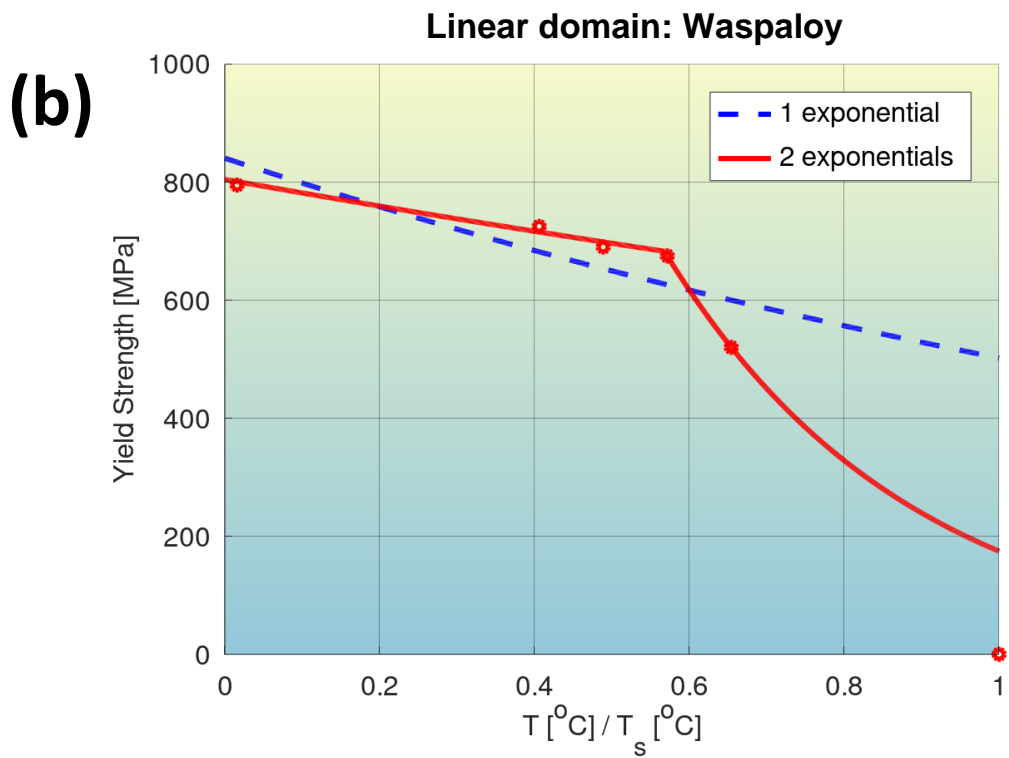

**Figure S6**: Quantification of modeling accuracy of the bilinear log model, for the composition No. 1 from Table S1 (Waspaloy), and comparison to that of a model with a single exponential.

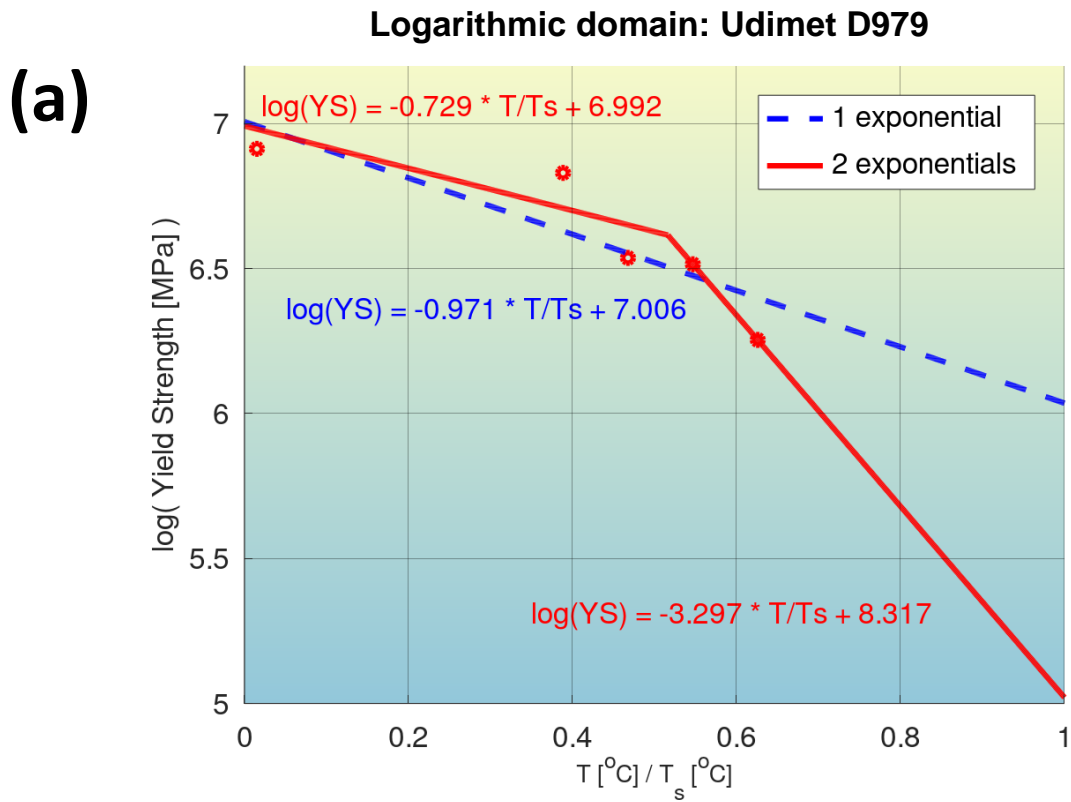

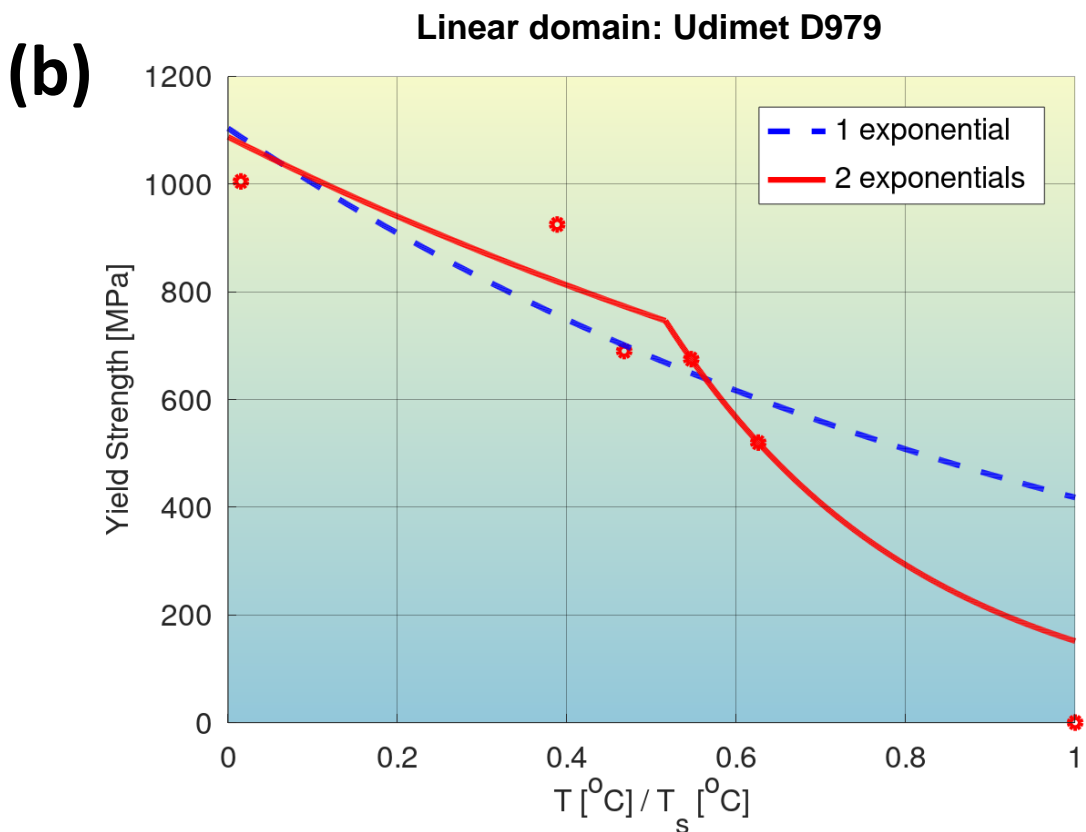

**Figure S7**: Quantification of modeling accuracy of the bilinear log model, for the composition No. 2 from Table S1 (Udimet D979), and comparison to that of a model with a single exponential.

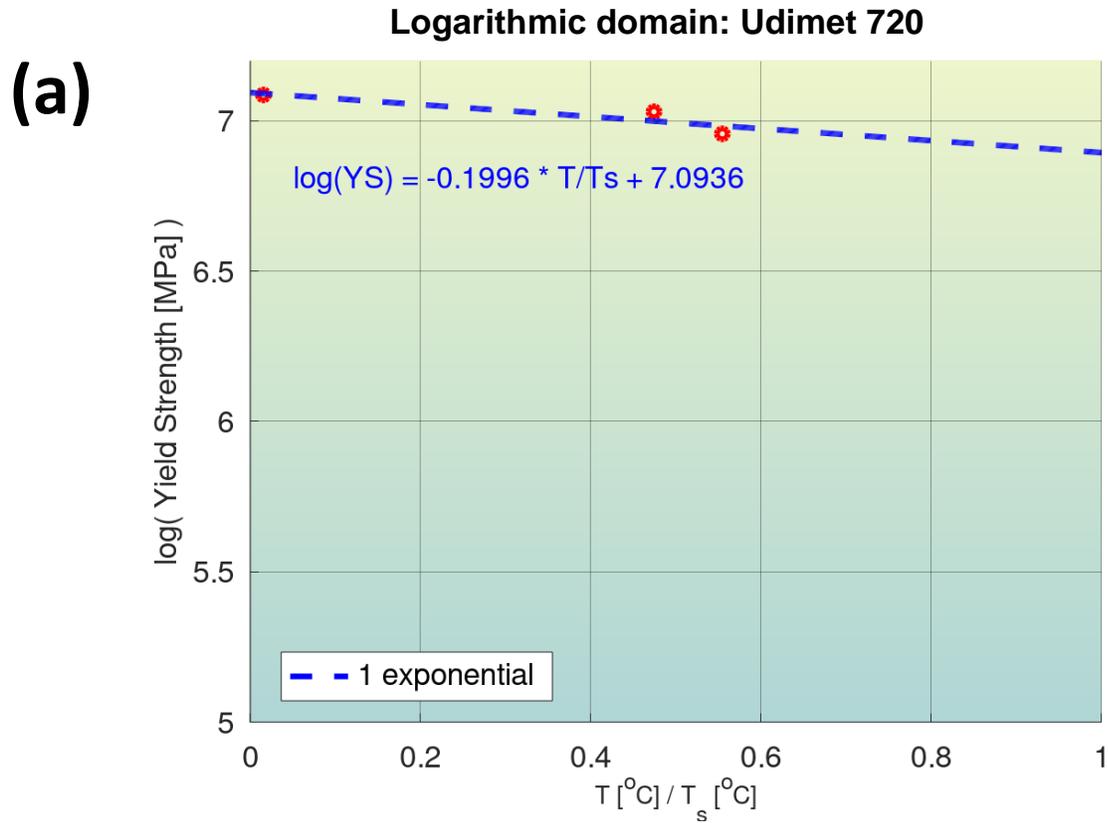

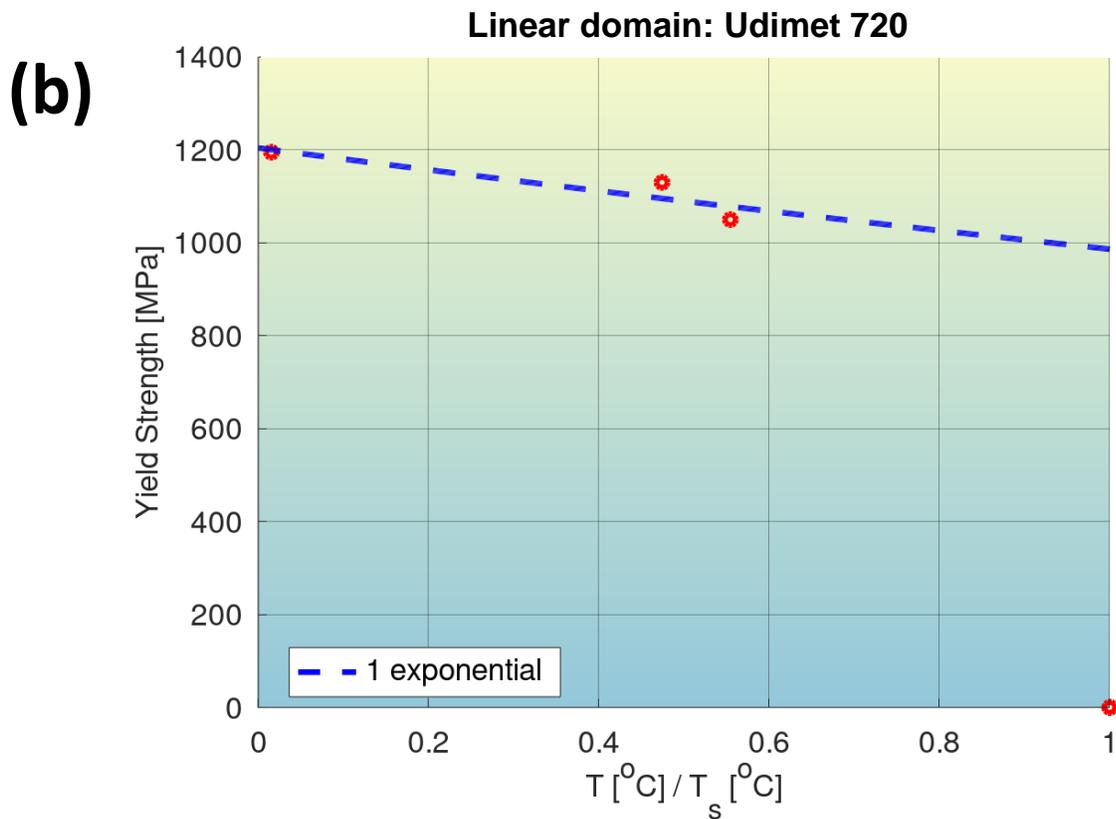

**Figure S8**: Quantification of modeling accuracy of the bilinear log model, for the composition No. 3 from Table S1 (Udimet 720), and comparison to that of a model with a single exponential.

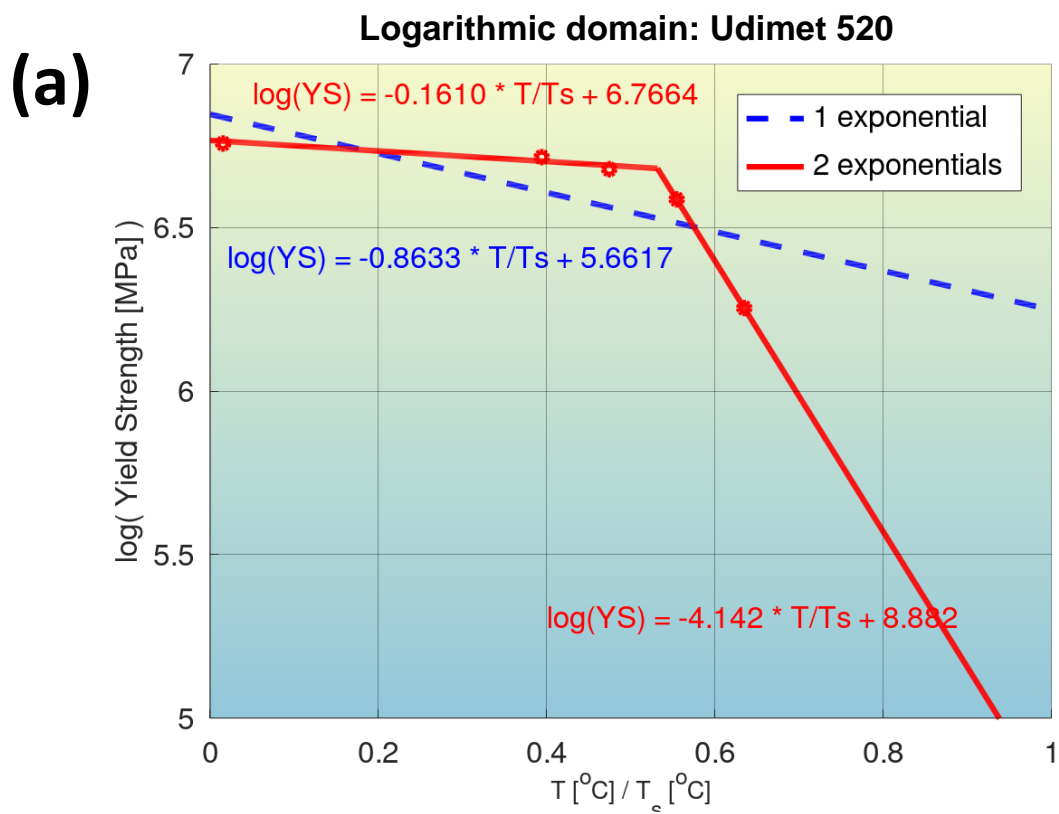

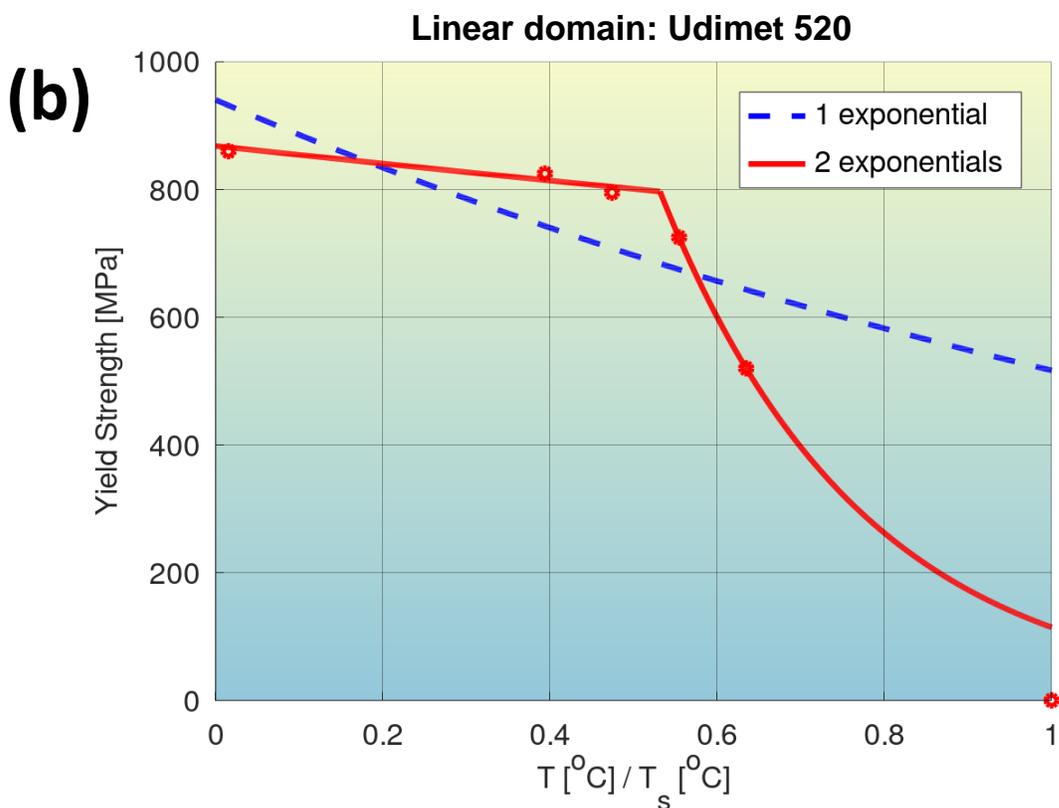

**Figure S9**: Quantification of modeling accuracy of the bilinear log model, for the composition No. 4 from Table S1 (Udiment 520), and comparison to that of a model with a single exponential.

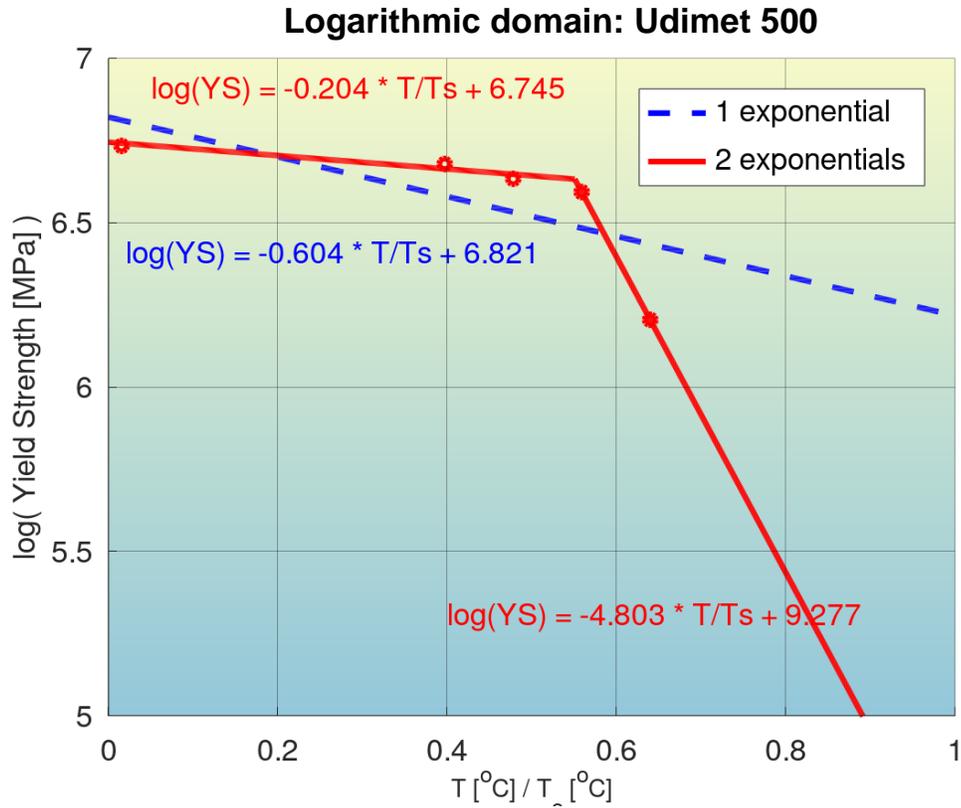

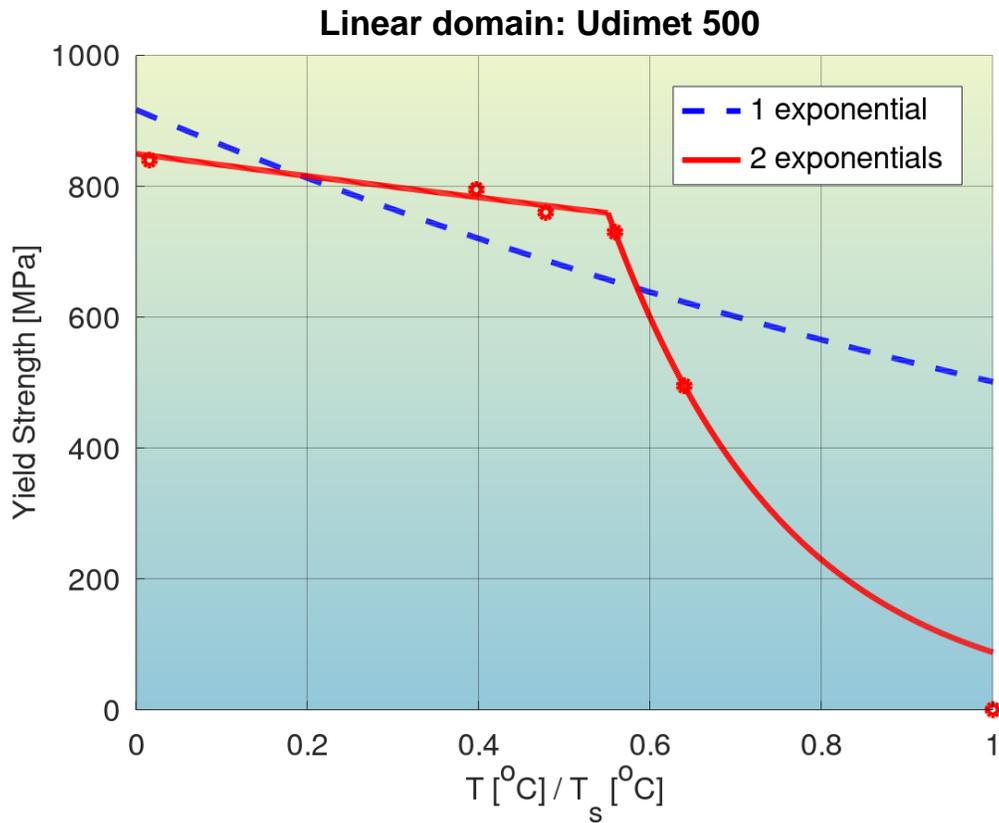

**Figure S10**: Quantification of modeling accuracy of the bilinear log model, for the composition No. 5 from Table S1 (Udiment 500), and comparison to that of a model with a single exponential.

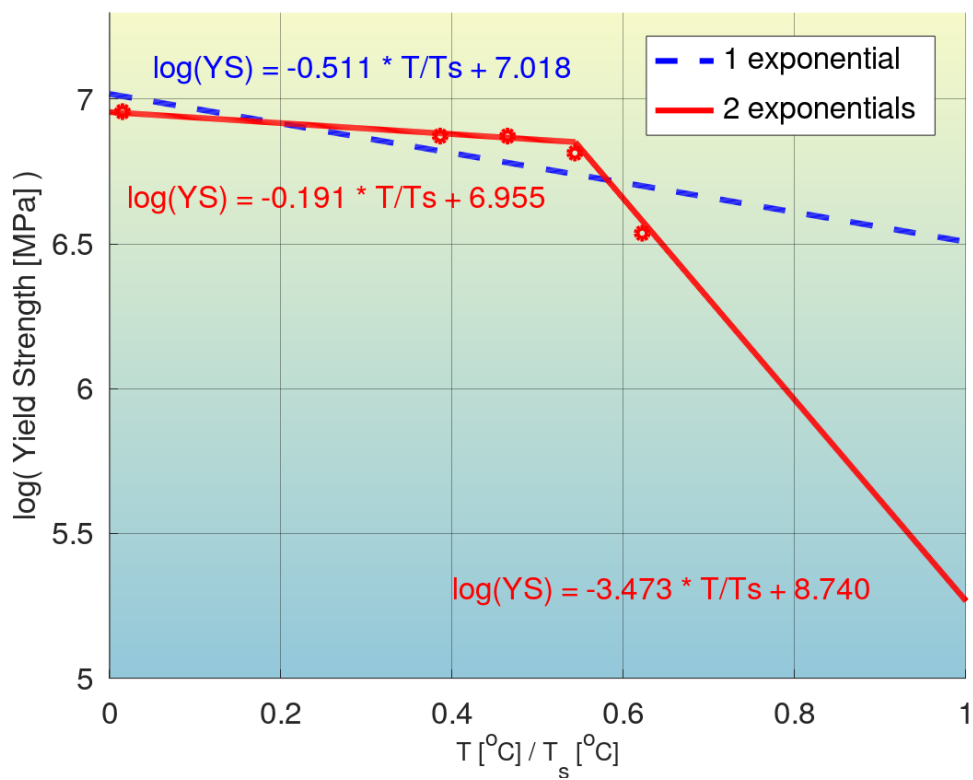

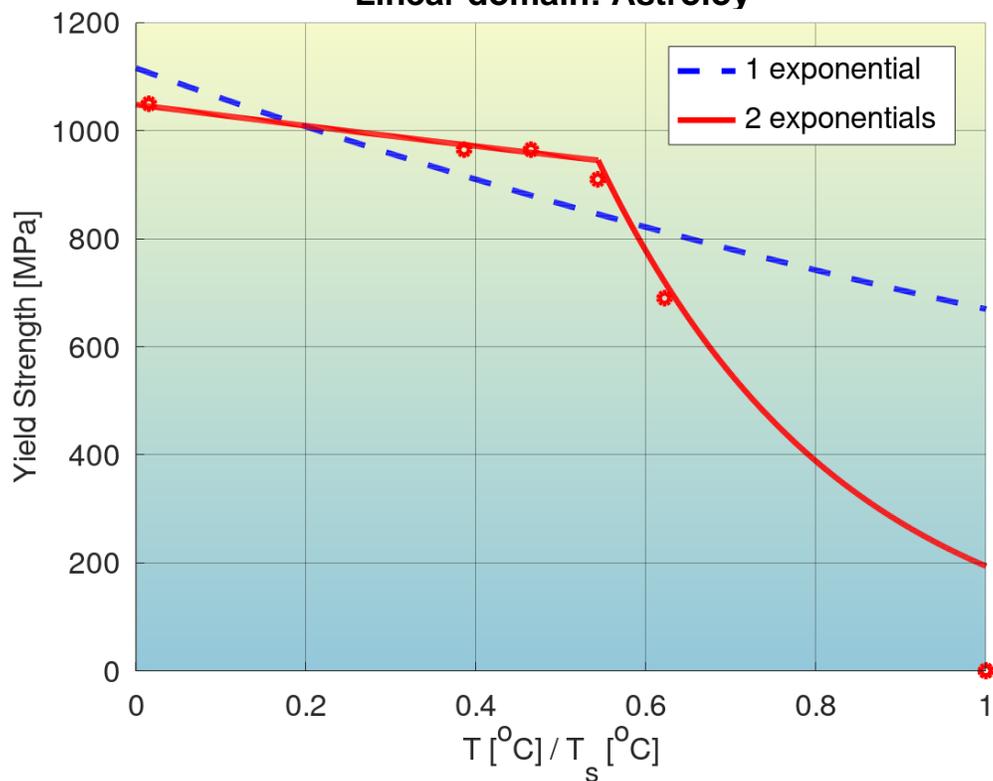

**Figure S11**: Quantification of modeling accuracy of the bilinear log model, for the composition No. 6 from Table S1 (Astroloy), and comparison to that of a model with a single exponential.

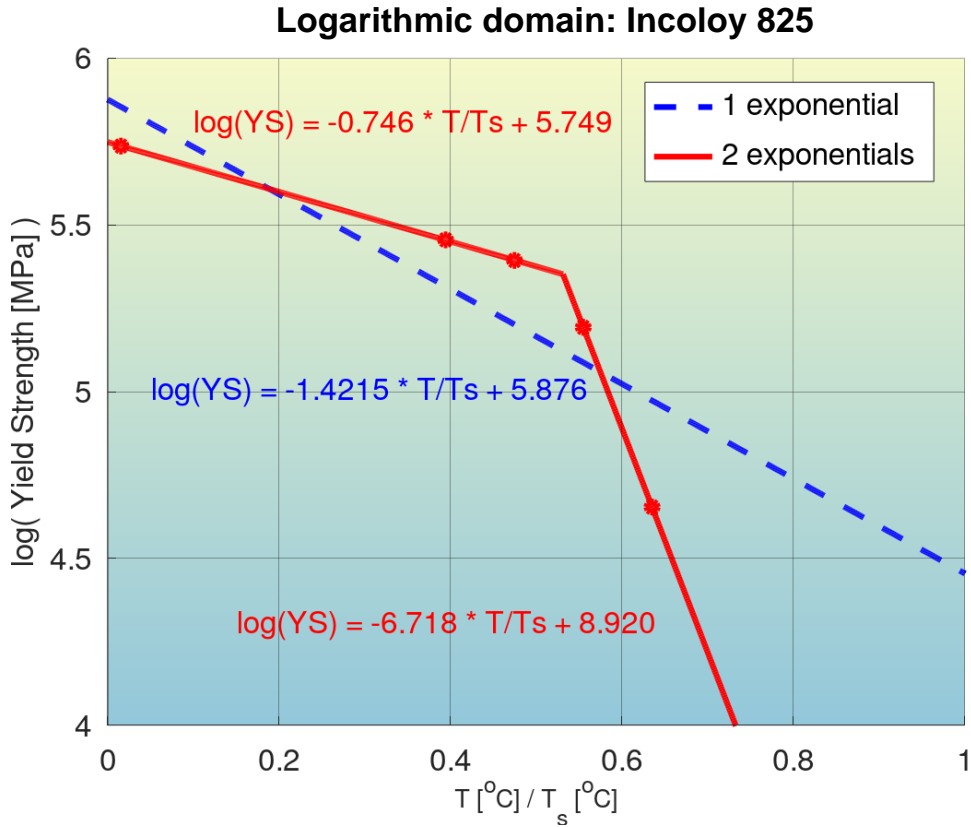

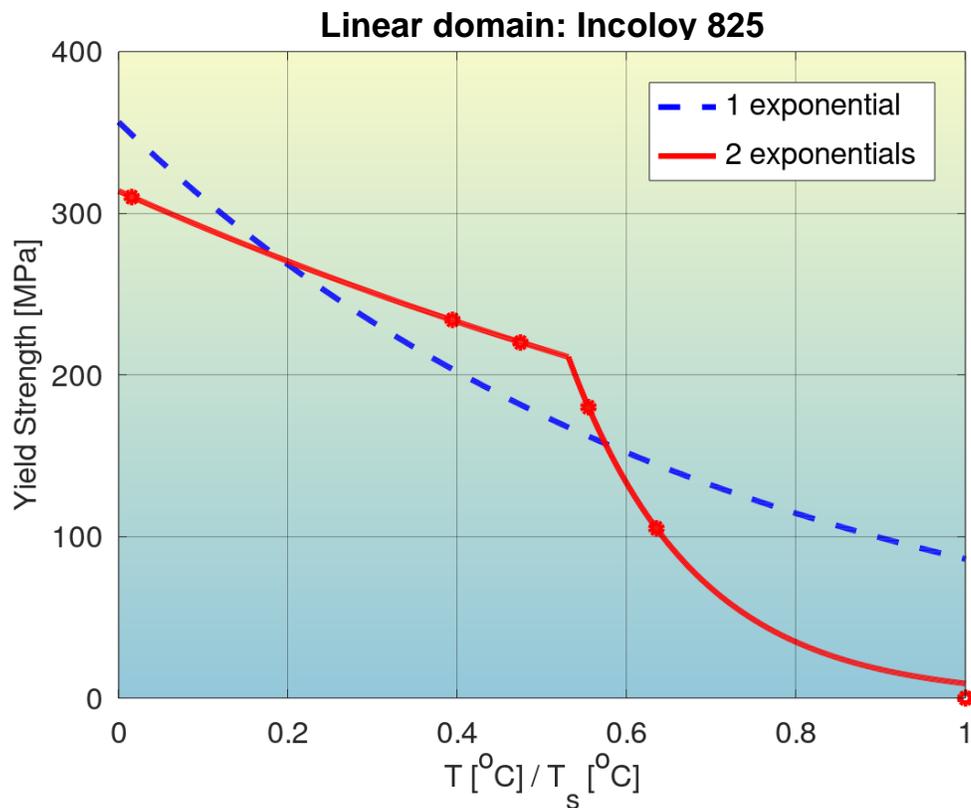

**Figure S12**: Quantification of modeling accuracy of the bilinear log model, for the composition No. 7 from Table S1 (Incoloy 825), and comparison to that of a model with a single exponential.

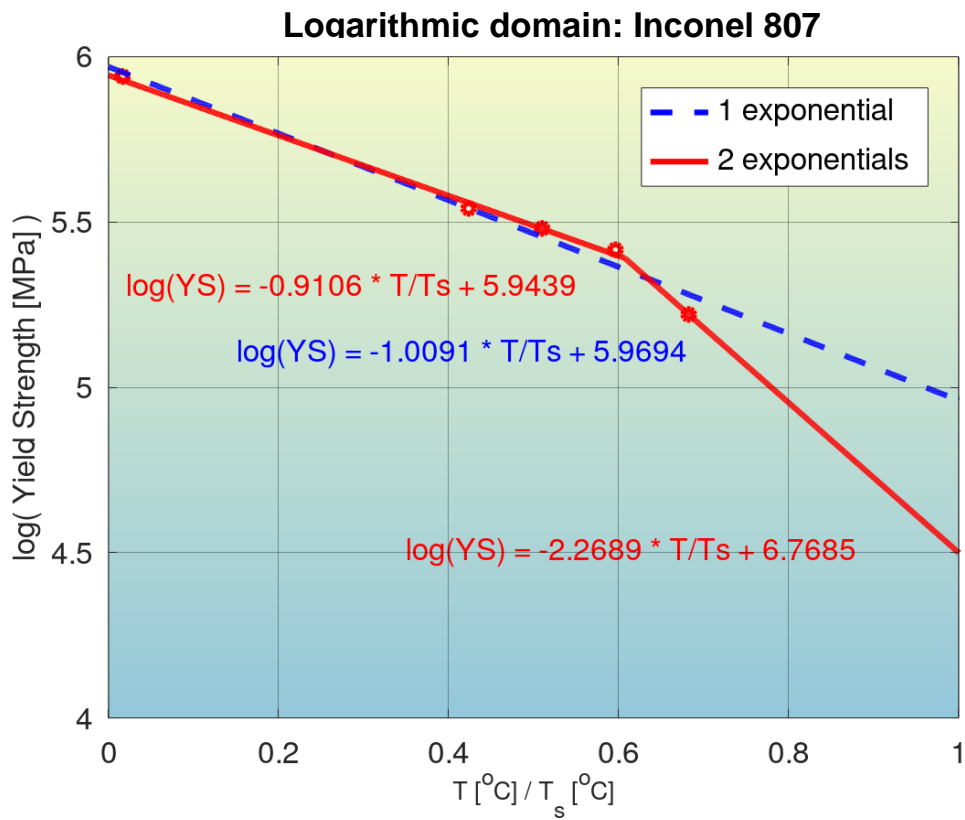

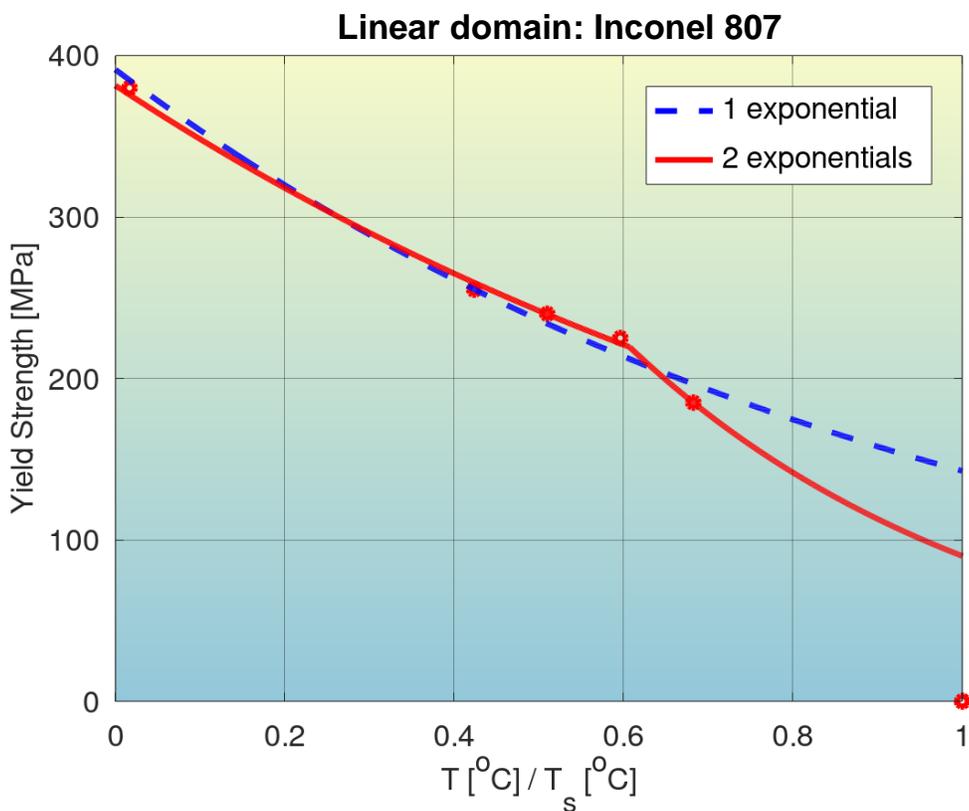

**Figure S13**: Quantification of modeling accuracy of the bilinear log model, for the composition No. 8 from Table S1 (Inconel 807), and comparison to that of a model with a single exponential.

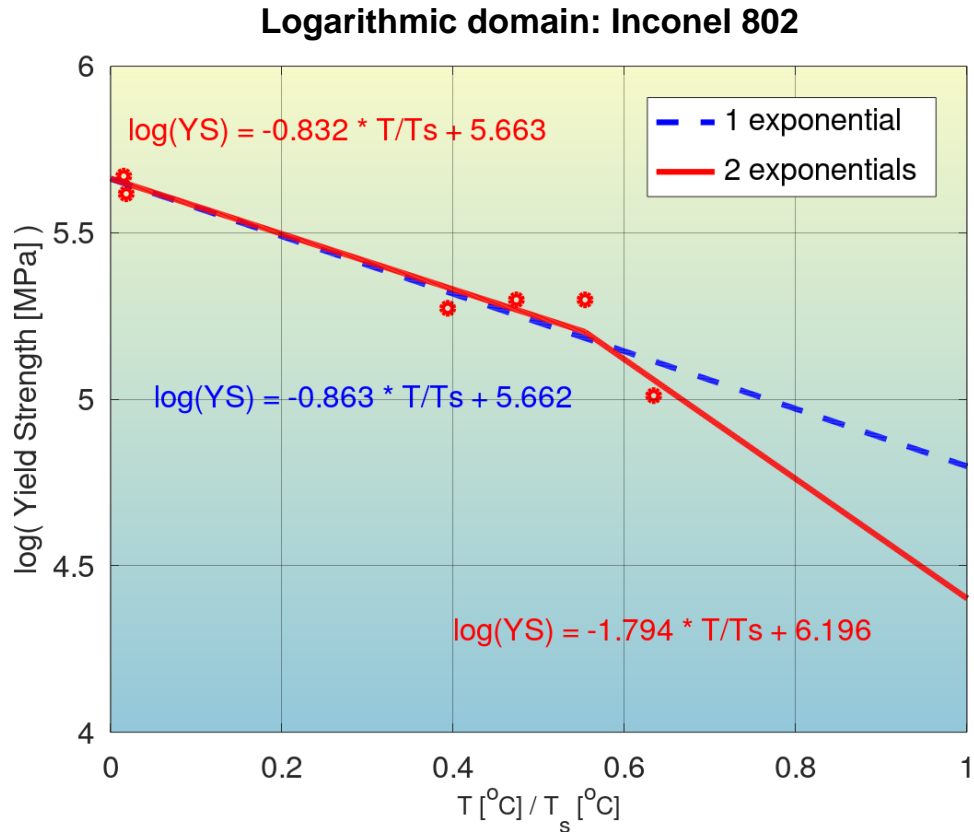

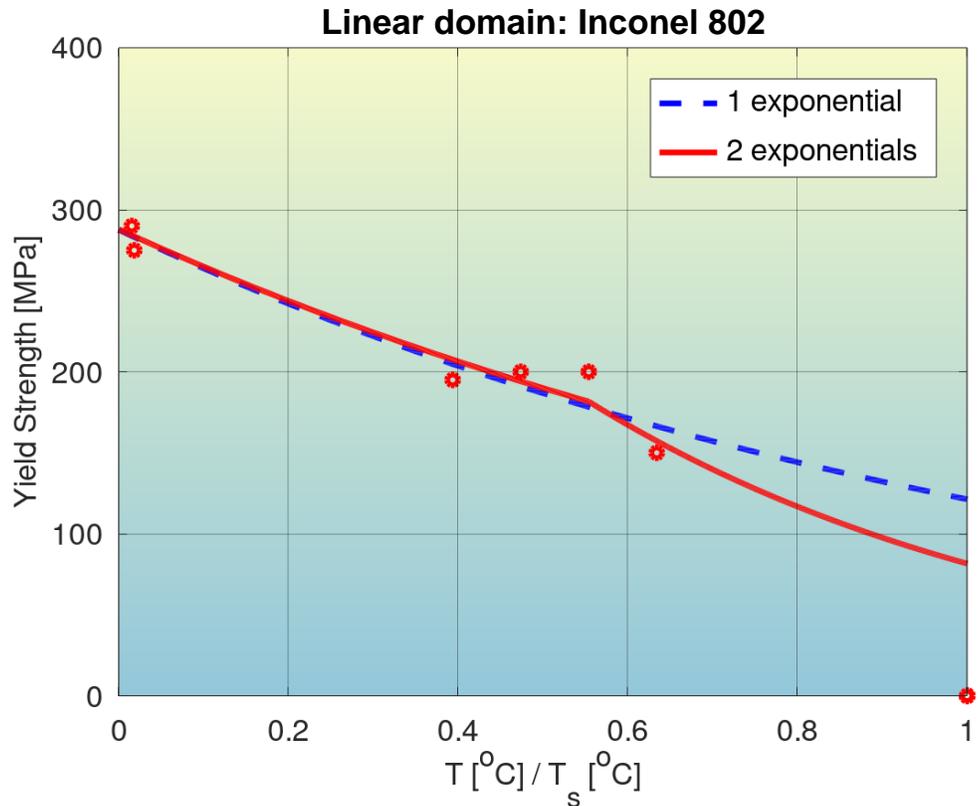

**Figure S14**: Quantification of modeling accuracy of the bilinear log model, for the composition No. 9 from Table S1 (Inconel 802), and comparison to that of a model with a single exponential.

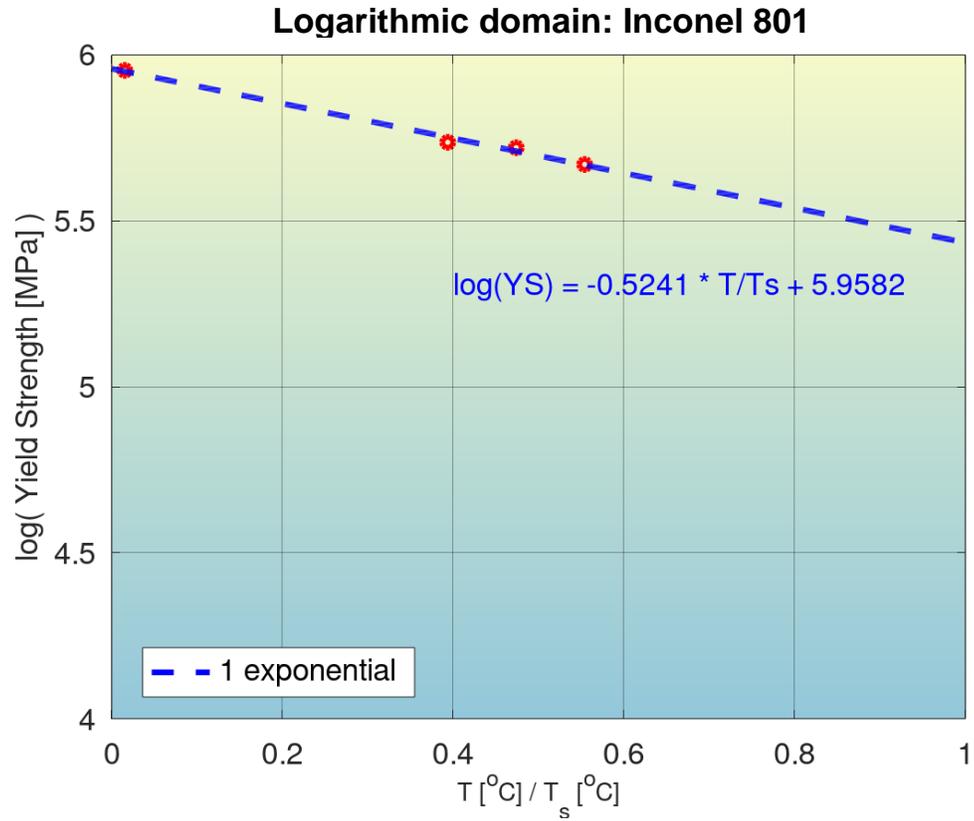

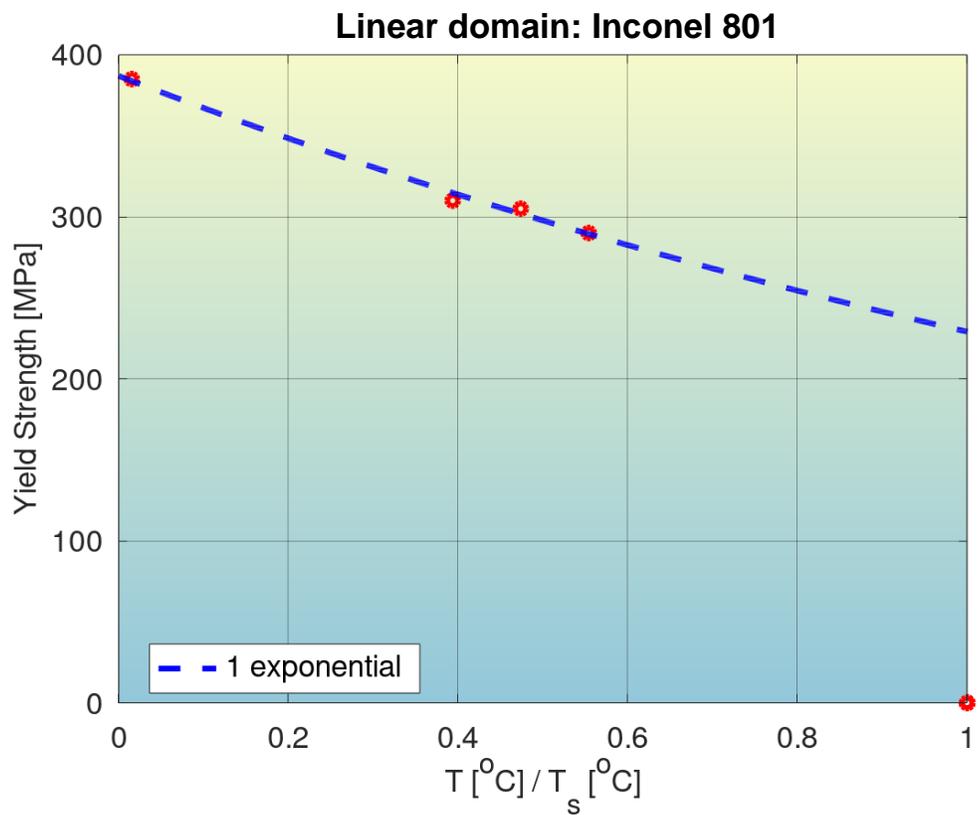

**Figure S15**: Quantification of modeling accuracy of the bilinear log model, for the composition No. 10 from Table S1 (Inconel 801), and comparison to that of a model with a single exponential.

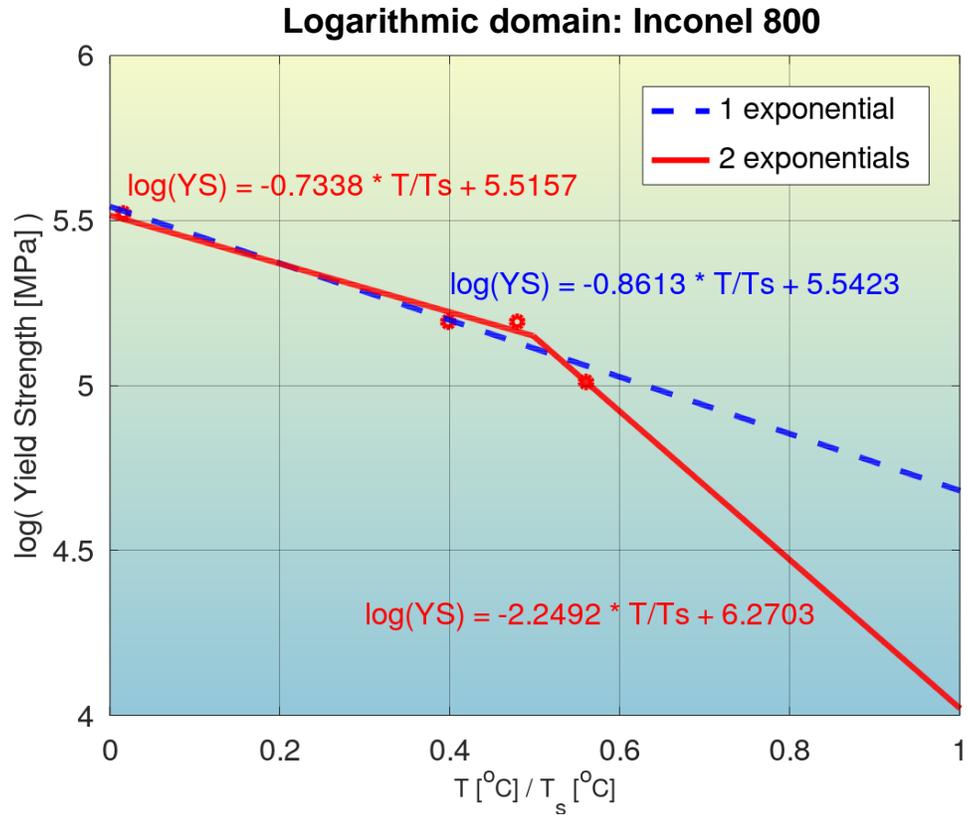

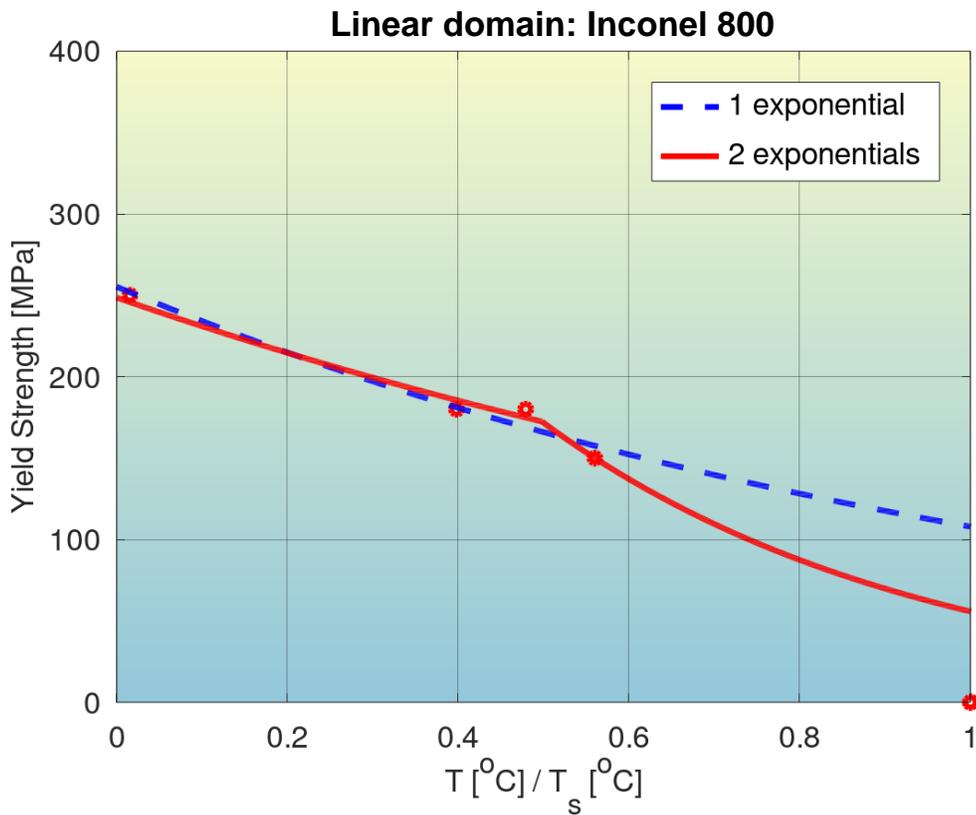

**Figure S16**: Quantification of modeling accuracy of the bilinear log model, for the composition No. 11 from Table S1 (Inconel 800), and comparison to that of a model with a single exponential.

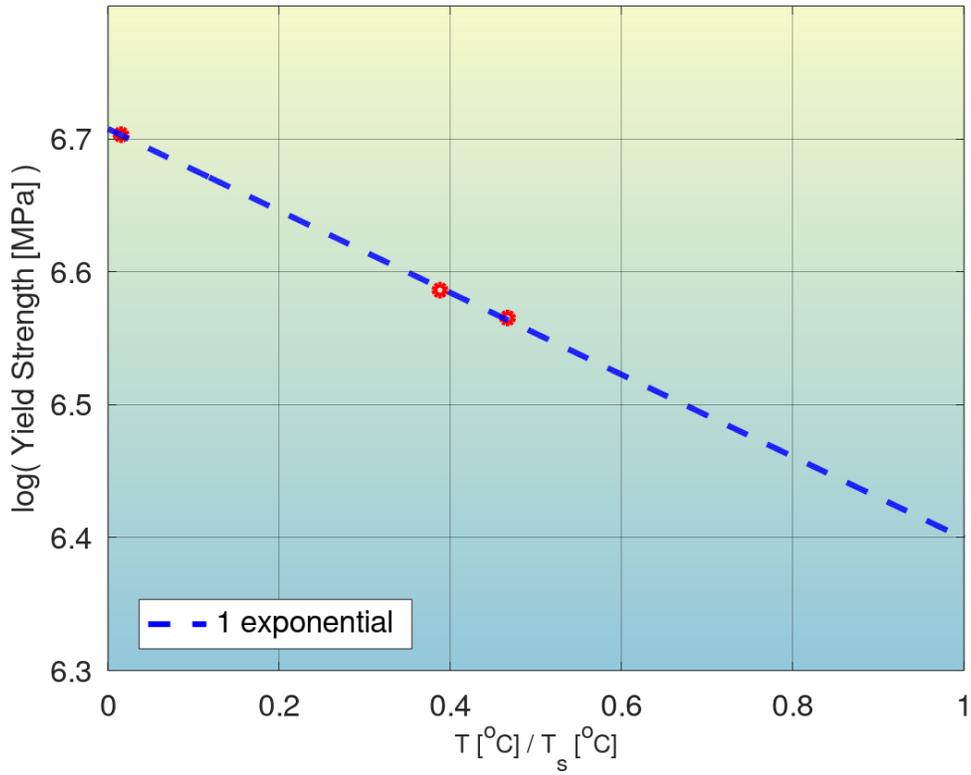

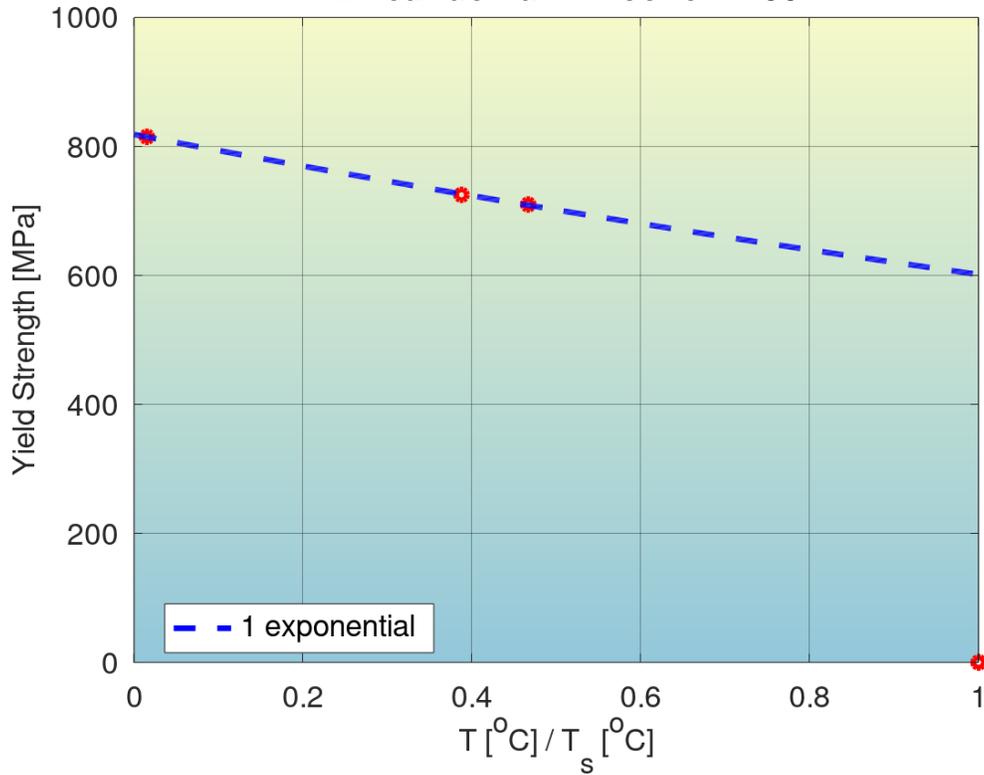

**Figure S17**: Quantification of modeling accuracy of the bilinear log model, for the composition No. 13 from Table S1 (Inconel X750), and comparison to that of a model with a single exponential.

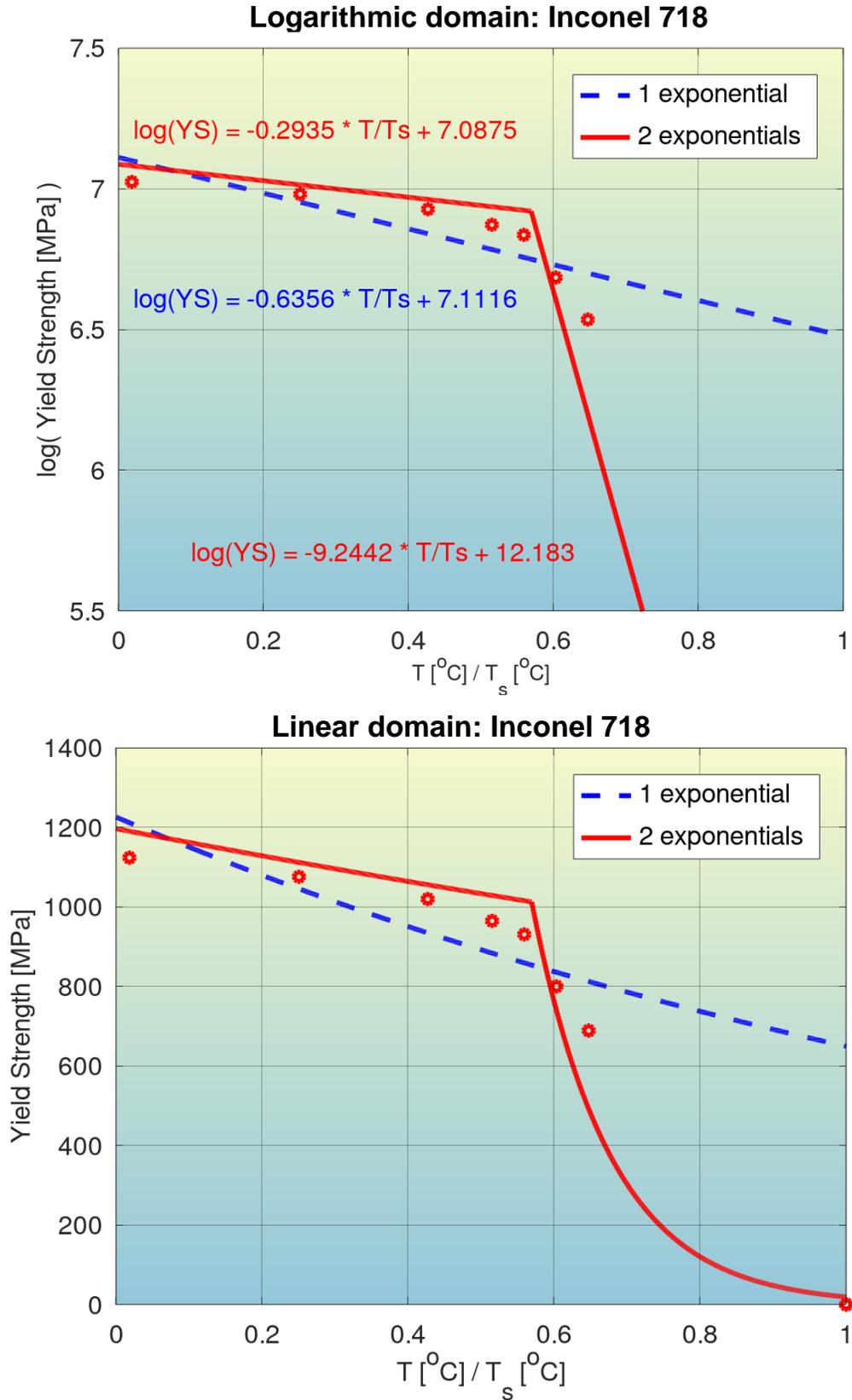

**Figure S18**: Quantification of modeling accuracy of the bilinear log model, for the composition No. 14 from Table S1 (Inconel 718), and comparison to that of a model with a single exponential.

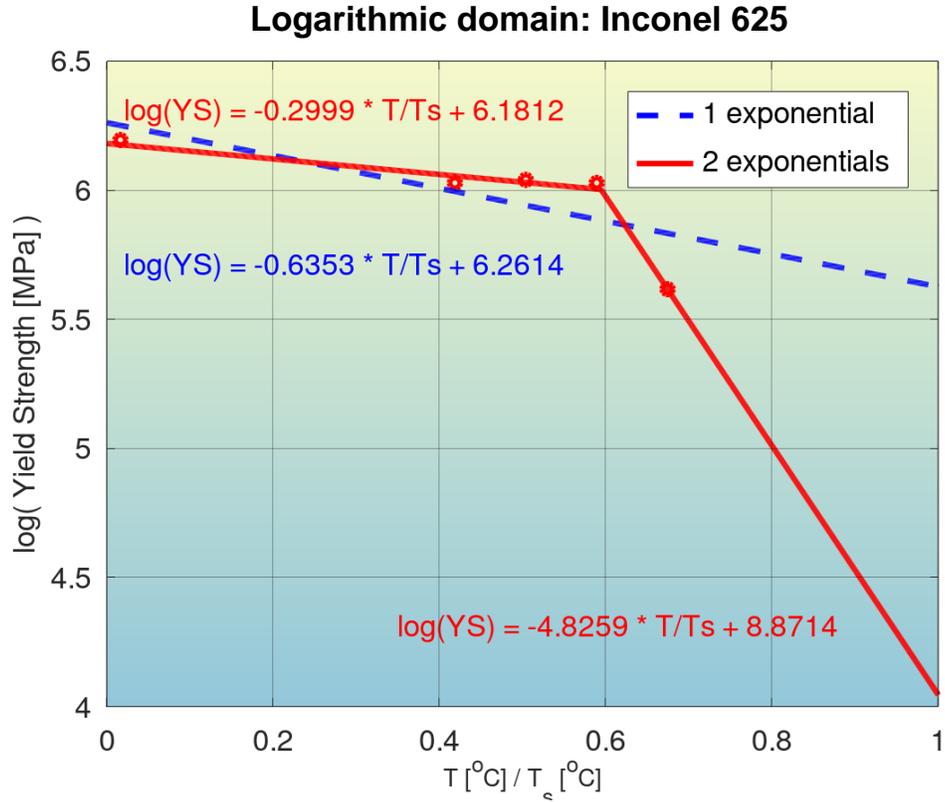

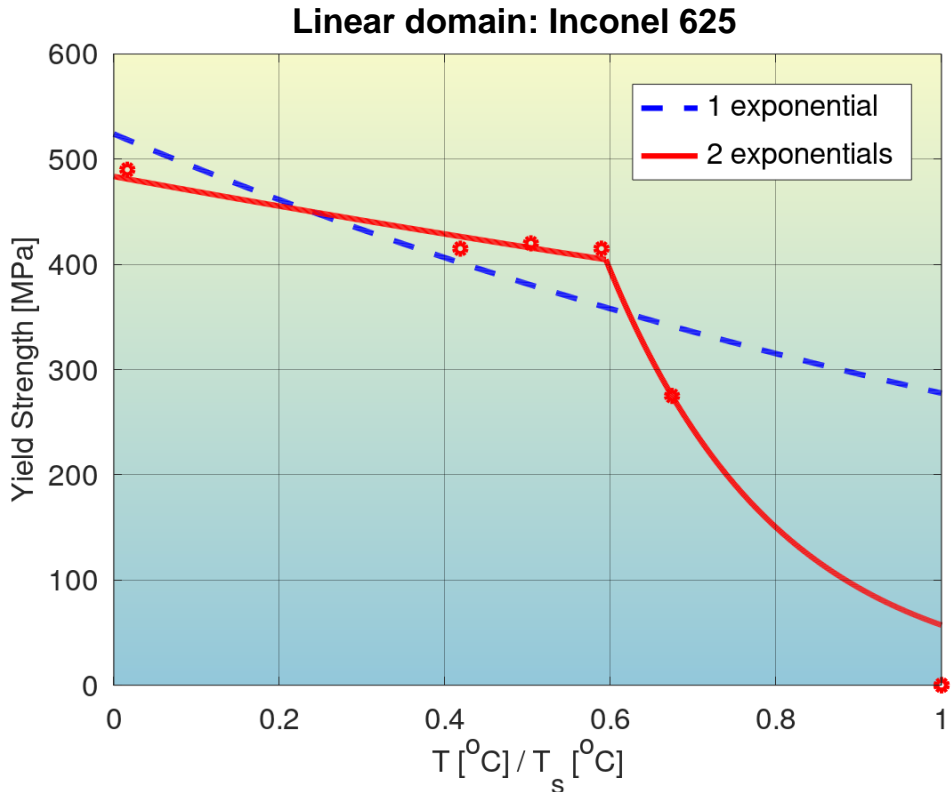

**Figure S19**: Quantification of modeling accuracy of the bilinear log model, for the composition No. 15 from Table S1 (Inconel 625), and comparison to that of a model with a single exponential. The outlier at room temperature has been excluded from modeling.

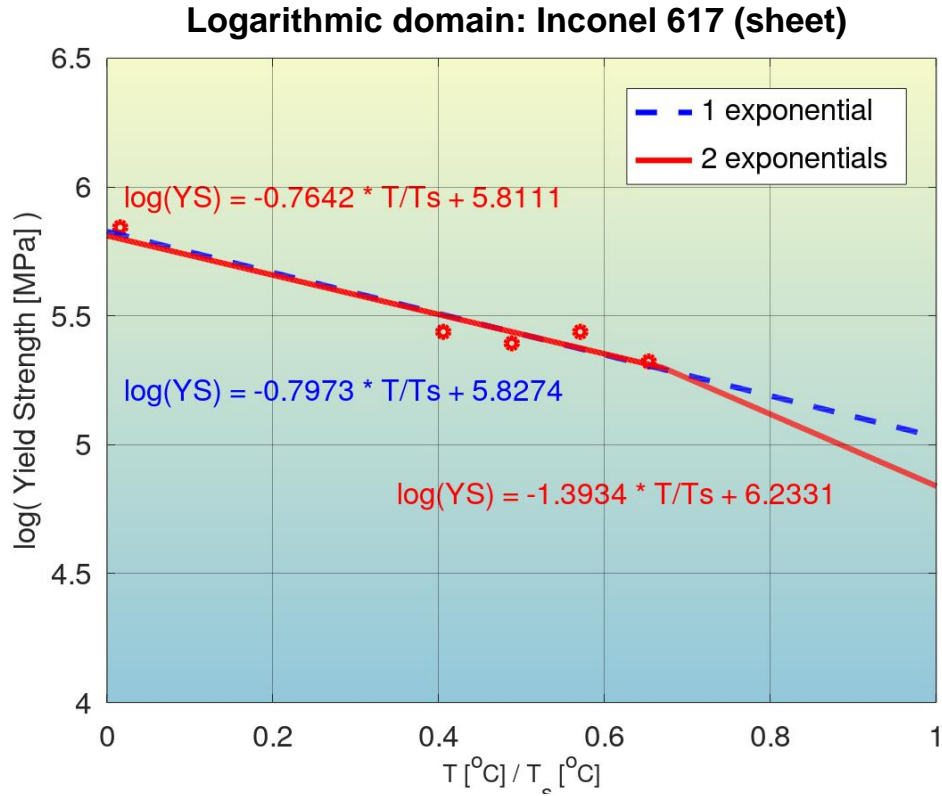

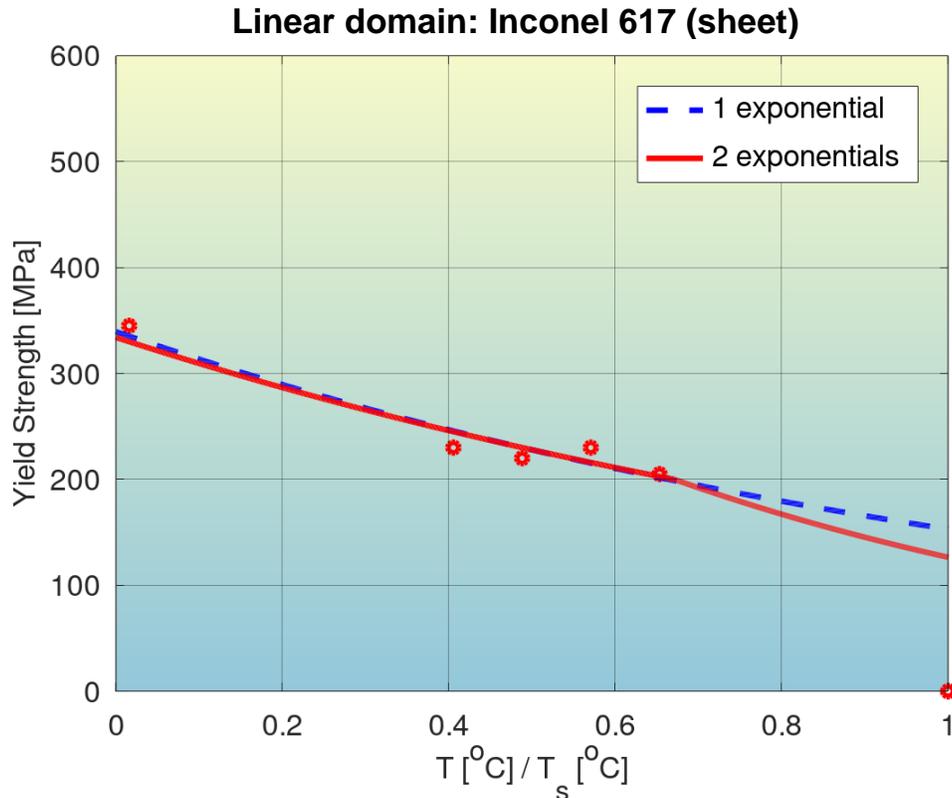

**Figure S20**: Quantification of modeling accuracy of the bilinear log model, for the composition No. 16 from Table S1 (Inconel 617 sheet), and comparison to that of a model with a single exponential. One outlier has been excluded from the modeling.

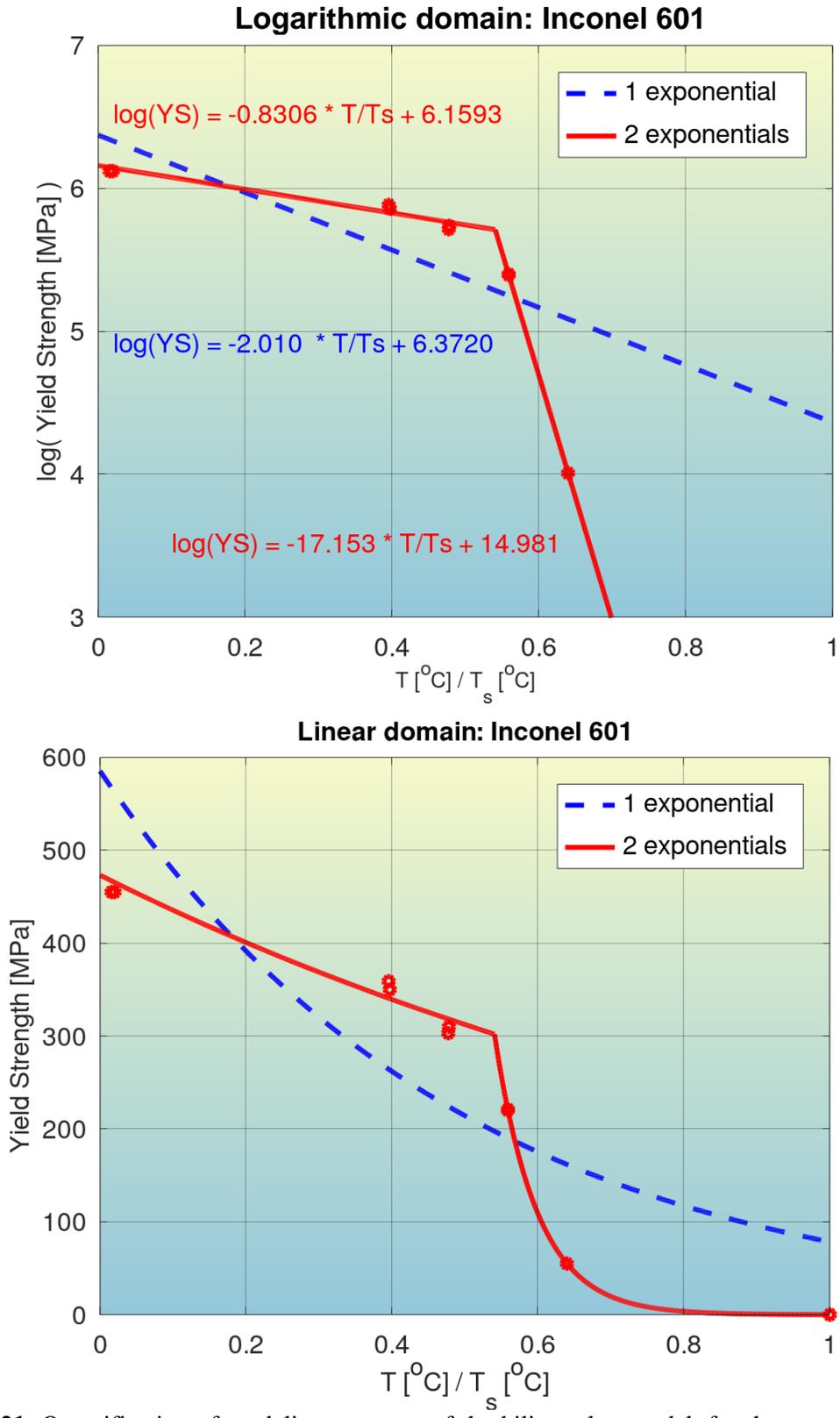

**Figure S21**: Quantification of modeling accuracy of the bilinear log model, for the composition No. 17 from Table S1 (Inconel 601), and comparison to that of a model with a single exponential. One outlier has been excluded from the modeling.

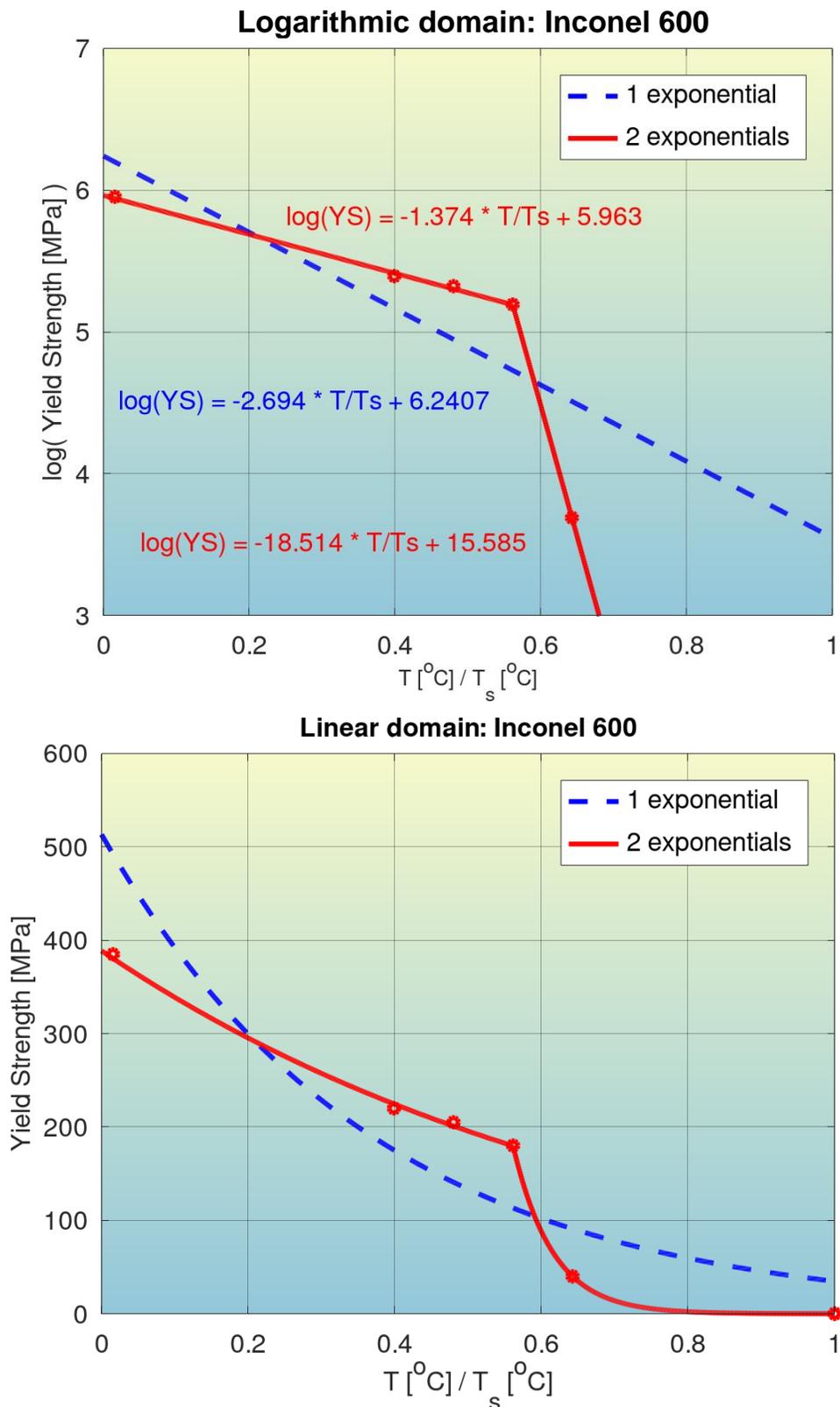

**Figure S22**: Quantification of modeling accuracy of the bilinear log model, for the composition No. 18 from Table S1 (Inconel 600), and comparison to that of a model with a single exponential. One outlier has been excluded from the modeling.

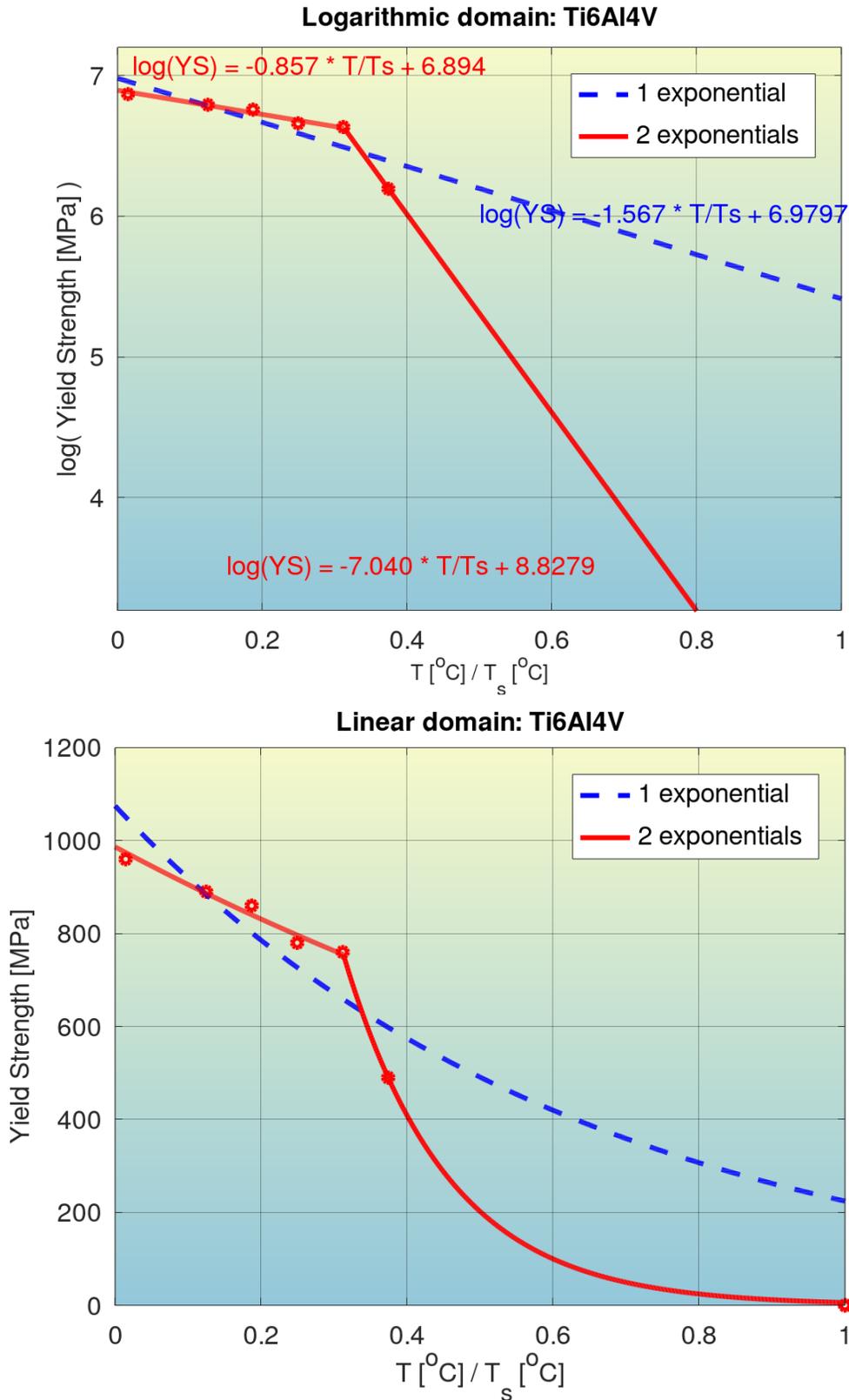

**Figure S23**: Quantification of modeling accuracy of the bilinear log model, for the composition No. 19 from Table S1 (Ti$_6$Al$_4$V), and comparison to that of a model with a single exponential. One outlier has been excluded from the modeling.

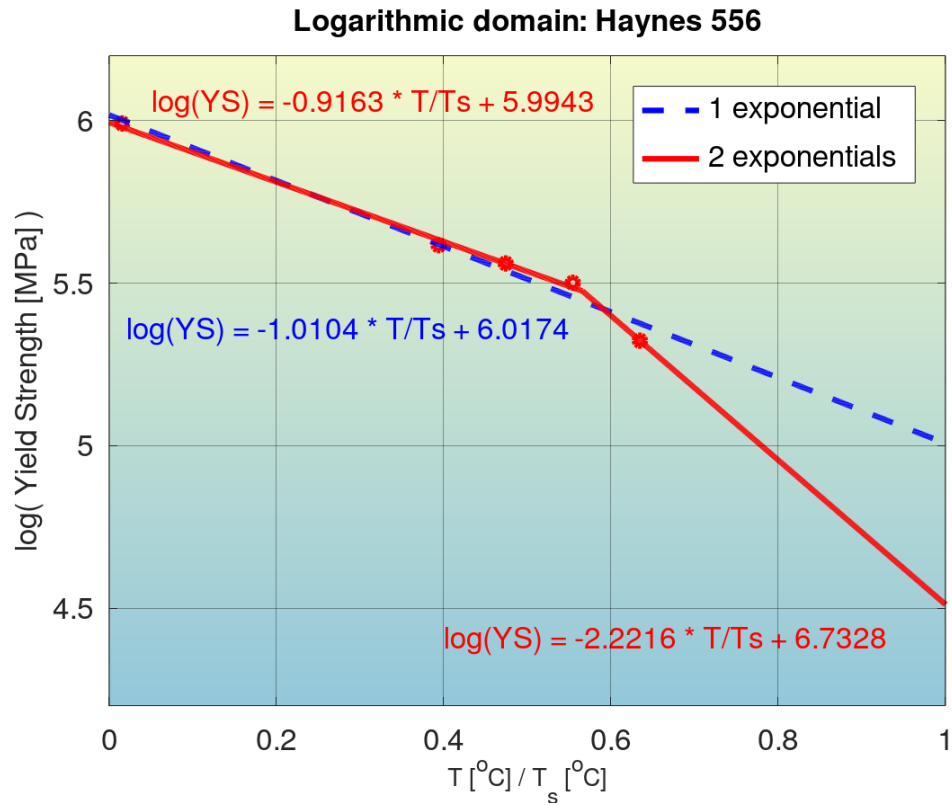

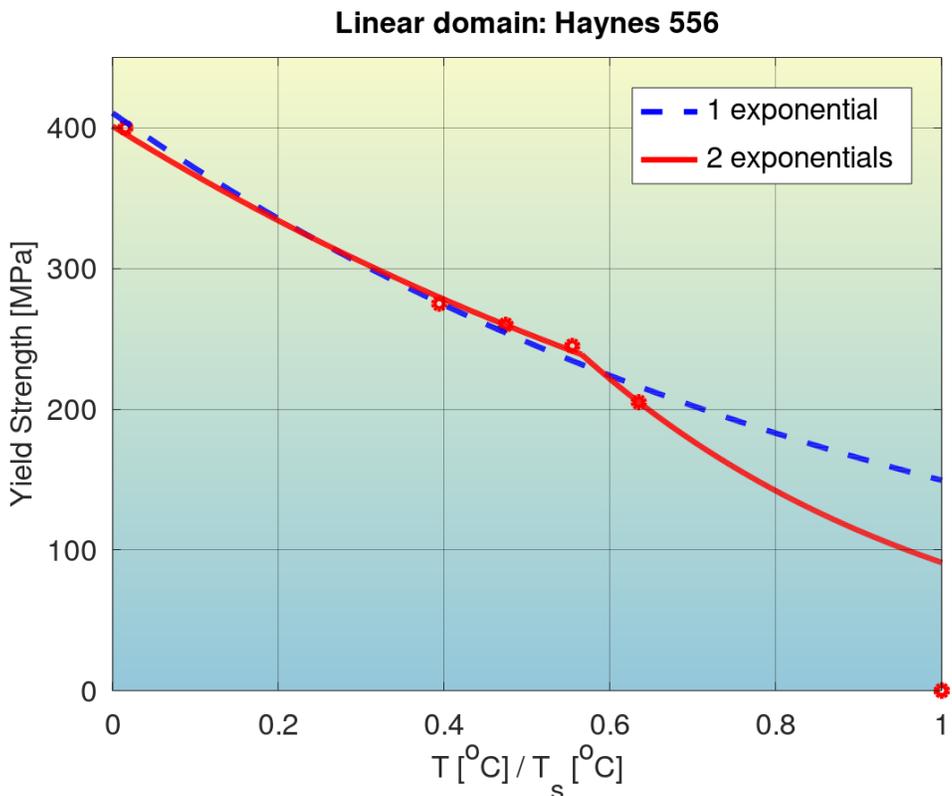

**Figure S24**: Quantification of modeling accuracy of the bilinear log model, for the composition No. 20 from Table S1 (Haynes 556), and comparison to that of a model with a single exponential. One outlier has been excluded from the modeling.

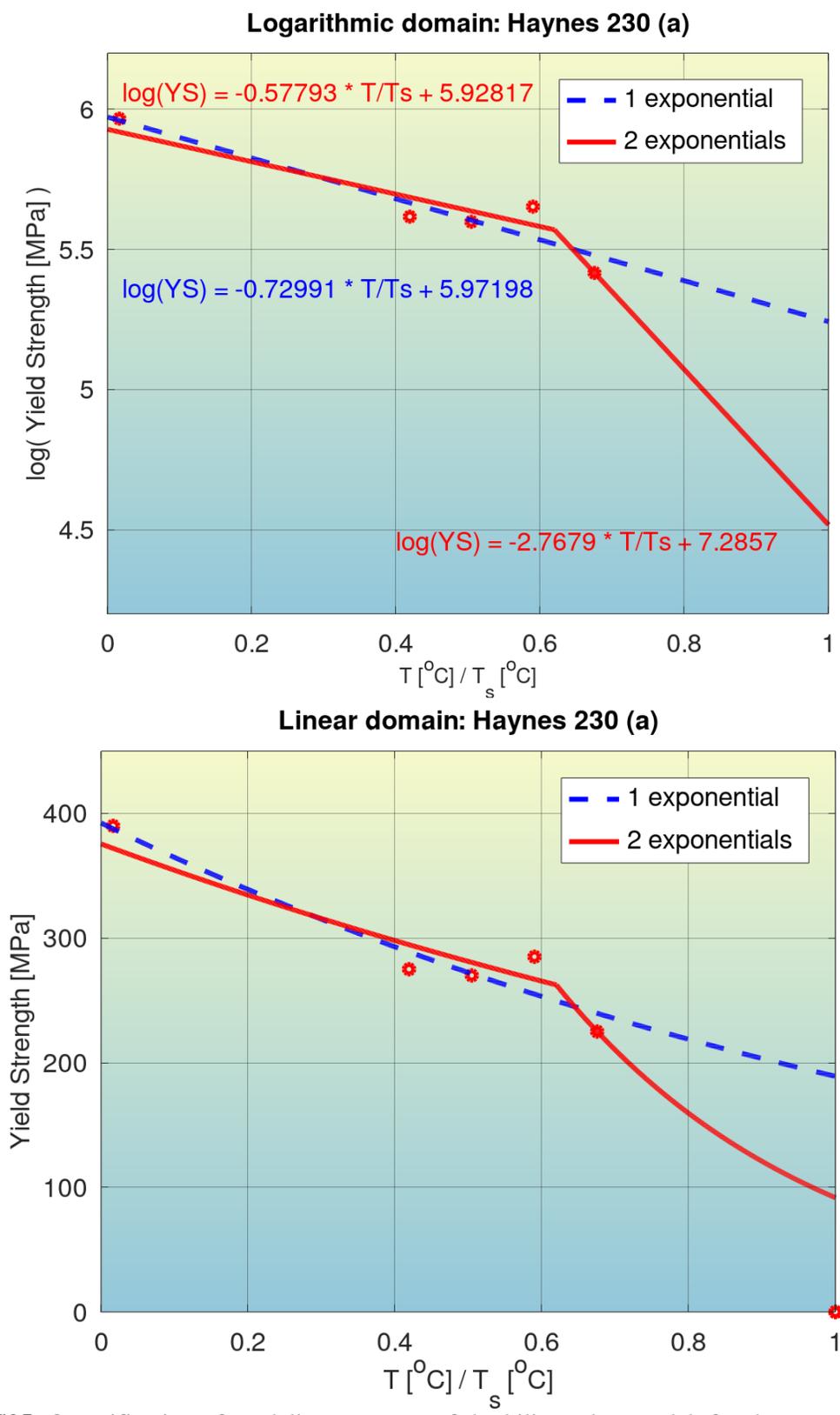

**Figure S25**: Quantification of modeling accuracy of the bilinear log model, for the composition No. 21 from Table S1 (Haynes 230 a), and comparison to that of a model with a single exponential. One outlier has been excluded from the modeling.

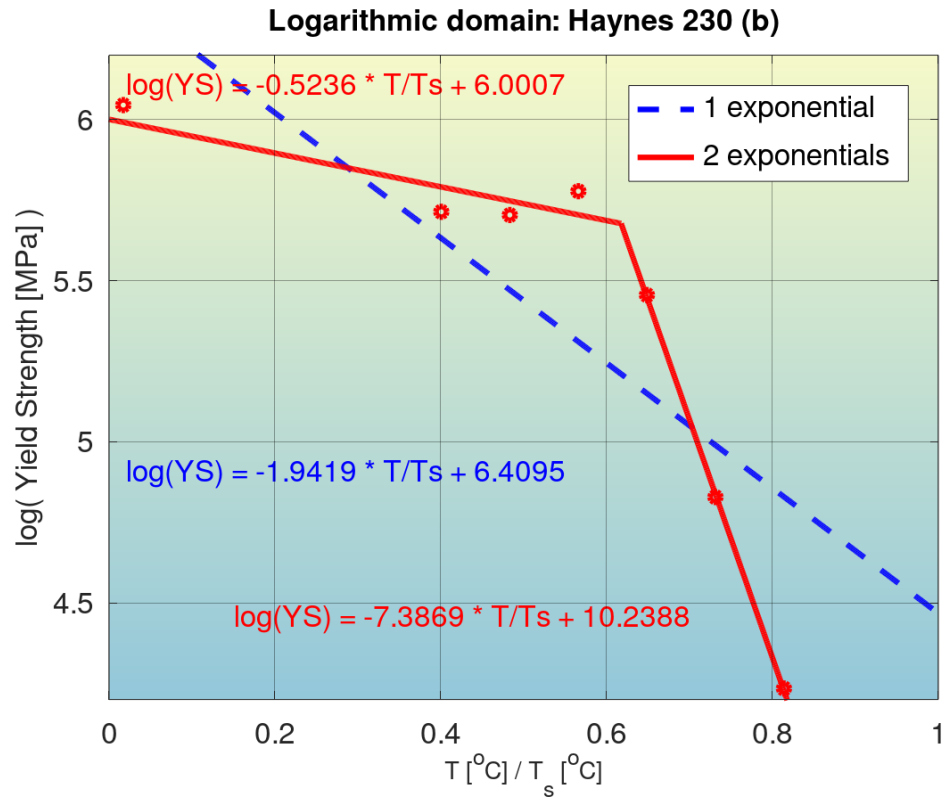

**Figure S26**: Quantification of modeling accuracy of the bilinear log model, for the composition No. 22 from Table S1 (Haynes 230 b), and comparison to that of a model with a single exponential. One outlier has been excluded from the modeling.

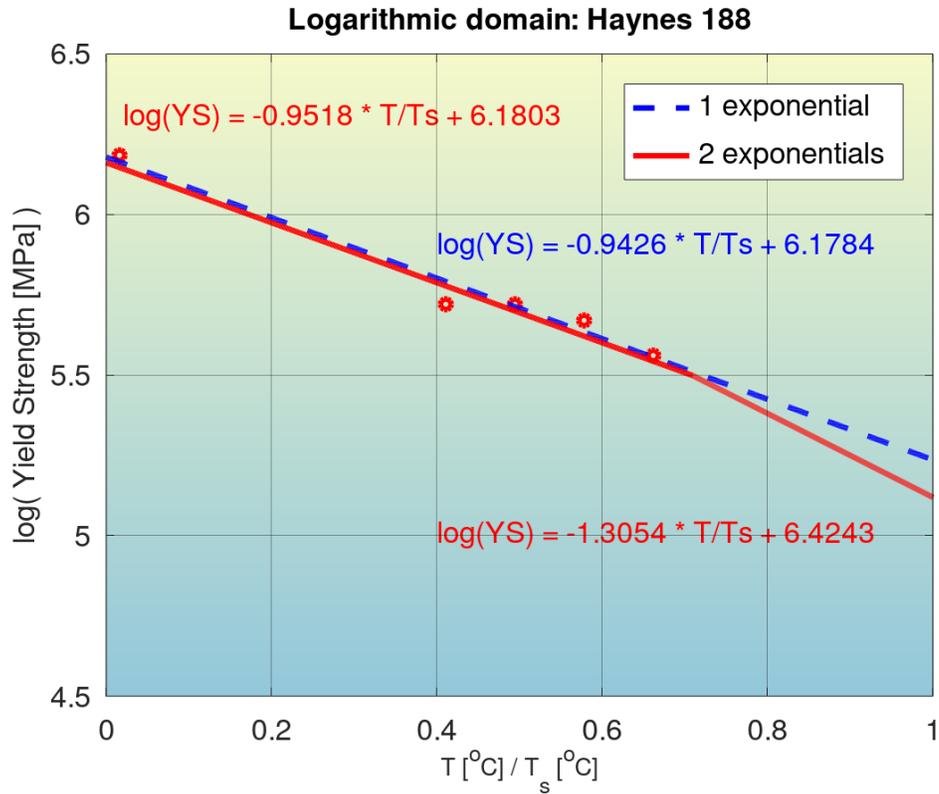

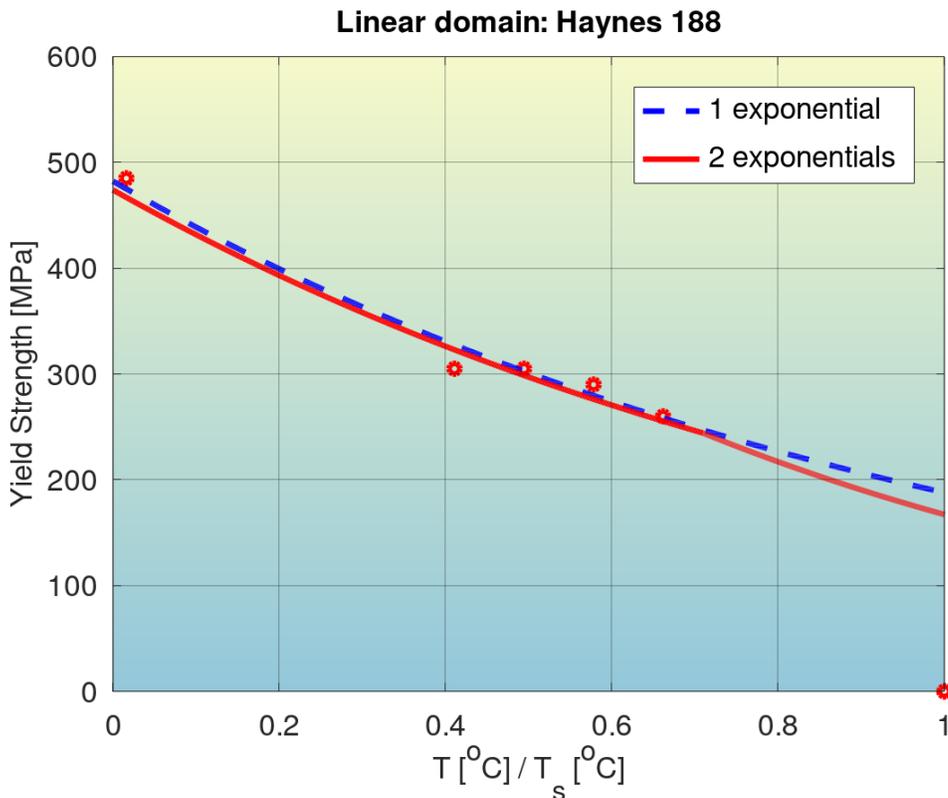

**Figure S27**: Quantification of modeling accuracy of the bilinear log model, for the composition No. 23 from Table S1 (Haynes 188), and comparison to that of a model with a single exponential. One outlier has been excluded from the modeling.

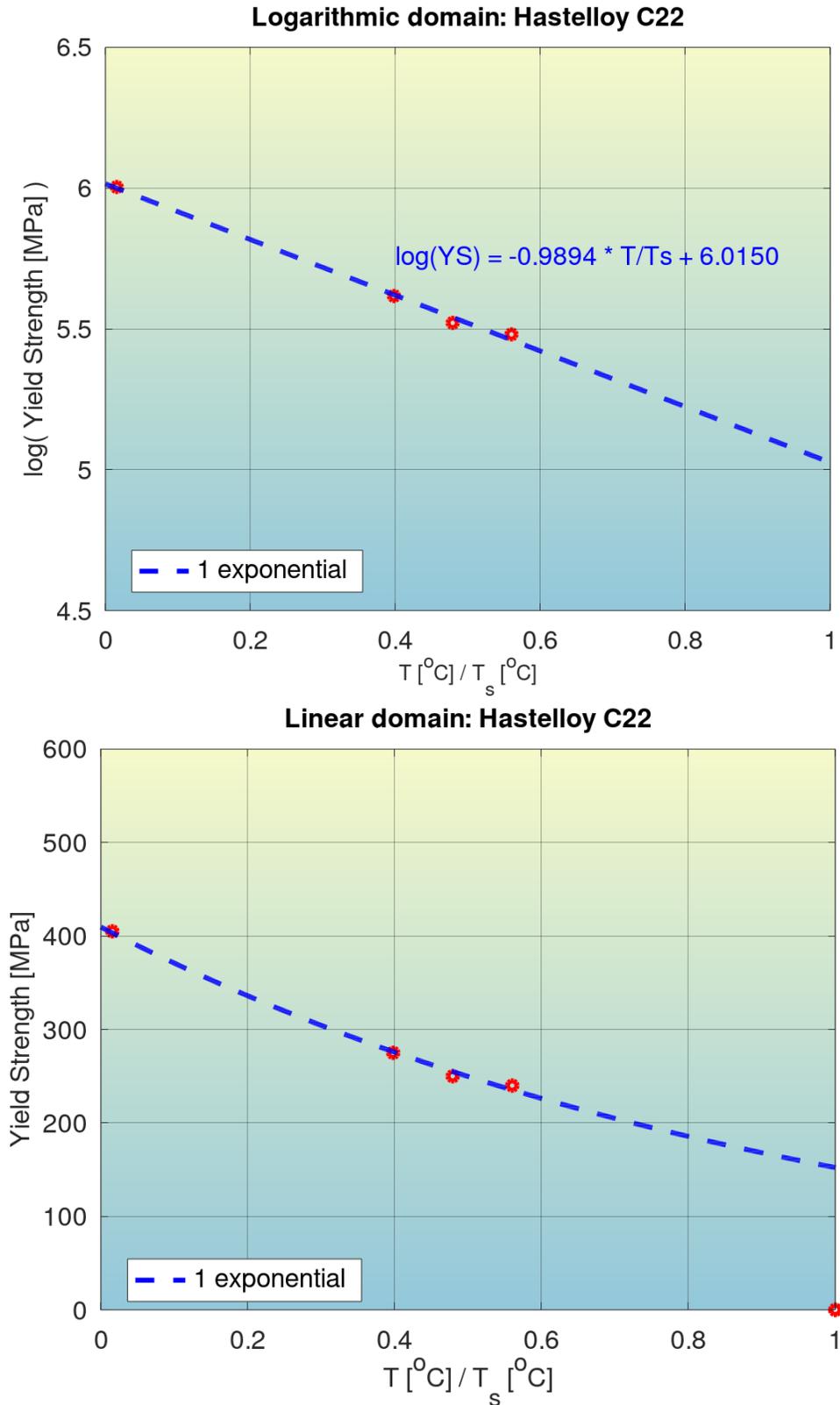

**Figure S28**: Quantification of modeling accuracy of the bilinear log model, for the composition No. 24 from Table S1 (Hastelloy C22), and comparison to that of a model with a single exponential. One outlier has been excluded from the modeling.

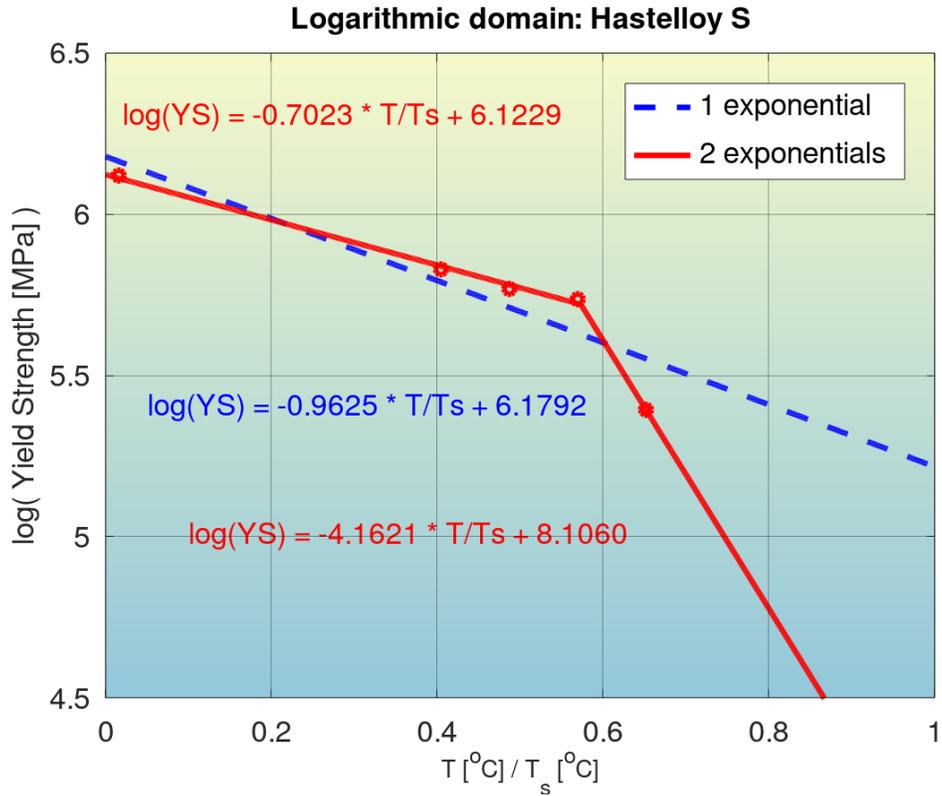

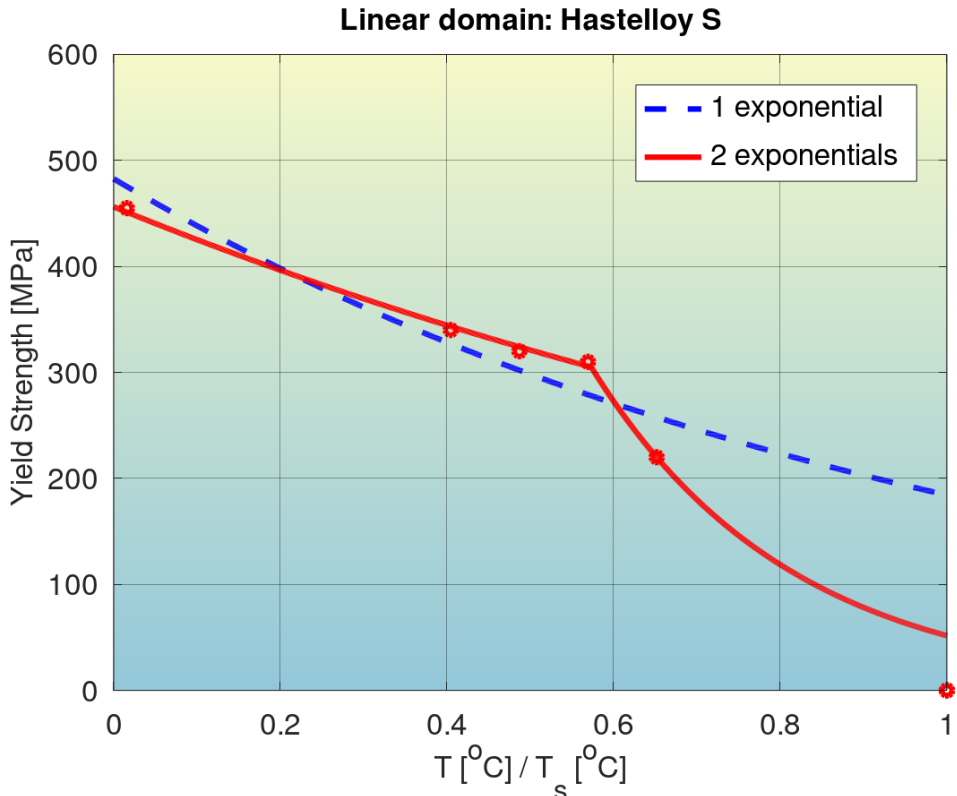

**Figure S29**: Quantification of modeling accuracy of the bilinear log model, for the composition No. 25 from Table S1 (Hastelloy S), and comparison to that of a model with a single exponential. One outlier has been excluded from the modeling.

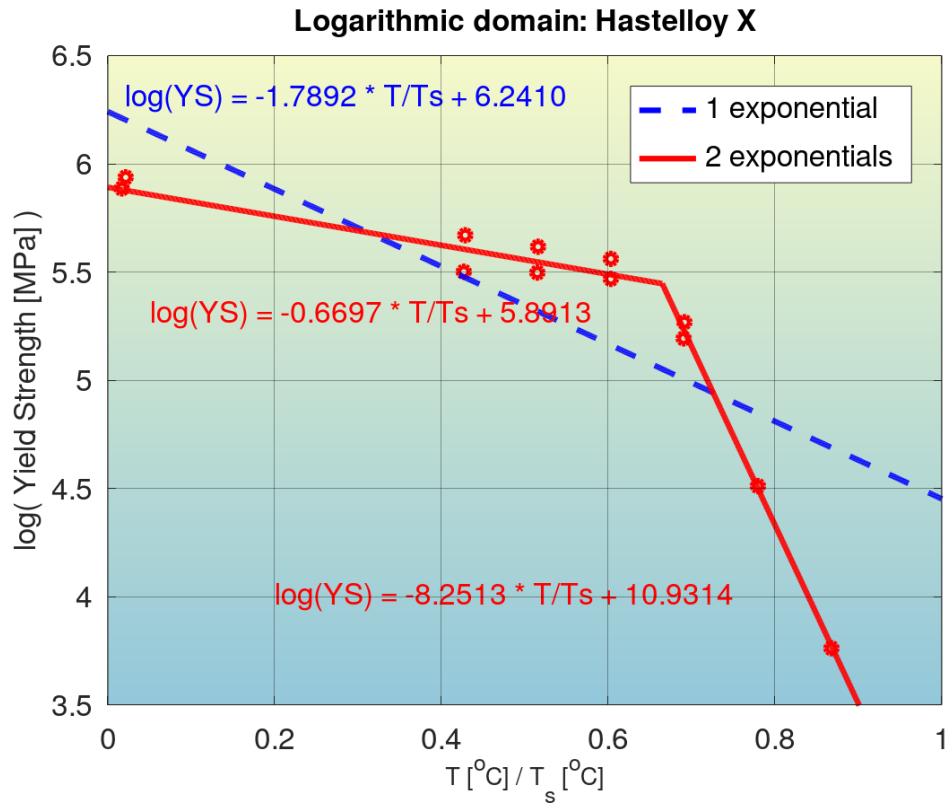

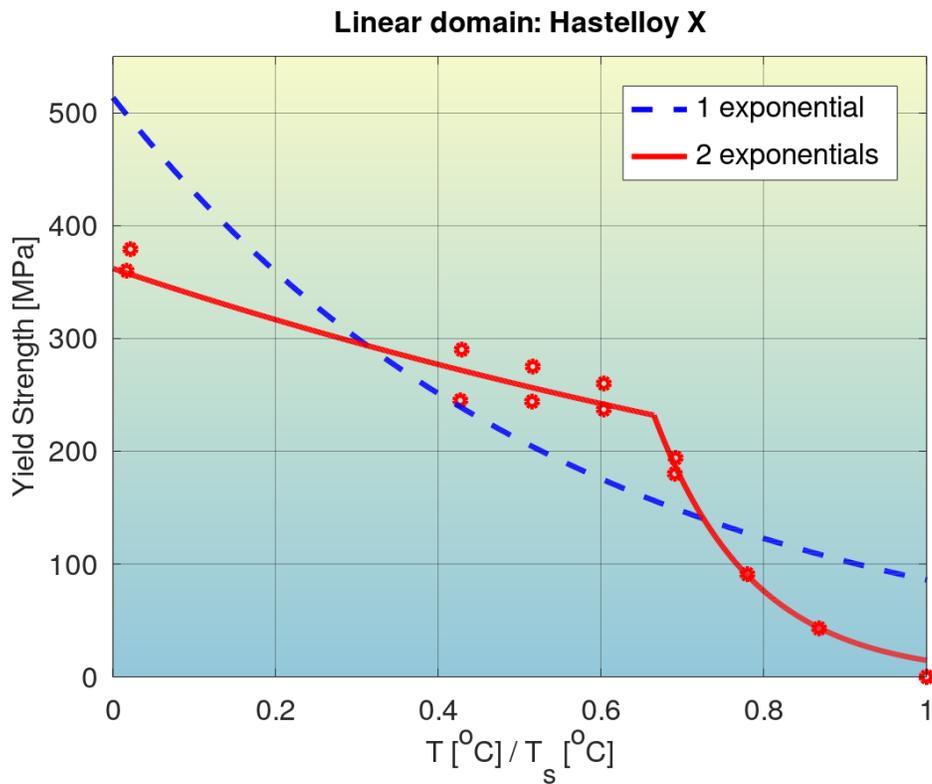

**Figure S30**: Quantification of modeling accuracy of the bilinear log model, for the composition No. 26 from Table S1 (Hastelloy X), and comparison to that of a model with a single exponential. One outlier has been excluded from the modeling.

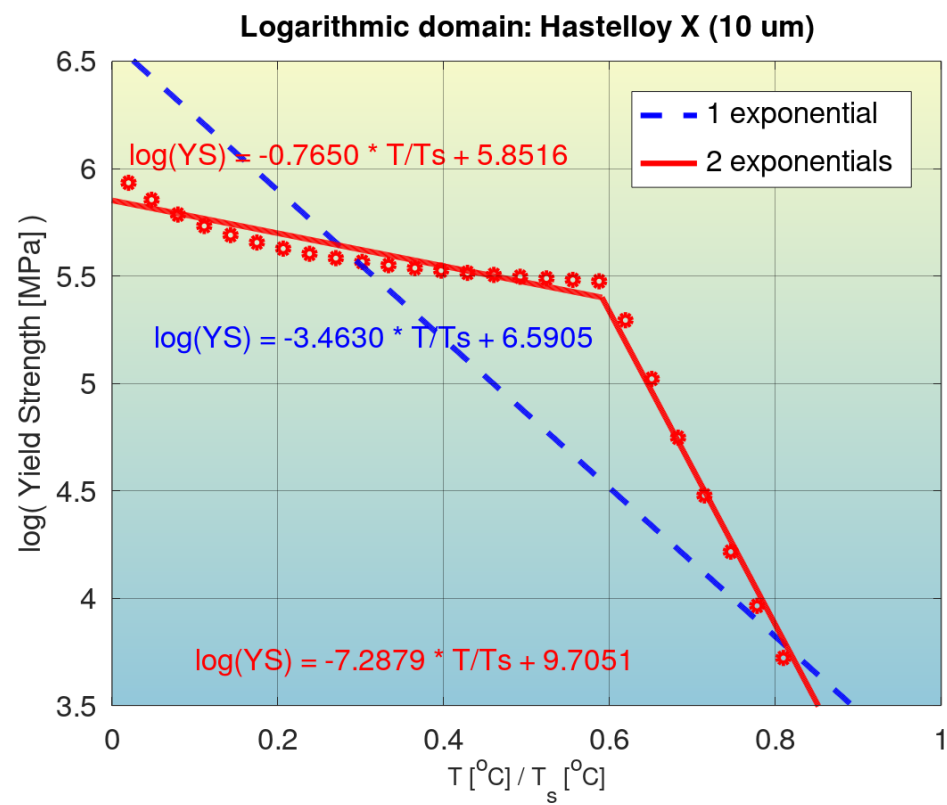

**Figure S31**: Quantification of modeling accuracy of the bilinear log model, for the composition No. 27 from Table S1 (Hastelloy X 10 μm), and comparison to that of a model with a single exponential. One outlier has been excluded from the modeling.

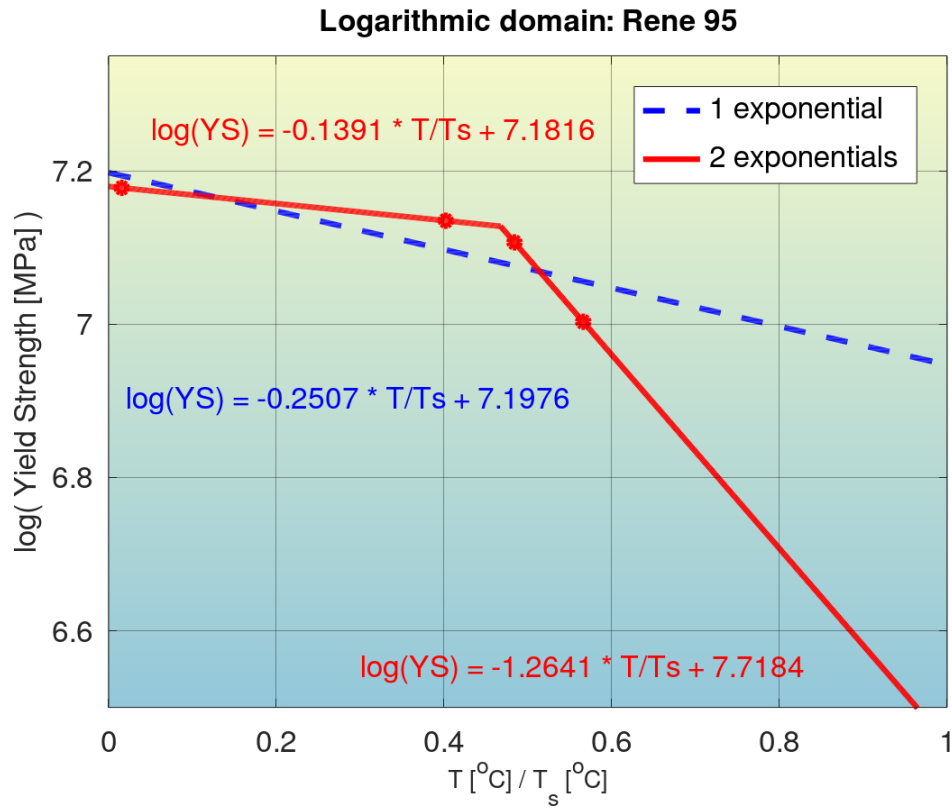

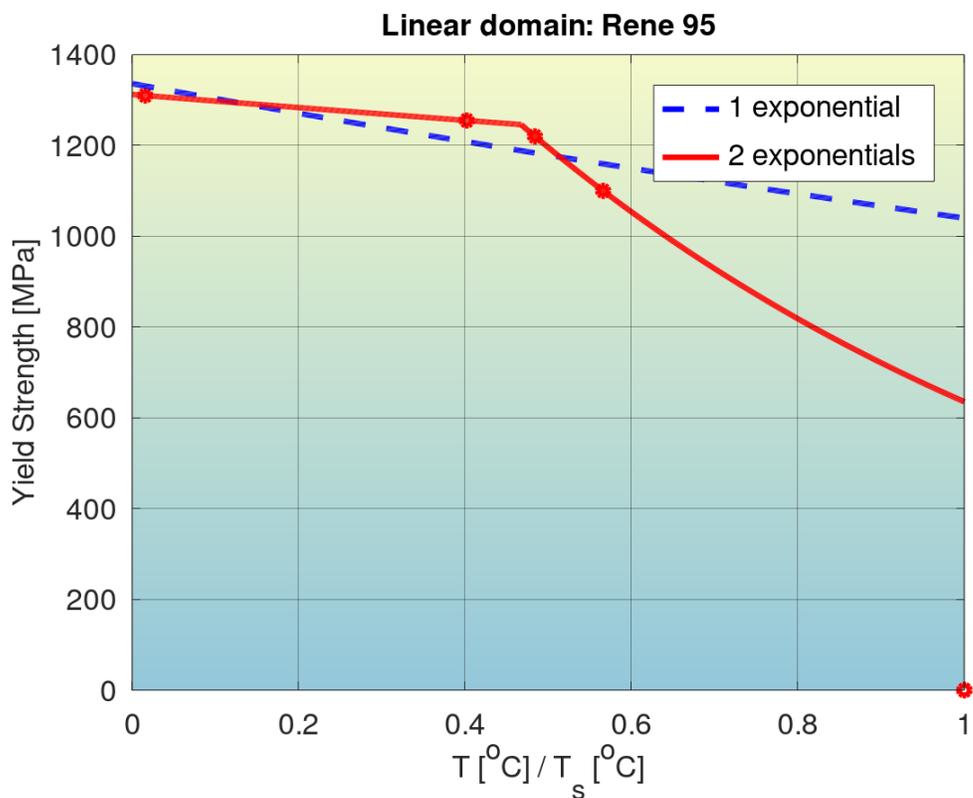

**Figure S32**: Quantification of modeling accuracy of the bilinear log model, for the composition No. 28 from Table S1 (Rene 95), and comparison to that of a model with a single exponential. One outlier has been excluded from the modeling.

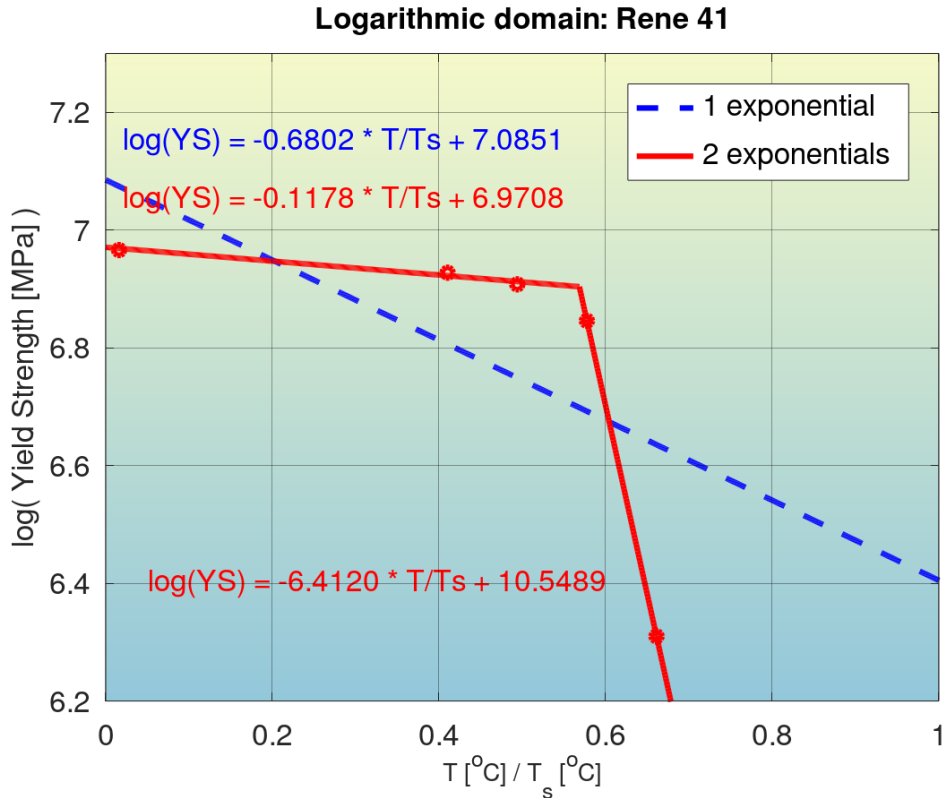

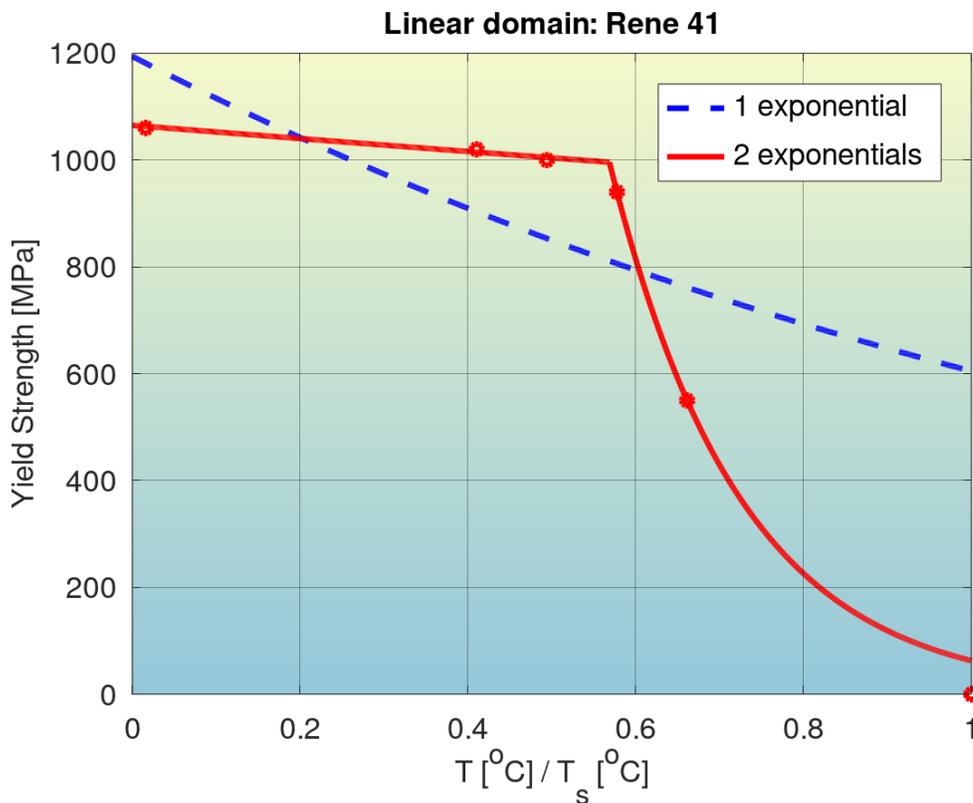

**Figure 33**: Quantification of modeling accuracy of the bilinear log model, for the composition No. 29 from Table S1 (Rene 41), and comparison to that of a model with a single exponential. One outlier has been excluded from the modeling.

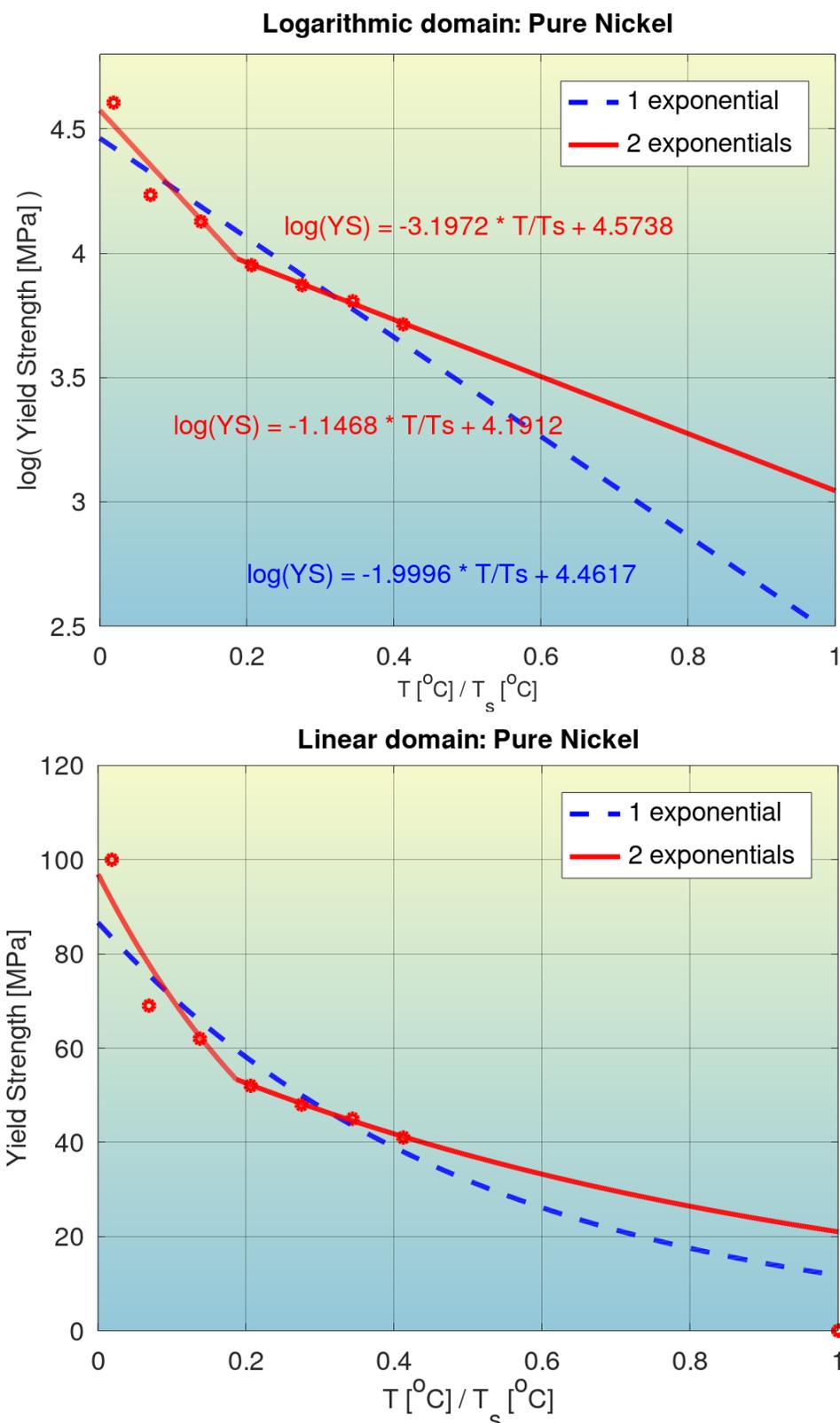

**Figure 34**: Quantification of modeling accuracy of the bilinear log model, for the composition No. 30 from Table S1 (Pure Nickel), and comparison to that of a model with a single exponential. One outlier has been excluded from the modeling.

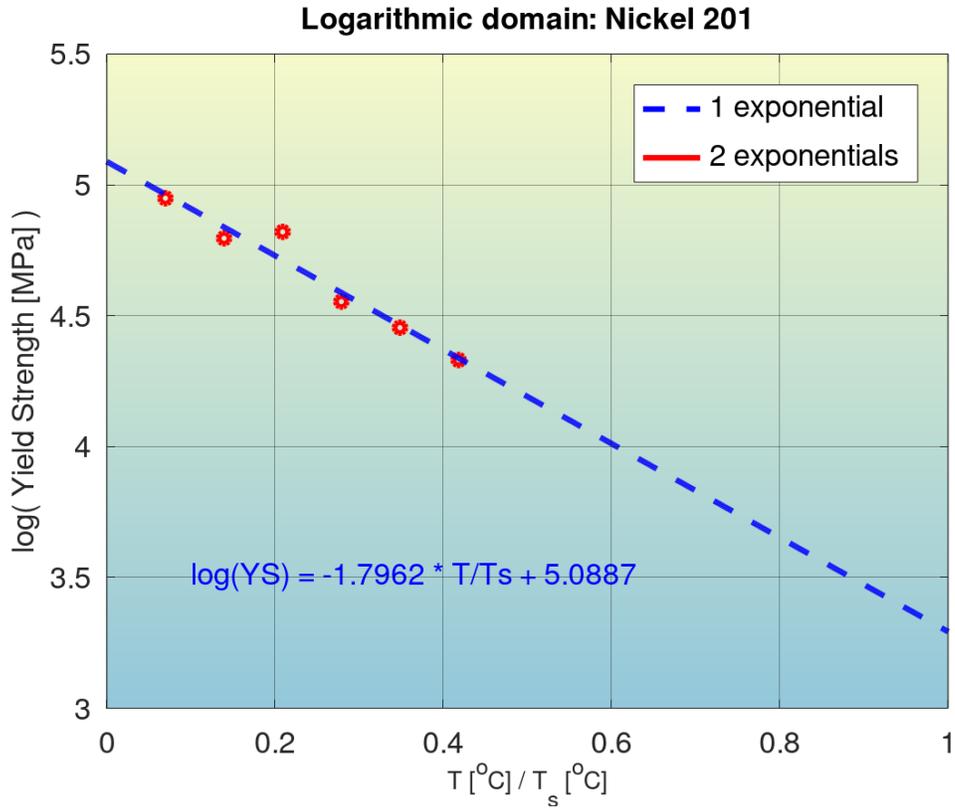

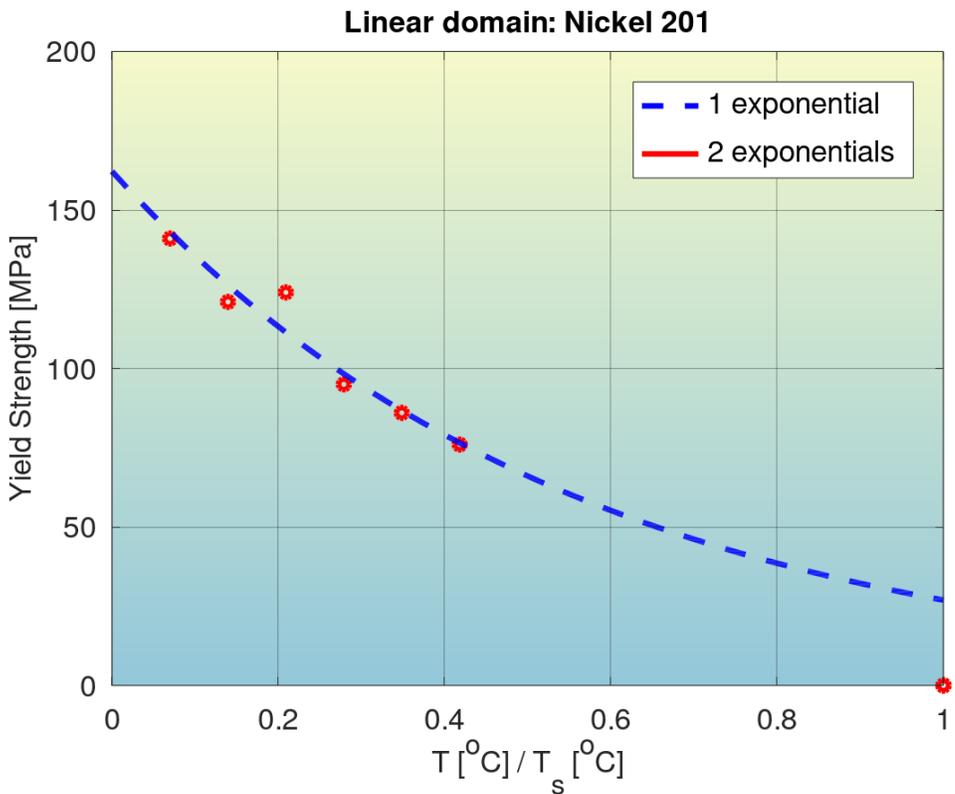

**Figure 35**: Quantification of modeling accuracy of the bilinear log model, for the composition No. 31 from Table S1 (Nickel 201), and comparison to that of a model with a single exponential. One outlier has been excluded from the modeling.

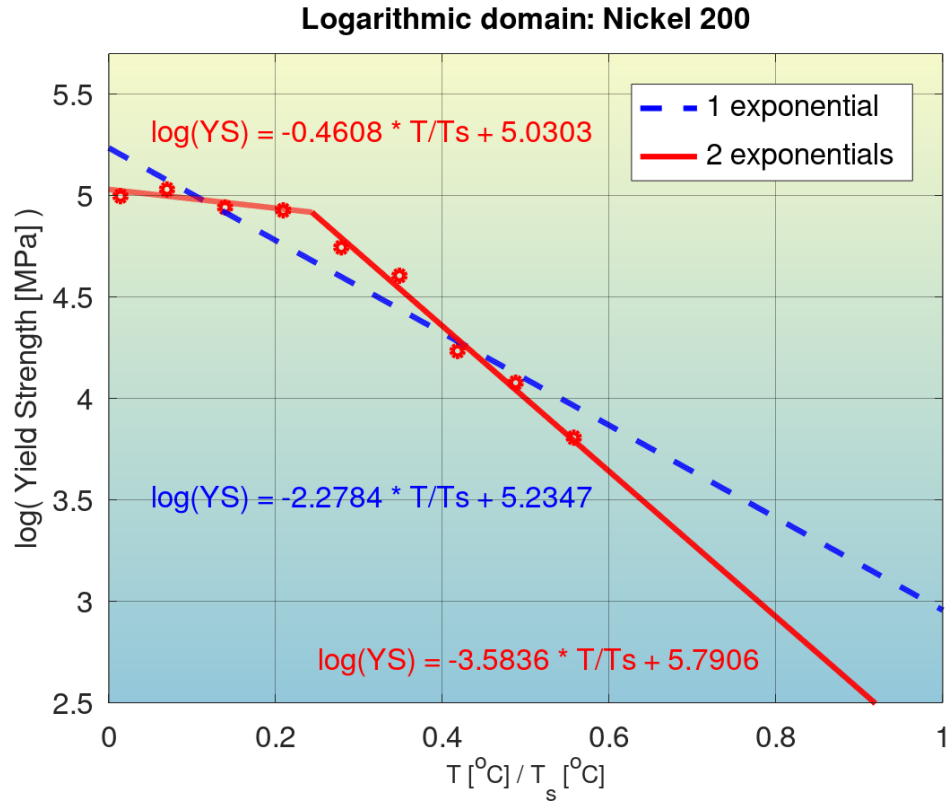

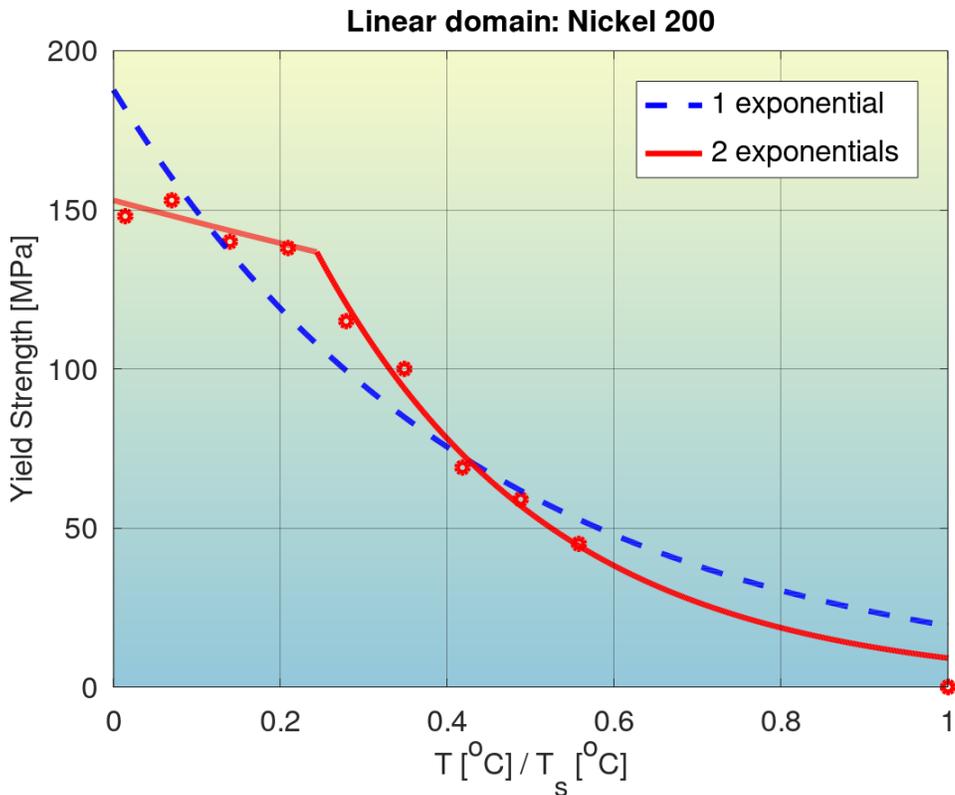

**Figure 36**: Quantification of modeling accuracy of the bilinear log model, for the composition No. 32 from Table S1 (Nickel 200), and comparison to that of a model with a single exponential. One outlier has been excluded from the modeling.

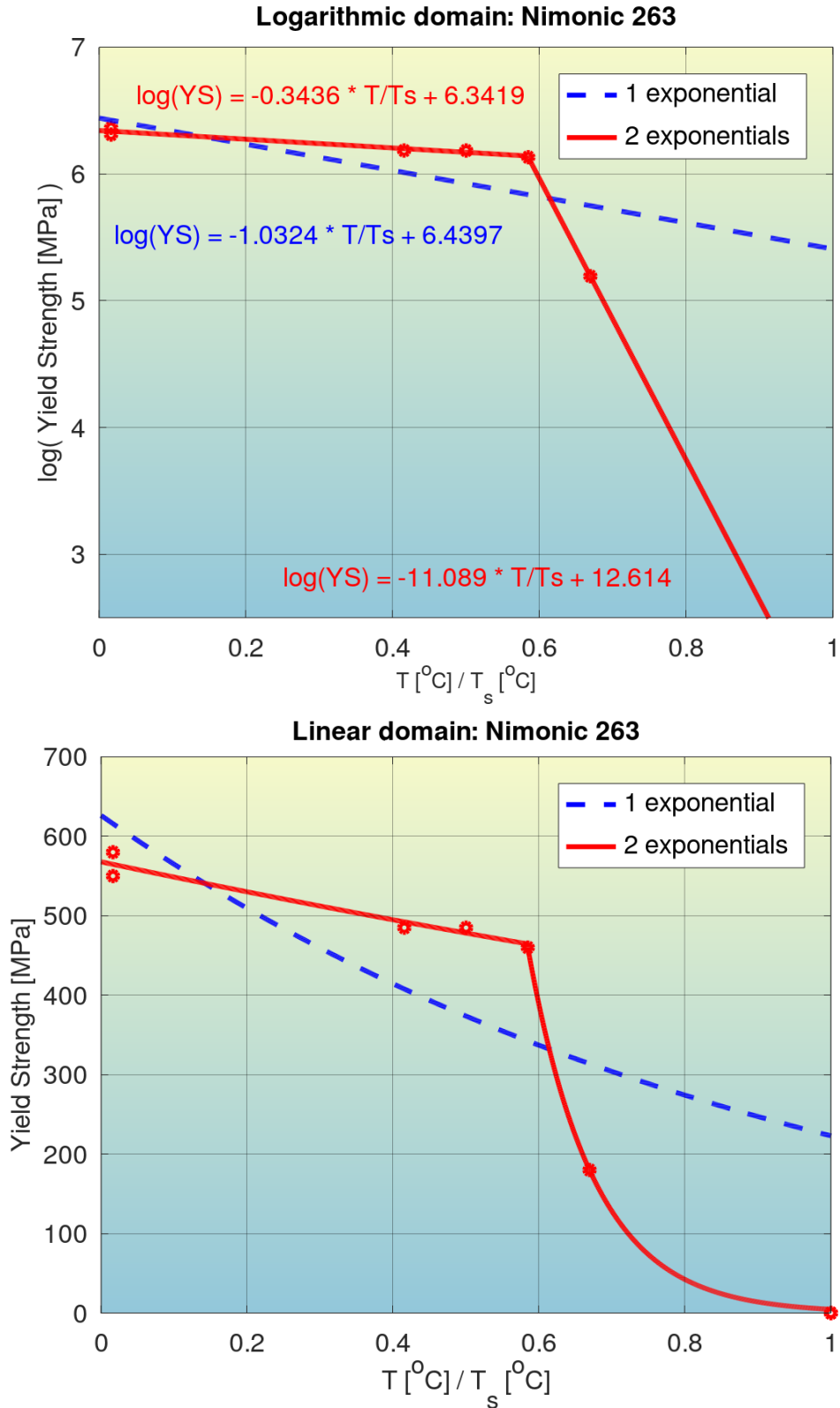

**Figure 37**: Quantification of modeling accuracy of the bilinear log model, for the composition No. 33 from Table S1 (Nimonic 263), and comparison to that of a model with a single exponential. One outlier has been excluded from the modeling.

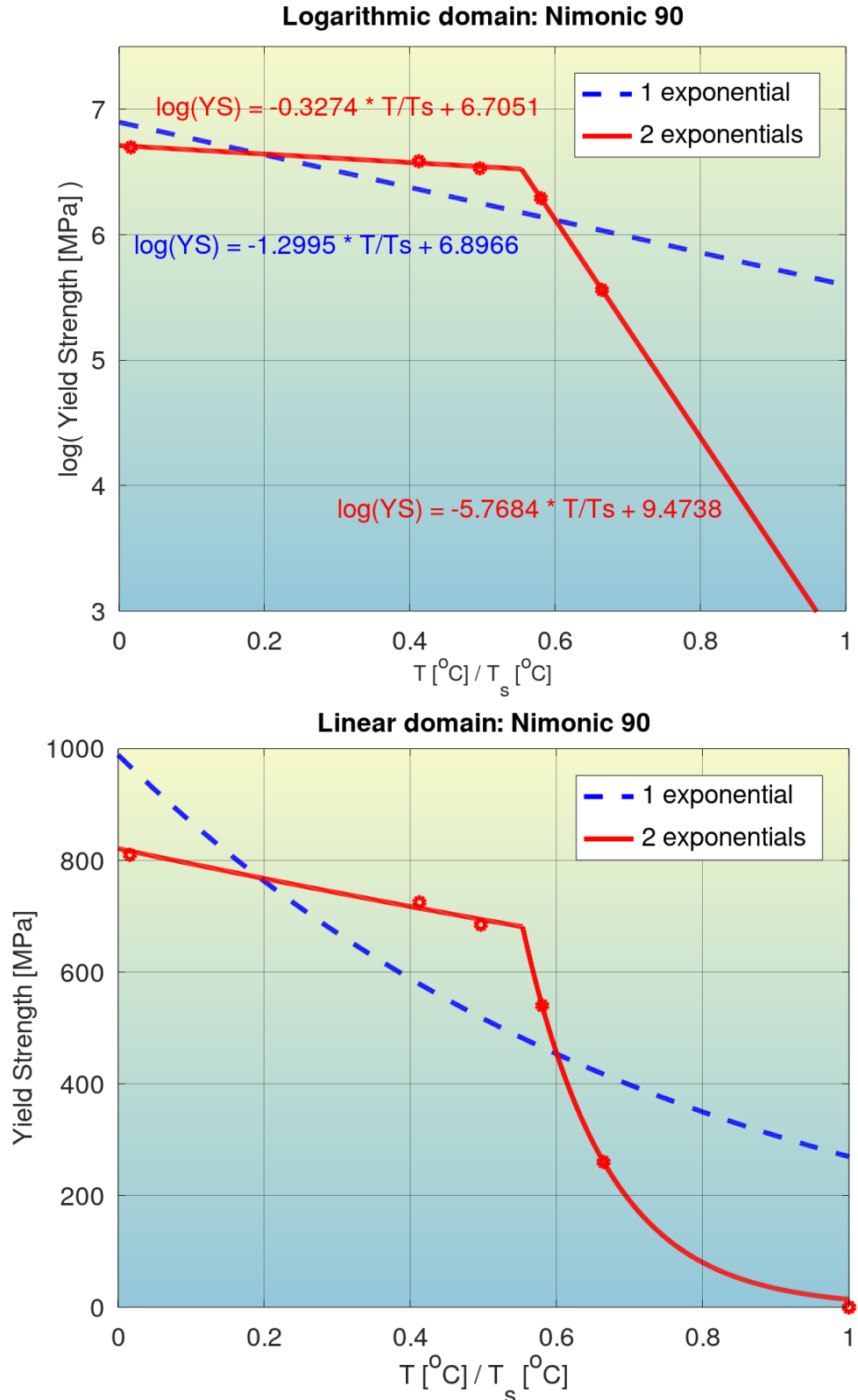

**Figure 38**: Quantification of modeling accuracy of the bilinear log model, for the composition No. 34 from Table S1 (Nimonic 90), and comparison to that of a model with a single exponential. One outlier has been excluded from the modeling.

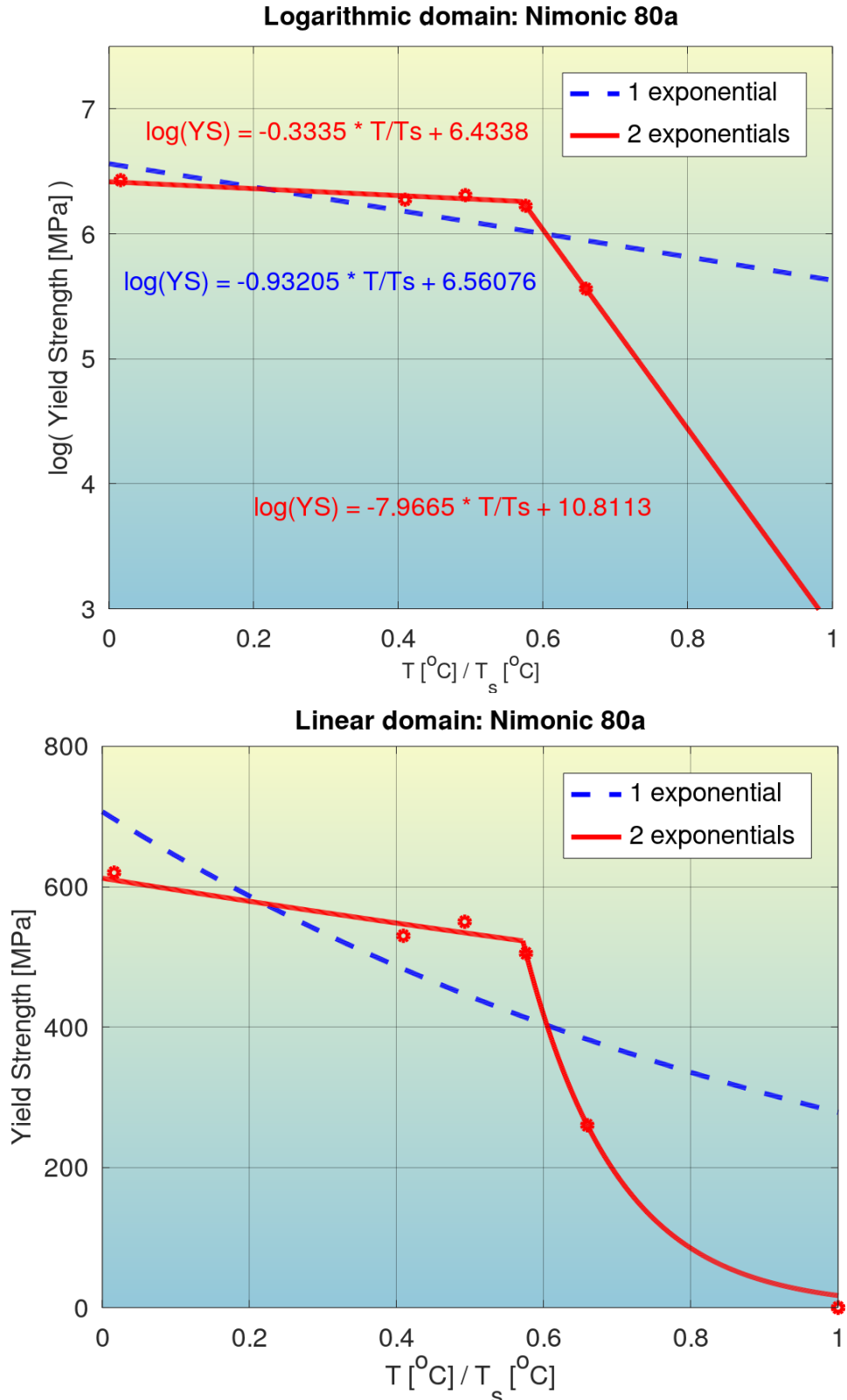

**Figure 39**: Quantification of modeling accuracy of the bilinear log model, for the composition No. 35 from Table S1 (Nimonic 80a), and comparison to that of a model with a single exponential. One outlier has been excluded from the modeling.

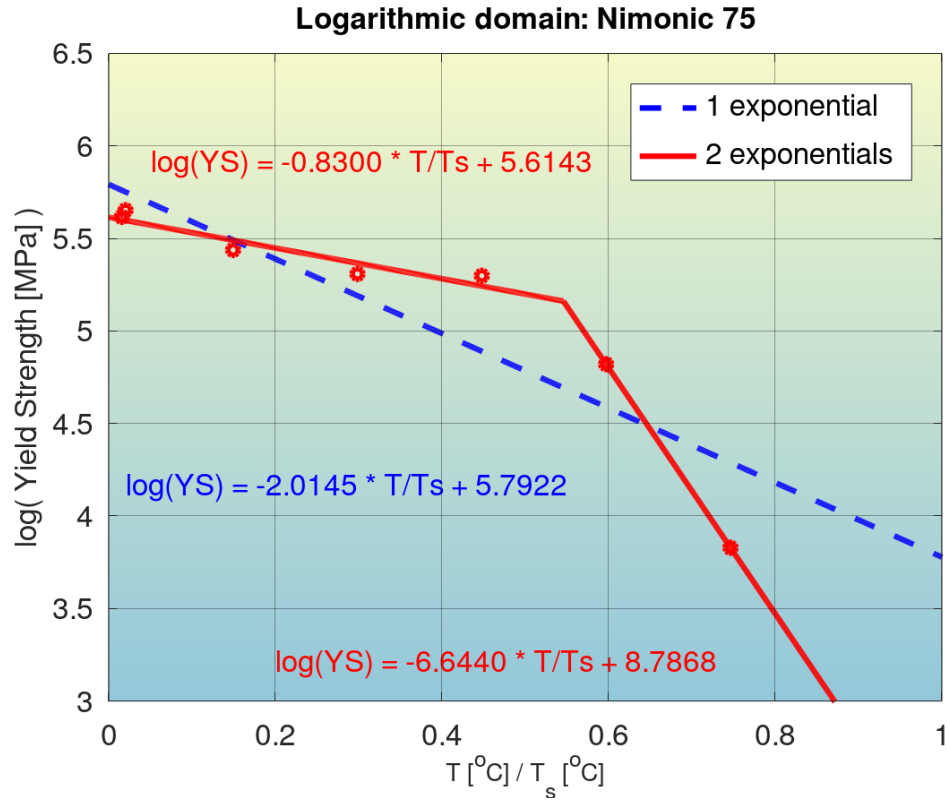

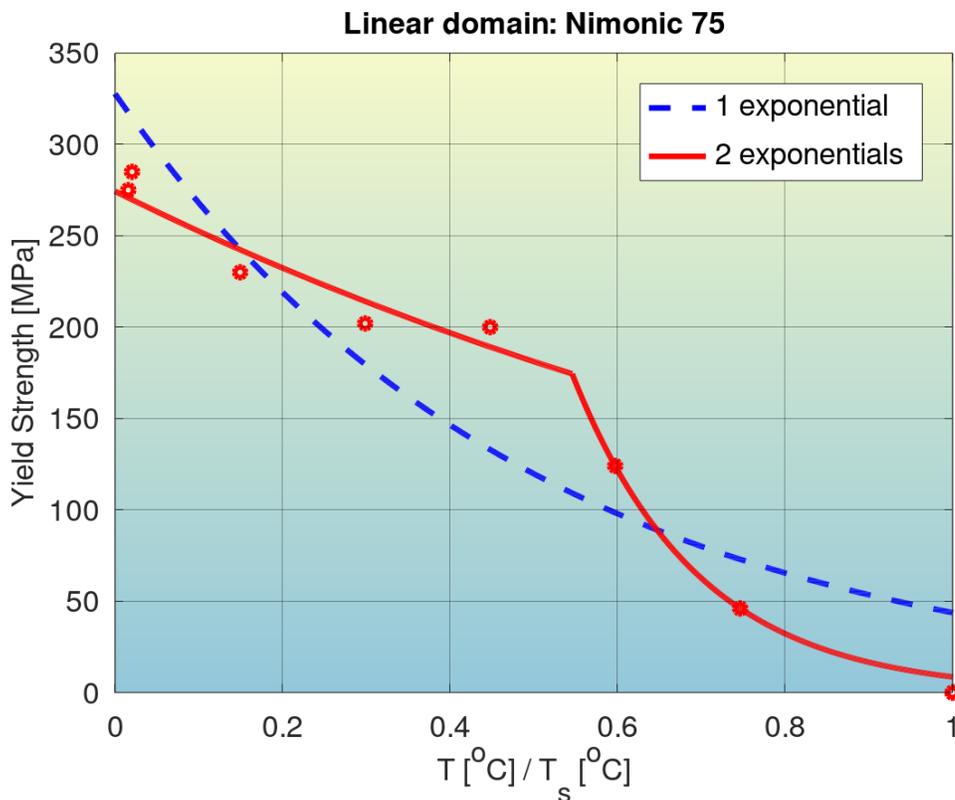

**Figure 40**: Quantification of modeling accuracy of the bilinear log model, for the composition No. 36 from Table S1 (Nimonic 75), and comparison to that of a model with a single exponential. One outlier has been excluded from the modeling.

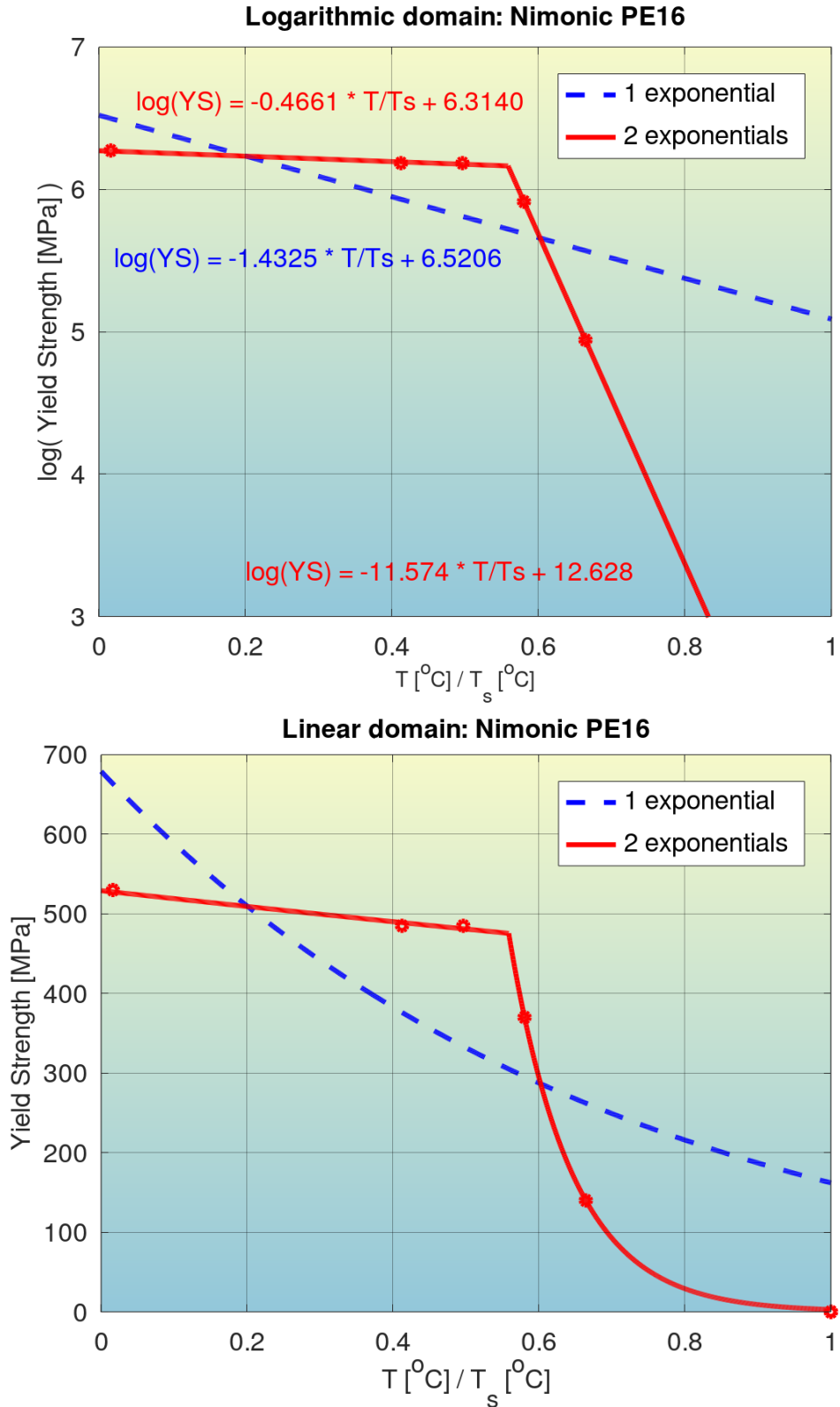

**Figure 41**: Quantification of modeling accuracy of the bilinear log model, for the composition No. 37 from Table S1 (Nimonic PE16), and comparison to that of a model with a single exponential. One outlier has been excluded from the modeling.

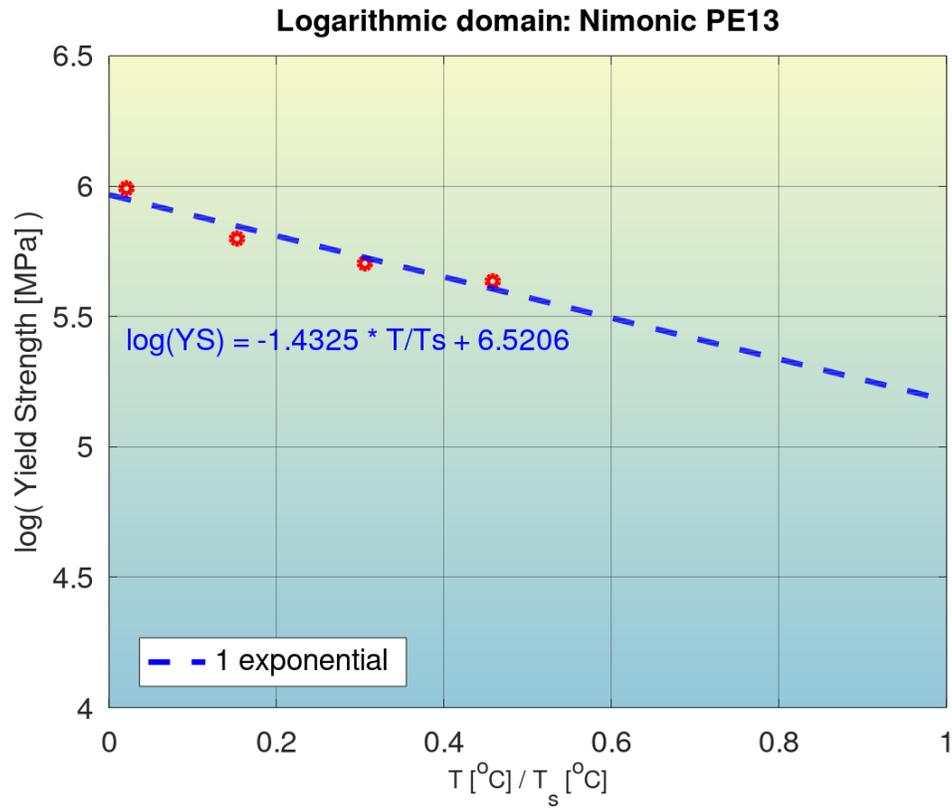

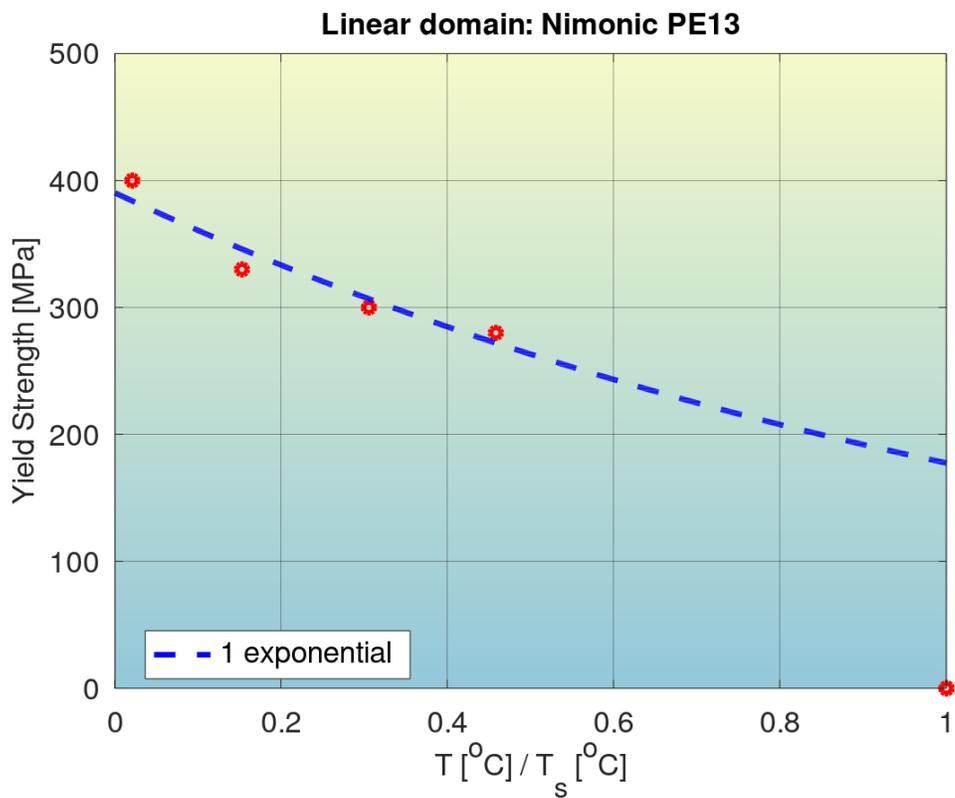

**Figure 42**: Quantification of modeling accuracy of the bilinear log model, for the composition No. 38 from Table S1 (Nimonic PE13), and comparison to that of a model with a single exponential. One outlier has been excluded from the modeling.

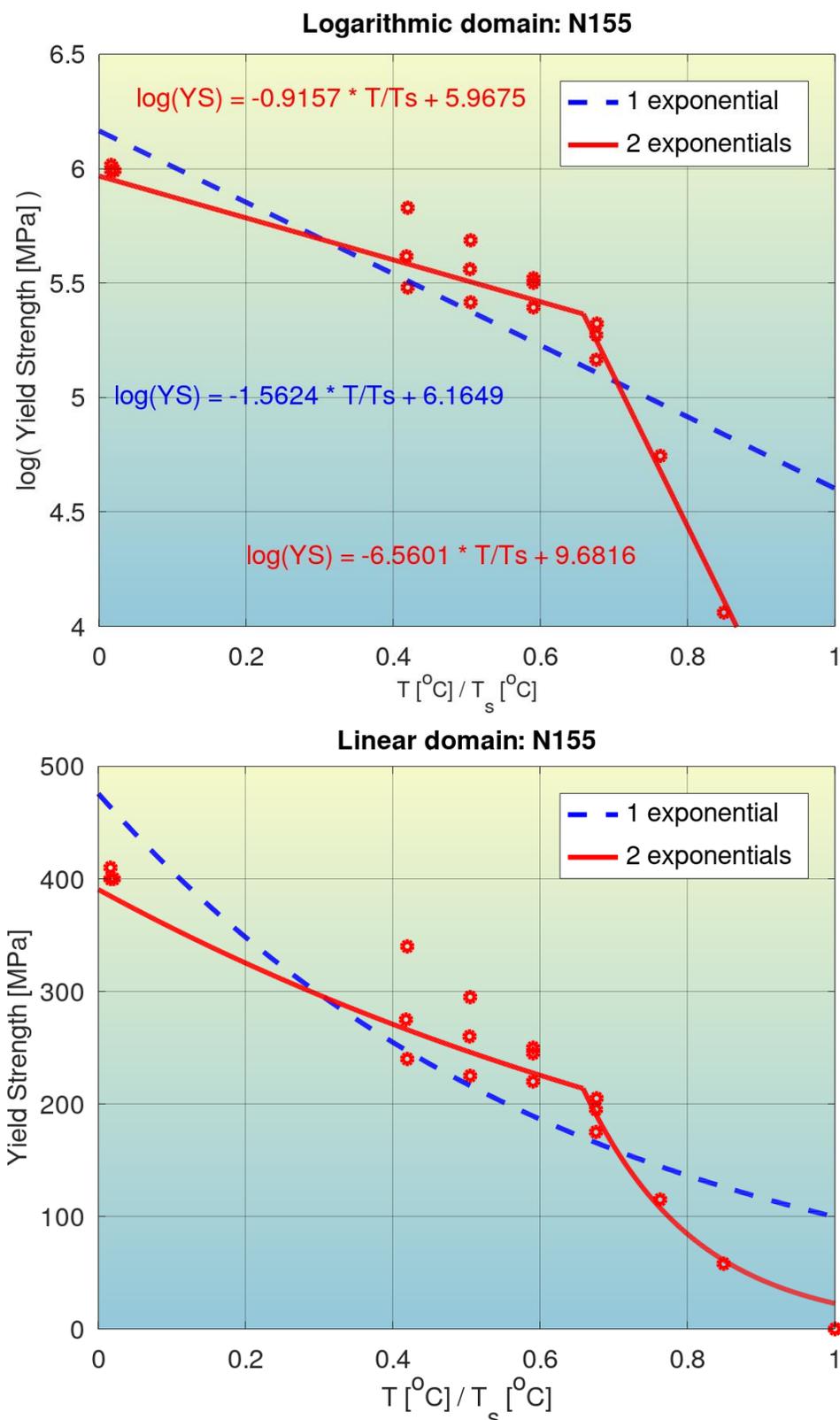

**Figure 43**: Quantification of modeling accuracy of the bilinear log model, for the composition No. 39 from Table S1 (N155), and comparison to that of a model with a single exponential. One outlier has been excluded from the modeling.

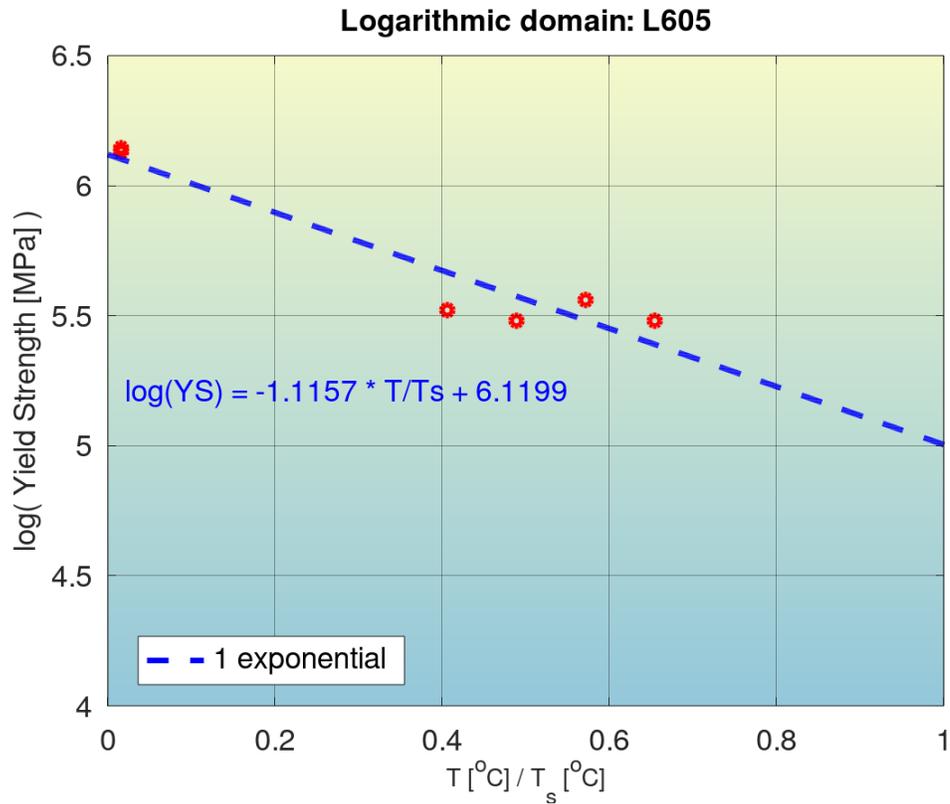

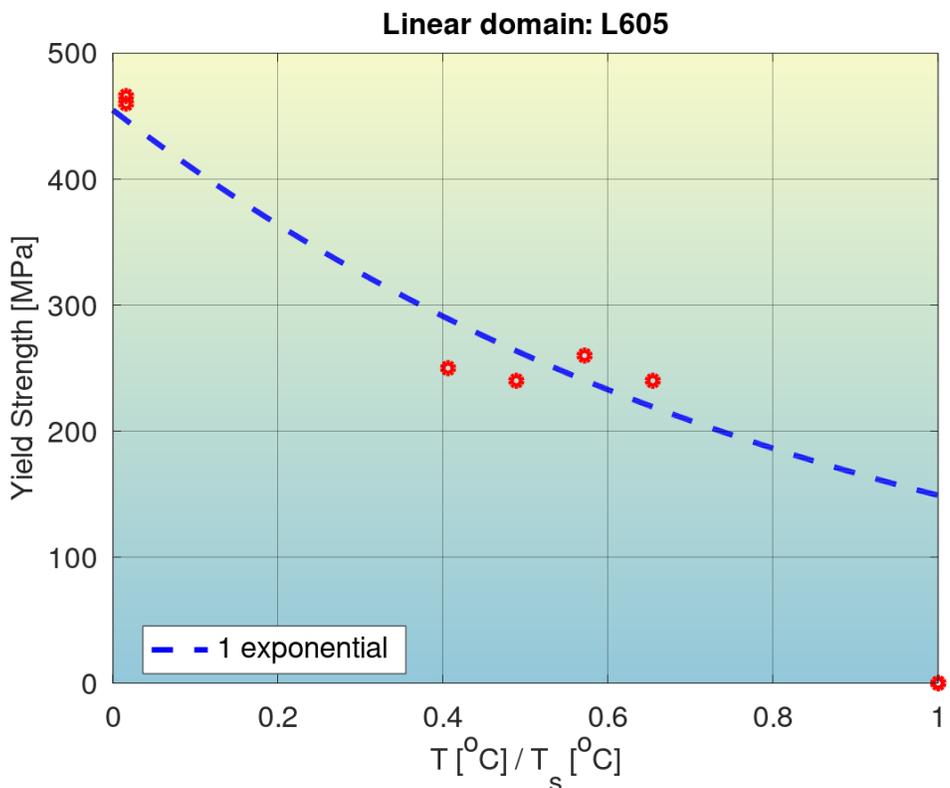

**Figure 44**: Quantification of modeling accuracy of the bilinear log model, for the composition No. 40 from Table S1 (L605), and comparison to that of a model with a single exponential. One outlier has been excluded from the modeling.

**Supplementary Notes**

1. *High-Level Classification of Alloy Materials for High-Temperature Applications*

Figure S1 offers a classification of alloy materials for fossil-energy power plants [40]. Ferrite, or α-Fe, refers to a specific body-centered-cubic (BCC) crystal structure of iron, which is observed below 912 °C (1,674 °F). Along with ferritic, martensitic, duplex and precipitation-hardened stainless steels, austenitic stainless steel is one of the five classes of stainless steels by a crystal structure. The primary crystalline structure of an austenitic stainless steel is the austenite phase [face-centered-cubic (FCC) γ-phase of iron (γ-Fe)], which prevents the steels from being hardenable by the heat treatment, and also makes them essentially non-magnetic. For a general overview of the Ni-base superalloys, refer to the Supplementary Note 2 below. As seen in Figure S1, the Ni-base superalloys Inconel 740, Haynes 230, and Inconel 617 offer superior mechanical properties at elevated temperatures, i.e., the maximum allowable stress, as compared to other engineering alloys [40].

2. *Brief Overview of Nickel-Based Superalloys*

For sample phase diagrams for Nickel-based superalloys, refer to [45].

1. The Gamma Phase (γ)

Most commercial Ni-based superalloys contain the following elements: C, Cr, Mo, W, Ni, Nb, Fe, Ti, Al, V, and Ta. The γ phase comprises the continuous matrix of the superalloy. The γ matrix is a solid-solution disordered FCC austenitic phase of the alloying elements, one that usually contains a high percentage of solid-solution elements (e.g., Co, Cr, Mo, and W). Ref. [46] provides a good overview of the metallurgy of superalloys, including the crystal structures and phases.

2. The Gamma Prime Phase (γ′)

The γ′ phase is an intermetallic precipitate and the principal strengthening phase in many Ni- and NiFe-based superalloys [46]. The intermetallic γ′ phase consists of $Ni_3Al Ni_3(Al,Ti)$ with an ordered FCC $L1_2$ crystal structure [46].

### 3. The Gamma Double Prime Phase ($\gamma''$)

The $\gamma''$ is the principal strengthening phase in Inconel 718 [46]. The $\gamma''$ phase normally consists of the $Ni_3Nb$ or $Ni_3V$ and is typically used to strengthen Ni-based superalloys at lower temperatures (below 650 ºC) relative to the $\gamma'$ phase. The crystal structure of the $\gamma''$ phase is ordered $D0_{22}$ body-centered tetragonal (BCT). The $\gamma''$ phase precipitates as 60 nm by 10-nm-anisotropic discs whose (001) plane aligns parallel to the {001} family for the $\gamma$ phase. Formation of these anisotropic phases is caused by the lattice mismatch between the BCT precipitates and the FCC austenitic $\gamma$ matrix phase. Such a lattice mismatch gives rise to high coherency strains, which contribute to the primary strengthening mechanisms [47].

### 4. The Eta Phase ($\eta$)

Between the temperatures of 600 °C and 850 °C, the $\gamma'$ phase can transform into the HCP ($D0_{24}$) $\eta$ phase. The $\eta$ phase consists of $Ni_3Ti$ (no solubility for other elements) and is found after the extended exposure in some FeNi-, Co-, and Ni-based superalloys with high Ti/Al ratios [46].

### 5. The Carbide Phases

Formation of secondary carbide phases, such as the cubic MC phase, the FCC $M_{23}C_6$ phase, the FCC $M_6C$ phase, or the hexagonal $M_7C_3$ phase, is common in Ni-based superalloys. Carbides present at grain boundaries have been used effectively to inhibit grain growth and grain-boundary sliding at elevated temperatures [48]. However, care must be taken to prevent continuous films along grain boundaries, which can cause the precipitate-free zone phenomenon near the boundaries, and increase notch sensitivity and intergranular cracking [46] .

### 6. The Topologically-Closed-Packaged (TCP) Phases

The TCP phases refer to a family of phases, including the $\sigma$ phase, the $\chi$ phase, the $\mu$ phase, and the Laves phase [46], which are not atomically close-packed. Instead, the TCP phases show closed-packed planes with an HCP stacking sequence. The TCP phases tend to be highly brittle and deplete the γ matrix of solute atoms, needed for the formation of strengthening intermetallics. The TCP phases tend to form as a result of kinetics after long exposure (thousands of hours) at elevated temperatures (> 750 °C) [47].

### 3. *Brief Overview of Cobalt-Based Superalloys*

As noted in [46], Co-based superalloys are usually strengthened by a combination of carbides and solid-solution hardeners.

1. The Gamma Phase (γ)

Analogous to the Ni-based superalloys, the γ phase comprises the continuous matrix of the superalloy. Developmental (i.e., non-commercial) superalloys contain the following elements: C, Cr, W, Ni, Ti, Al, Ir, and Ta [49], [50].

2. The Gamma Prime Phase (γ')

Similar to the Ni-based superalloys, the γ' phase constitutes the precipitate used to strengthen the Co-based superalloys [49], [51].

3. The Carbide Phases

As prevalent with the carbide formation, the appearance of carbide phases in Co-based superalloys does provide precipitation hardening, but decreases low-temperature ductility [50].

4. The Topologically Closed-Packaged Phases

The TCP phases may appear in some developmental Co-based superalloys. But the goal is to avoid such phases.

### 4. *Brief Overview of Iron-Based Superalloys*

Use of steels for applications necessitating superalloys may be of interest, because certain steels have exhibited creep and oxidation resistance analogous to that of Ni-based superalloys, while being much less expensive to manufacture.

1. The Gamma Phase (γ)

Similar to the Ni-based superalloys, the Fe-based superalloys feature a γ matrix phase of the austenite iron with an FCC microstructure [52].

2.   The Gamma Prime Phase (γ′)

As for the Ni- and Co-based superalloys, the γ′ phase is introduced in the form of precipitates to strengthen the Fe-based superalloys.

5.   *Qualitative Overview of Strengthening Mechanisms in Ni-Based Superalloys across Temperature*

Figure S2 and Figure **S3** provide a qualitative overview of the strengthening mechanisms in Ni-based superalloys across temperature. At lower and medium temperature ranges, the superalloys benefit from precipitation strengthening (from γ′ to γ″). With increasing the temperature, there is an earlier onset of precipitate coarsening and often times transformation into metastable phases, or formation of new ones. In turn, the effect of solid-solution strengthening can become more significant, as well as the presence of ideally homogenously dispersed, fine carbides in the matrix and globular, larger ones along the grain boundaries.

6.   *Towards Quantitative Overview of Strengthening Mechanisms in Superalloys across Temperature*

The total yield strength is the outcome of an interplay of the strengthening contributions from mainly the secondary phases and the solute atoms in the matrix. Generally, there is a linear relationship between the high-temperature strength and γ′ volume fraction [53]. Commonly, the volume fraction varies between 25 – 60 %, with modern superalloys being on the upper end of this range. Besides the influence of the volume fraction, finer γ′ precipitates may impart greater strength than coarser precipitates, at a constant volume fraction. However, as the temperature varies, different dislocation-interaction mechanisms may be activated, favoring larger precipitates, with higher lattice misfits.

The γ′-precipitation hardened superalloys exhibit unique positive temperature dependence with respect to the flow stress [54]. Between -200 ℃ and 800 ℃, the flow stress tends to be reversible. This feature is related to the cross-slip of screw dislocations and the associated formation of Kear-Wilsdorf locks on {100} planes [55], [56]. Similarly, lowering of the anti-phase boundary (APB) energy and thermally activated cross-slip on cube planes defines the location of the peak shown in Figure S4. The momentary flow stress and the peak location depend on the solute content (primarily the Al content, thus the volume fraction) [57].

*7. Further Specifics on the Superalloys Listed in the Supplementary Table S1*

2. Udimet D979: Here, there are no grain-boundary carbides of Type M23C6 or M6C due to a low carbon content.

3. Udimet 720: We believe that the volume fraction, the average size of the primary precipitate, and the average size of the secondary/aged precipitate correspond to the standard treatment.

6. Astroloy: Most research here pertains to the powder-metallurgy material.

8. Inconel 807: No data available, from what we can tell.

9. Inconel 802: No data available, from what we can tell.

10. Inconel 801: No data available, from what we can tell.

11. Inconel 800: Most commonly presented as variation 800H/HT, which have higher design stresses allowable by the American Society of Mechanical Engineers (ASME), due to tighter C and Al + Ti contents.

14. Inconel 718: Here, we have chosen Inconel 718 with traditional two-step aging: $720^{\circ}$C for 8 hours and $620^{\circ}$C for 8 hours.

24. Hastelloy C22: We understand that Haynes has been emphasizing the point of age-hardenability with a long-range ordering (LRO) $Ni_2Mo$ phase. LRO has been observed for some intermetallic phases, such as Laves or Pt2Mo. It is believed that the increased volume fraction of the Ni2Mo phase, upon long-term aging, contributes to the improved high-temperature strength of the Hastelloy C22 superalloy.

25. Hastelloy S: We have not identified any conclusive dataset thus far.

27. Hastelloy X: We have not identified any conclusive dataset thus far.

38. Nimonic PE13: Here we seem to be looking at a case of limited publications.

# Supplementary References